\documentclass[nofootinbib,showpacs,showkeys]{revtex4}
\usepackage[utf8]{inputenc}
\usepackage{amsfonts}
\usepackage{amsmath}
\usepackage{amsthm}
\usepackage{amscd}
\usepackage{epsfig}
\usepackage{epstopdf}
\usepackage{graphicx}
  \graphicspath{{/}{figures/}}
  \DeclareGraphicsExtensions{.pdf,.png}
\usepackage{comment}
\usepackage{enumitem}

\newcounter{mycomment}

\SetLabelAlign{parleft}{\parbox[t]{\labelwidth}{\raggedleft\textbf{#1}}}
\setlist[description]{align=parleft,labelindent=0.0ex,labelwidth=7ex}

\newtheorem{Definition}{Definition}
\newtheorem{Theorem}{Theorem}

\newcommand*{\ICHAIN}{\ensuremath{(\mathrm{ITD}-\mathrm{IMF})\mathrm{chain}_{1}}}

\begin{document}


\title{The Chern-Simons current in systems of DNA-RNA transcriptions}


\author{Salvatore Capozziello}
\email{capozziello@na.infn.it}
\affiliation{Dipartimento di Fisica, Universit\`a di Napoli ''Federico II'', Via Cinthia, I-80126, Napoli, Italy,}
\affiliation{Istituto Nazionale di Fisica Nucleare (INFN), Sez. di Napoli, Via Cinthia, Napoli, Italy,}
\affiliation{Gran Sasso Science Institute, Via F. Crispi 7,  I-67100, L'Aquila, Italy.}

\author{Richard Pincak}
\email{pincak@saske.sk}
\affiliation{Institute of Experimental Physics, Slovak Academy of Sciences, Watsonova 47, 043 53 Kosice, Slovak Republic,}
\affiliation{Bogoliubov Laboratory of Theoretical Physics, Joint Institute for Nuclear Research, 141980 Dubna, Moscow region, Russia.}

\author{Kabin Kanjamapornkul}
\email{kabinsky@hotmail.com}
\affiliation{Department of Survey Engineering, Faculty of Engineering, Chulalongkorn University, 254 Phyathai Road,
Bangkok, Thailand. }

\author{Emmanuel N. Saridakis}
\email{Emmanuel_Saridakis@baylor.edu}
\affiliation{Department of
Physics, National Technical University of Athens, Zografou Campus
GR 157 73, Athens, Greece,}
\affiliation{CASPER, Physics Department, Baylor University, Waco, TX 76798-7310, USA}

\begin{abstract}
A Chern-Simons current, coming from ghost  and anti-ghost fields of supersymmetry theory, can be  used to define a spectrum of gene expression in new  time series data where a spinor field, as alternative representation of a gene, is adopted  instead of using the standard alphabet sequence of bases $A, T, C, G, U$.    After a general discussion on the use of supersymmetry in biological systems, we give examples of the use of supersymmetry for living organism, discuss the codon and anti-codon ghost fields and develop an algebraic construction for the trash DNA, the DNA area which does not  seem active in biological systems. As a general result, all hidden states of codon can be computed by Chern-Simons 3 forms. Finally,
we plot a time series of genetic variations of viral glycoprotein gene and host T-cell receptor gene  by using a gene tensor correlation network related to the Chern-Simons current.
An  empirical analysis of genetic shift, in host cell receptor genes with separated cluster of gene and  genetic drift in viral gene, is obtained by using a tensor correlation plot over time series data derived as  the empirical mode decomposition of Chern-Simons current. 
\end{abstract}
\date{\today}

\pacs{11.15.Yc,  11.30.Pb, 87.14gn, 87.14gk}

\keywords{Chern Simons currents; spinor fields; ghost fields; time series; genetic code.}

\maketitle

\section{Introduction}\label{sec:intro}

One of the greatest scientific achievements of present epoch  is to explain  the genetic code of living organisms \cite{Stanley,genetic_code}  and, in particular,  of  viruses \cite{virus_a,virus_gene}. This goal could be realized by using graphical drawing and  biochemical processes for the detection of  gene samples. However, one of the problems in the genetic engineering is the prediction of the gene variation and the representation of the  genetic code alphabet. This issue emerges  in the plotting graphs related to the connection curvature  of the docking processes. Specifically, the docking process is important  in the genes of the protein structure and could be adopted instead of using a very long alphabet notation as the string sequence and the comparison of  the sequences of docking. In this perspective, methods of Quantum Field Theory, and related results,  can be of high interest. In particular, 
the connection \cite{simon} and the  supersymmetry curvature  \cite{witten_int} derived for  ribozymes \cite{ribo} are extremely important issues  to explain both the origin of life \cite{life} and the reason why all the living molecules have only right-hand symmetry. Specifically, the left-hand symmetry  of ribozyme curvature  and HIV gene \cite{ribo2}  are  very active research areas today. The equilibrium between the supersymmetry and the mirror symmetry of the left-hand and the right-hand DNA and RNA, amino acid and nuclid acid molecules can be explained by AdS in the Yang-Mill theory and the Chern-Simon currents in biology as the curvature of the spectrum in the genetic code of the protein curvature. In this perspective, methods and results of General Relativity and Quantum Field Theory can assume a main role  in biology.

Bioinformatics uses  new algorithms to predict the genes in the genetic database of various organisms \cite{alphabet}: such algorithms  match with the metabolism of enzymes as a protein-protein interaction without any unifying equation for the explanation of   living organisms.
  
At present time,   scientists and engineers  study the genetical structure by using the standard alphabet code $A,T,C,G,U$ as a sequence of  strings for the representation of genetic code for various organisms without any exact definition of a new time series of  genetic code \cite{code,code2017} in the standard time series modeling. With this representation \cite{protein}, it is difficult to calculate the genetic variation \cite{h1n1_variation,h7n9} and to perform some intensive calculations by algebraic and geometric tools within a self-consistent mathematical theory \cite{supermath,superla,bv}, namely in the context of  superstrings, M-theory and G-theory \cite{g, alireza}.

 Several researchers are still performing empirical data analysis of the genetic variation \cite{amino} and are detecting the pattern matching over the gene sequence by using algorithm over an alphabet code of the genetic code as their time series representation. One of the main problems in the virology and the time series prediction is how can we predict the genetic variation and the gene structure in the viral particle and other organisms. In the context of   new representation,  the issue is how  we can explain the intuition behind the definition of a new time series data of gene, involved for example in    the Batalin -Vilkovisky (BV)-cohomology \cite{bv} of DNA, and the viral gene structure.
 The Chern-Simons current and the anomaly \cite{anomaly} over the superspace \cite{chern} of  cell membrane can be applied to diagnose a new gene disease, the cloning technology and the gene therapy for using a virus as a vector to cancel an HIV disease \cite{cdot}. As we will see below, the method can be improved also in view of describing the trash area of DNA.

In the modern researches of  algebraic geometry \cite{Massey} for  time series data, there exists another approach which uses the spinor field \cite{bv,e8} in the Kolmogorov space of the time series data \cite{Kolmogorov} over the genetic code to represent the gene structure as the ghost \cite{ghost,witten} and the anti-ghost fields of the codon and the anti-codon. This is achieved  in the frameworks of  supersymmetry \cite{super,super2,super3} and  G-theory \cite{anomaly,g}. Result of the construction show that all the calculations over codon can be assumed as a new superspace of the time series representation of the gene structure \cite{superpoint,Kolmogorov} .

 Here, we introduce a new representation of the genetic code in the  time series for a string and a D-brane \cite{string5,Pincak10,pincak11,pincak12} modeling  by applying a spinor field to a superspace in a time series data \cite{Pincak13,cohomo7} over the viral glycoprotein \cite{v3} and a viral replication gene \cite{viral_replication}.
 The method allows to develop supersymmetry for living organisms. In particular, it is possible to control the anomalies in codon and anti-codon ghost fields and construct an algebraic approach for the trash DNA, which is considered one of the big puzzle of modern biology. Concrete applications on specific genetic codes are developed.

The paper is organized as follows. Section II is an overview of the present multidisciplinary approach. We  sketch briefly the problem of transcription and translation in biological systems as well as the essential feature of supersymmetry. In Section III, we specify the basic definition of the spinor field in the genetic code and how the concept of an exact sequence of the cohomology is related to the data in a time series of a chemical reaction between 2 proteins.  We define a Chern-Simons current for biology in a time series data of codon and anti-codon by using extra-dimensions of an underlying topological space. We derive also  a new  master equation that can be used for a V3 loop in an HIV attach to the CD4 gene of host cell. We fix real values of 64 codons in the Chern-Simons current by using an explicit formula which we obtain by using the  well-known  theoretical physics adopted for strings. Furthermore, we algebraically discussed the so called "Trash" DNA that biologists cannot frame in  the standard scheme of protein production. In Section IV, we develop the computation of all codon hidden states  by the Chern-Simons 3-forms in DNA.   The specific methodology is discussed in  Section V. Here  we calculate the Chern-Simons current over a time series of the genetic code of V3 loop with 3 species of the HIV virus and 3 species of CD4 in the rabbit host cell. The results of the calculation are obtained by using the data analysis of $\ICHAIN(n)$ \cite{Kolmogorov} and considering  a time series in terms of quaternionic projective space. We  find an evolution feedback between an antigene drift in V3 gene and an antigene shift in the CD4 gene from. In Section VI, we discuss the results of the proof and consider about a time series in terms of the quaternionic projective space over the V3 and the CD4 gene. Discussion and conclusions are reported in Section VII. In Appendix A, we provide a source of the time series of the V3 gene  adopted for the  HIV virus.

\section{Supersymmetry and biological systems: an overview}

Alan Turing, who  invented the first computer \cite {bio},   conjectured  that  biological systems and living organisms  can be fully explained by using algebraic operators  coming from  theoretical physics \cite{bio}.  For example,  he had in mind some quantum computer where DNA could be used as a permanent memory storage medium. 
In order to model out this kind of models, an approach could be   to adopt the {\it sheaf cohomology}  for the resolution  of 
basis coming from local sections of DNA, RNA, proteins, cells, tissues  and living organisms  in  the framework of  the so called  {\it Grothendieck topology} \cite{grothen} over predefinite  energy states.  For example an autonomous transition state for an adaptive behavior with respect to a given  environment can be defined by using  spinor fields for a gene expression state and   ghost  and anti-ghost fields for hidden states  over cells of living organism framed in a $E_{8}\times E_{8}$ unified theory coming  from  D-brane and string theory.  

For example, Chern-Simons currents, induced from system of DNA-RNA transcriptions,  are interesting phenomena adopted in  gravitational physics which could have interesting counterparts also in biology. This  "gravitational" analogy  could have impact  in human immunosystem \cite{immune} giving rise to a useful   representation of codons in human genome.
The modified gravitational field,  derived from the  Chern-Simons currents, can be useful in biology to explain the source of   connections  over  protein-docking states: in this perspective  adopting  cohomology in biology can be useful as a  new tool for plotting genes with spinor field in time series data.  In particular, the trash area of DNA, with repeated inactive genes, can be  represented by  Chern-Simons currents with extended 
structures of knot states in a Laurent  polynomial of knots: such a structure  is   predefined in an equivalent class, modulo hidden states of genetic code.

The canonical structure of inactive genes can be a partition of Coxeter numbers $h$ \cite{coxeter} in a repeated transition  of hidden states  in a gene wave function. This current can be given  in complex numbers starting from the below analysis of hidden currents and can be the answer of the fact that we have   non-active states.
In an equilibrium state of undocking  protein-protein interactions, antigene and antibody, induced 
from a system of DNA-RNA transcription, cannot successfully defense with respect to  HIV infection of undocking equilibrium states. In order to figure out these situations,  we use a new model,  based on analogue  gravitational effects, where    sheaf cohomology and BV cohomology are considered in biology. 
The equilibrium state  of HIV docking with a host cell can be realized as the curvature in left and right supersymmetry.  
 An exact sequence of equilibrium states of protein docking is  induced by the infinite sequence of cohomology over a
biological system from the local section of the gene expression in the  global section of the immunosystem. The Mayer-Vietoris sequence \cite{mayer} of relative cohomology can be used to explain to  constant chemical  equilibrium  and to induce a hidden space of free energy in unoriented supermanifold: the model explains the   protein folding  inserting, by a copy and paste mechanism,  hidden states of  gene in the trash DNA.

It is possible to demonstrate  that this puzzle of feedback loop from docking to undocking states    with hidden directions of interaction can be represented by  8 hidden states in a $E_{8}\times E_{8}$ model of gene considering  spinor fields.   The approach consists in adopting both the 
cohomolgy theory  and the Chern-Simons theory with  unoriented extradimensional supermanifold to study  the supersymmetry  of  hidden states in   time series data of spinor fields.
 The duality of unoriented supermanifold  over living organisms with extradimensions can 
 break the  chiral supersymmety over the {\it central dogma of molecular biology} \cite{dogma}. The superpace of DNA and the genetic code  can be modeled as a  loop space with  the central dogma sieve  in the Groethodieck topology as the adjoint 
 left- right group action over the Hopf fibration of gene expression. With this definition in mind, we can explain inactive area of trash DNA by using the feedback loop in unoriented extradimension to
change  the protein docking from non-equilibrium to equilibrium  state considering, for example,  the  receptor protein with HIV   viral glycoprotein.

In biology, this analogue  gravitational field  could contribute to explain the origin of life as a "gravitational"   effect  to the 
 immunosystem and the HIV. Ghost  and anti-ghost fields  are the   source of  connection over 
 spinor fields in genetic code representation of codon and anticodon where  hidden states of gene are contained in a unitary theory.

\begin{figure}[!t]
 \centering
\includegraphics[width=0.5\textwidth]{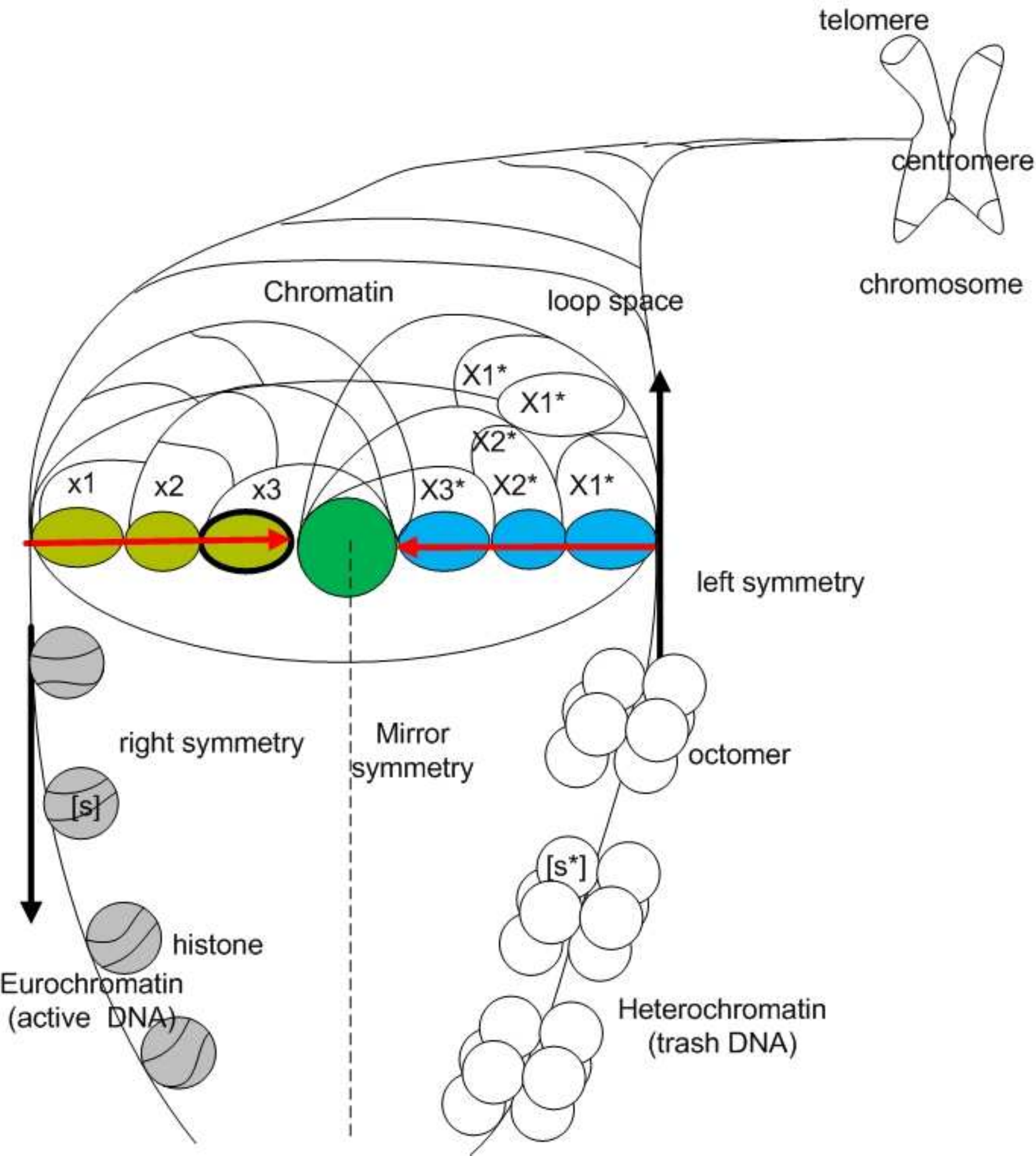}
 \caption{ The picture shows the loop space in time series data of genetic code as the DNA in chromosome. The loop space comes from the structure of chromatin, the  so called heterochromatin. The octomer is a source of hidden 8 states of spinor field in time series data. The octomer area  is a region  of trash DNA.\label{fig1}}

 \end{figure}
 
One of the great discovery  of last decades was the  central dogma. Specifically, it states that the  mapping
 to explain transcription and translation processes is one way. Genetic codes are encoding 20 amino acids with start and stop alphabet into 64 codons with only  3 alphabets  over the genetic code. Up to now,
 there still exists many  open questions that need   to be answered: for example,  why do all organism share a common molecular structure of cell and DNA based sequences where they share  the same codon table with only 3 alphabets, in common, of  20 amino acids only? how to construct or built up a new living organism from these alphabets?  And why 23 pairs of human chromosomes, and not  24 pairs, with the   lack of one pairs in XY?   Might the last  one be a dual state of pairs of chromosomes   in order to allow the presence of XY and all 23 chromosomes?
In present time, scientists \cite{alphabet} try to synthesize artificial living organism in laboratory starting 
 from  bacterial DNA.  Nucleic acids  and amino acids in living organism exist only in right-handed  symmetry of chiral molecule, no left-handed symmetry and no more than 20 amino acids,  in nature, can  become a living organism in a cell including parasitism living-like organism or viral particles. Recent studies found that  the left-handed  symmetry can be produced from $X-Y$ synthetic pairs based on a  new codon table of 4 synthesis  alphabets. This  left-direction amino acids, built up by synthesis,  still cannot become  real living organisms according to cloning technology.

Furthermore,  biologists try to  understand the origin of life  by using evolution mechanisms induced by the integration of  genes of  all species and gene expressions in common patterns of unify genomic superspace. They found that plant and human contain a lot of trash or junk DNA and contain  only 2\% of genes in the active area of DNA from all chromosomes.  In order to answer why almost $98\%$ of human genome and all living organism have no function in DNA  is today one of the  most active research area. Before the advance of nanotechnologies and quantum theory, researchers realized that the  trash area of DNA  can have some functions like the structural DNA    in epigenetic theory of proteins. From recent studies, it comes out that the  trash area is  involved with the causes of cancer, viral infections, gene diseases etc. Scientists   found new functions for  this area and understood  that this is not junk DNA  from the  discovery of transposon and retrotransposon. This area is also useful for authenticate the signature of  different organisms in the same specie with different amounts and repeated pattern of inactive gene. Some inactive gene can be also  similar gene  to other species in active region. 
One of the most important functional area in trash DNA is the {\it telomere} \cite{telomere}.  For example, ancer cells are spread though other infected cells to generate   extra telomerase activity
 than normal cells by using the mechanism of inserting and deleting inactive gene: this is the  so called transposon and retrotransposon. It was therefore proposed that cancer or HIV infection might be treated by protein transcription from the telomere in trash DNA. 
Some inherited diseases are now known to be caused by telomerase defects, such as  the anemia and inherited diseases where  the function of trash DNA is evident. Therefore this junk DNA is not a junk anymore as biologists stated in the past.
 On the other hand, it is a source of various types of hidden states for control area of active gene and it can be useful to develop new  drugs  for cancer  treatment.
 If this area is malfunctioning, it will causes also all inherited diseases and all types of cancers including HIV and other viral diseases. 
 Recently, scientists discovered   various  nonfunctional RNA more than mRNA, sRNA and tRNA. Also some  
 functional RNA with enzymatic property like the so called ribozyme, RNAi, miRNA have been discovered \cite{rna}.
 They react like retrovirus or  retrotranspon where    points of DNA sequence dock for transcription or 
insert their genome copy into the trash area.The mechanism  inserts and deletes genes and causes that  genes repeatedly  jump transition 
state into some feedback control loop of the new gene node. Any node  can be in a gene tensor network that can be used  to treat cancer.
 All these fact are evidences that the central dogma rule can be broken.
 Furthermore, trash DNA can work as the hidden part of DNA that reacts with the DNA active part by indirect structural proteins, the so called histone proteins:   for these systems, the hidden state can evolve into the the  anti-self dual 2 forms of hidden states in trash DNA.
To shorten the  notation,  we will use $D$ for DNA, $R$ for RNA, $P$ for protein in this paper.

 In molecular biological theory of living organism, the source of codon in central dogma might be involved with retrotransposon and transposon. It is a hidden transition state of gene to loop  back in feedback path of central dogma: this gives rise  to  recombinations again and again as an  infinite cohomology sequence. The cohomology for biology is a source of  partition functions in docking and undocking states of proteins and genes.  
 
For example,   retrotransposons can be represented as ghost  and anti-ghost fields of adjoint functors in sheaf cohomology of gene expression in equilibrium state.  The docking of retrotransposons can induce feedback loops in the  space of genes of the time series of the DNA sequence. In particular,  trash DNA can be  the source of various  transition states  of gene from  ground states to excited states,  or the signal communication between cells and  organisms of different species. An active gene coming from a species can be a transposon part into  other species being the  transition state  between 2 hidden states of the  genome. When this area is malfunctioning in human genome,  cancer, viral inflections and other diseases come out.

The  ground state of inactive gene in the loop space of time series data of genetic code is represented in  Fig. \ref{fig1}.
The  active part is jumping into this area  in form of transposon and retrotransponson state of inactive gene.  It  can give rise to a  transition back to to the active part for replication processes. For example, the  HIV transposon can be  used for reversed transcription into the DNA of a host cell.  This means that  HIV can break the supersymmetry of central dogma with infinite loops of $0_{D-brane}\rightarrow\mathcal{O}_{{R}^{\ast}} \rightarrow \mathcal{O}_{D }\rightarrow \mathcal{O}_{R^{\ast}}\rightarrow 0_{anti-D-brane}$ and $ \mathcal{O}_{R^{\ast}}\rightarrow \mathcal{O}_{D}\rightarrow \mathcal{O}_{R}\rightarrow \mathcal{O}_{P}$. This  recursive loop can destroy the core central dogma process of host cell until the host cell dies.
 
All transitions are induced from the 64 codons present in the 3 alphabets as triplet states in the genetic code. The puzzle is  to answer why, in  livings organisms,  only 20 amino acids    with right symmetry exists for translation processes without any   left symmetry. For example HIV contains 2 identical RNA genomes with identical  genetic code. In our model,  we will demonstrate   that  HIV is a bound isolated state  of retrotransposon 2 forms
  with its dual $[s_{6}]-[s_{8}]$. When the pair is isolated in human DNA, it  fuses to DNA in reversed direction by a wrap state with unoriented supermanifold of living organism. It gives rise to an asexual reproduction with singlet inactive states of transposon area in trash DNA. 
  This  trash DNA interacts with  enzymes to insert some transposon or
 retrotransposon with  twistor from  right to left symmetry and backward  into DNA  knot states then it reconnects the hidden states of copies of repeated components of transposon and retrotransposon. This 
transition element  communicates to the new gene tensor network a signal between  species.  The  evolution of the spinor field 
is a gene communication protocol for the transition between different species:  for example it can give rise to the 
 adaptation to the environment  to allow that species to  survive.  
 
 Furthermore, physiology of human  junk
  DNA area  is one of the active research both in molecular biology  and in quantum computing. Hidden state of time series data of DNA can be
 realized as an underlying spherical $S^{7}$ manifold of octomers: 8 histone proteins induced an   entanglement state in trash  DNA area. It is a source of 
loop space in spinor fields of time series data of genetic code. This phenomenon is analog to the storage of memory in the DNA genetic code.
The  histone protein can be realized as  a M\"obius map of octomers, i.e. 8 spherical histone proteins 
as sources of $[s_{i}^{\ast}], i=12,3,\cdots $, 8 dual states of spinor fields of time series over the superspace of DNA. It  reacts as a passive feed back loop of hidden control protein layers induced from twist pairs of hydrogen bonding between 3 types 
of spinor fields in time series data of central dogma. The equilibrium of  docking between protein, RNA and DNA comes from  twistors in the  hydrogen bonds. They can be realized to control equilibrium point in  the central dogma. This situation can be explained by using  the   supersymmetry of central dogma. It is
  a  cohomology sequence like $0_{D-brane}\rightarrow \mathcal{O}_{P^{\ast}}\rightarrow \mathcal{O}_{D }
\rightarrow \mathcal{O}_{R} \rightarrow \mathcal{O}_{P} \rightarrow 0_{anti-D-brane}$ which can become  infinite.  

\begin{figure}[!t]
 \centering
\includegraphics[width=0.5\textwidth]{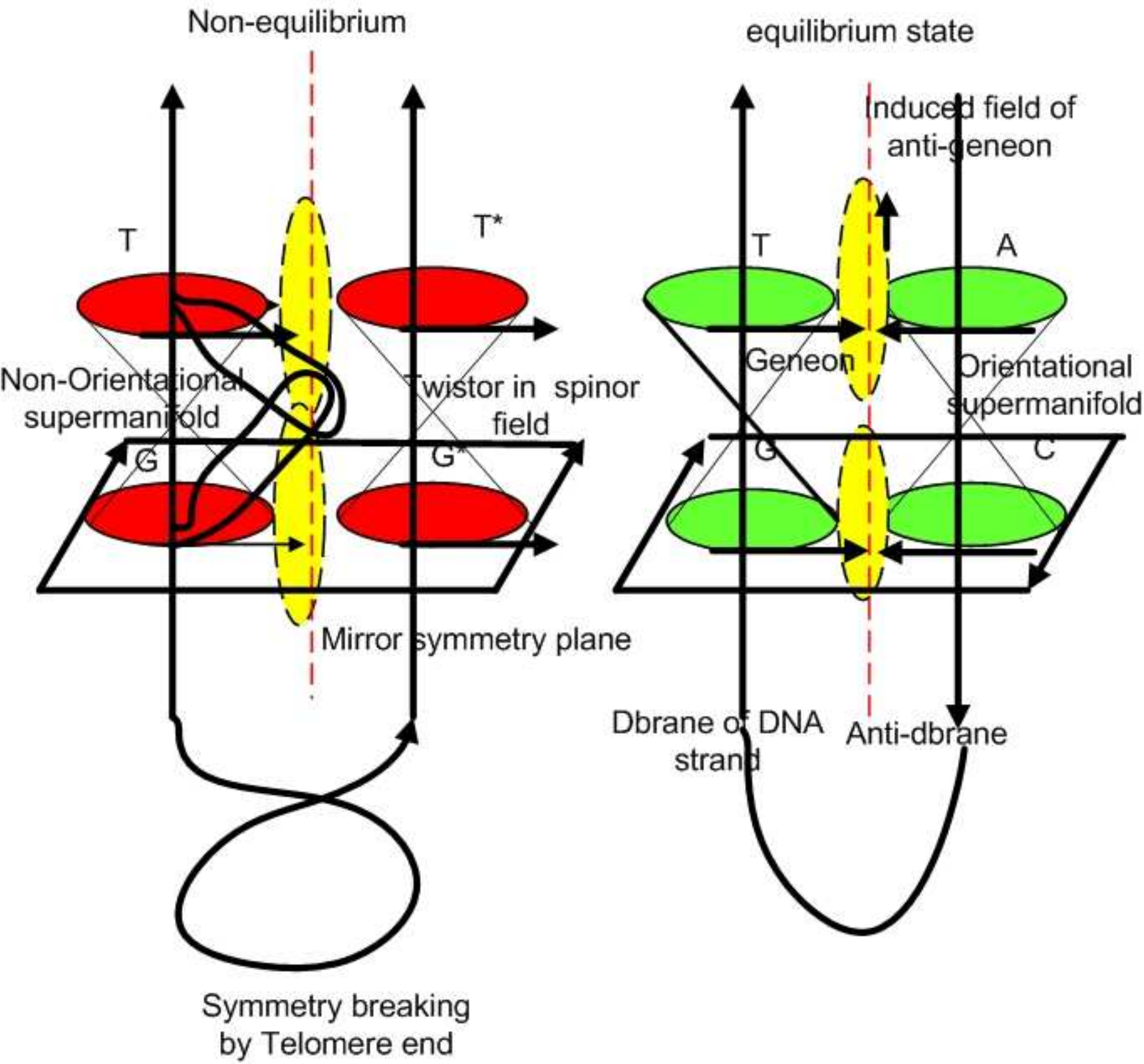}
\caption{ The interaction of geneon and anti-geneon as a D-brane and anti-D-brane model of protein docking and undocking states induced from the adaptive behavior of gene expression shown in the left drawing. When DNA twists by telomere, they glue 3 ends and twist back to 5 ends.  DNA results in the  parallel plane of D-brane and anti-D-brane in both passive plane of geneon-anti-geneon in the undocking feedback loop. In the   docking, the   curvature is changing  with nonactive gene expression of retransposon and transposon in an unoriented coupling related to the  transition of twisor pairing  in trash DNA. See  the right drawing. \label{fig_geneon}   }
 \end{figure}

\begin{figure}[!t]
 \centering
\includegraphics[width=0.48\textwidth]{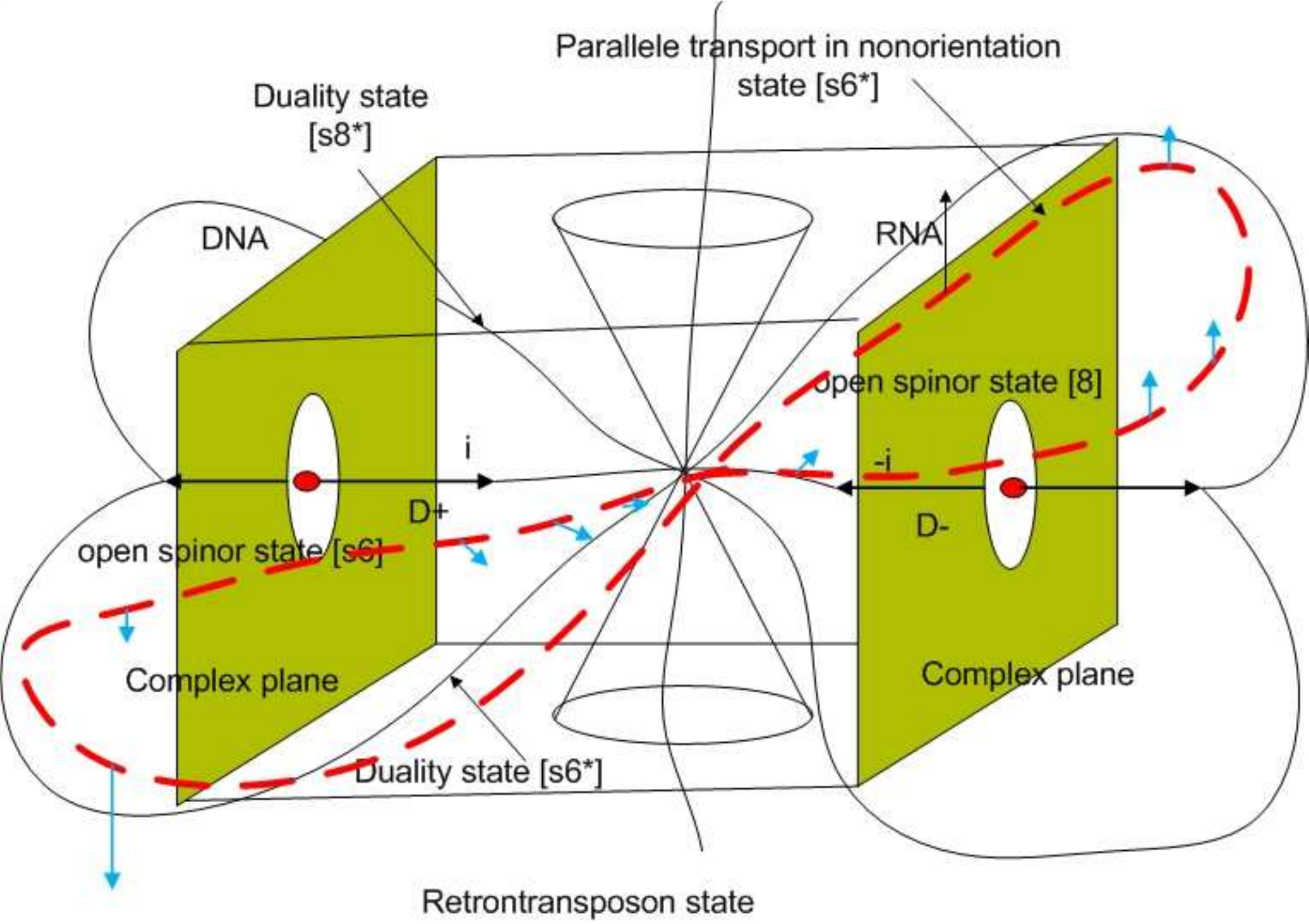}
\includegraphics[width=0.48\textwidth]{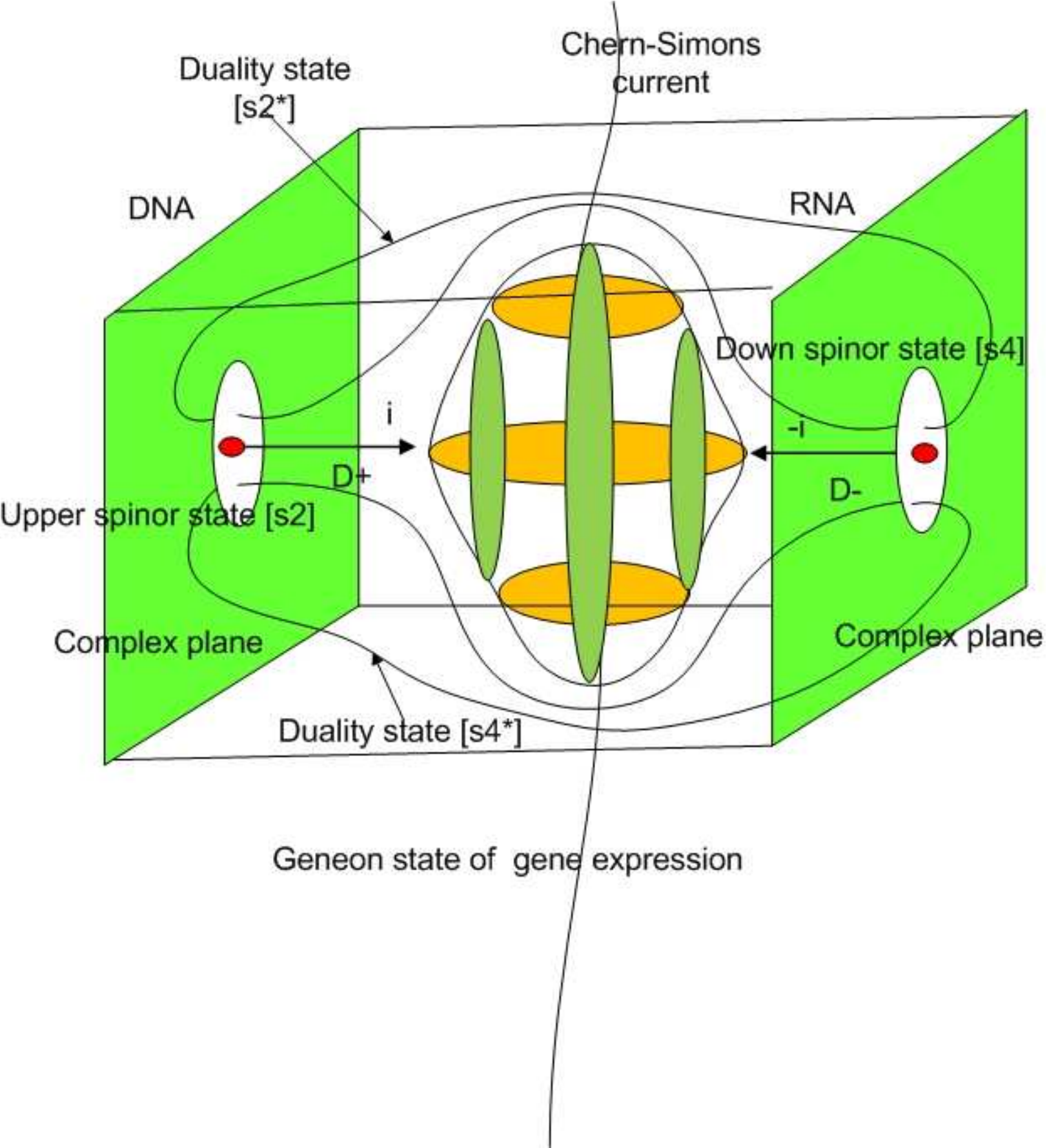}
 
\caption{ On the left,  the interaction of  retrotranspon can be visualized as the entanglement state in the loop space of time series data with extradimension over the Hopf fibration. We notice that the induced Chern-Simons current in this area is quantized into non-separable fiber of covering space of links in  knots of transposon twistor form as self dual field. The twist plane of extradimension induces a spinor field as the
hidden state of retrotransposon around the octomer, i.e. the  8 histone protein where the  junk DNA  curls around the spherical Hopf fibration shape. This model  is a fiber of non-oritented supermanifold over the superspace of time series data of the genetic code in the active area. The twistor shape  of M\"obius strip of parallel transport of free electron trajectory is a  model of quantization of hydroghen bonding inside pairs based in  the central dogma objects. On the right, the orientation area of  fiber
space of DNA, it is an active part of the state with spherical shape of a single histone protein in the active zone of DNA, that is  the  
2\% area of gene expression  in the  human genome. We notice that the induced Chern-Simons current in this area is quantized into separeted fiber in the covering space of the sphere. \label{fig_retrotransposon1}   }
 \end{figure}

In summary,  supersymmetry seems to be a very powerful tool to describe structures and evolutions of biological systems, first of all in relation to the DNA dynamics.
In particular, DNA can be  realized as a loop space with associate open string or interaction between D-brane and anti-D-brane of protein-protein interaction in docking and undocking state of  information flow  in  genetic code by using central dogma as short cohomology exact sequence.  The mathematical approach can be a
  $E_{8}\times E_{8}$ unified theory  for time series data. 
  It can be proved the  existence of hidden 8 states  of time series data, that is  $[s_{i}],i=1,2,3\cdots 8$. Here $[s_{2}]$ is the maximum state and $[s_{4}]$ is the minimum state. These 8 states are optimized state and exists in 8 dimesions on $S^{7}$, the manifold of  spinor fields in time series data.
 The  algebraic construction of trash DNA is based on the Kolmogorov space of time series data related to the underlying
 genetic code  in  modified Wilson loop and gravitational field of Chern-Simons current. We can use a  chart  over the homogeneous coordinates in tangent and co-tangent space of the unoriented supermanifold \cite{super} of living organism. This allows us  to construct the Hopf fibration of $S^{7}$ as a fiber  over the  Kolmogorov space $X_{t},Y_{t}$ with spin invariant CPT property of Dirac  matrix $\gamma_{5}$ for left and right chiral supersymmetry. 
  The codon and anti-codon  span a tangent space starting from  basis and dual basis of predefined  8 states$s_{1},s_{2},\cdots s_{8}$ (we call closed shell, s-orbital of spinor fields in time series data) with their dual hidden state of trash DNA in dual tangent space (we call p-orbital of spinor field in time series data for open shell.)
To explain the source of Chern-Simons current in genetic code, we use the   de Rahm coholomogy of central dogma as a main tool. It  comes from entanglement states of fields that give rise to  transcriptions and translations.
  The connection over the  DNA fiber space, denoted by $A_{i}$ (as in  gravitational field in Chern-Simons theory), is with respect to an equivalent class of evolution of  genetic variations. It gives rise to   the  curvature in  proteins.  
  The cohomology of integration process allows to insert and delete genes of transposon and intron over all species in living organism including viral particles (see Figs. \ref{fig_geneon}, \ref{fig_retrotransposon1} for details of induced spinor fields).
  
  After this quick and incomplete overview pointing out how supersimmetry can be adopted to describe biological systems, let us consider in details the role of spinors in the time series of genetic code.


\section{The role of  spinor fields in the time series of  genetic code}

Let $\mathbb{H}P^{1}$ be a quaternionic projective space. Let $H_{0}(X_{t})$ be a pointed space of DNA alphabet sequence  with $[A],[T],[C],[G]$ as an equivalent class of $[A],[T],[C],[G]\in H_{0}(X_{t,DNA}):= \Phi_{i}(X_{t})$ a ghost field with parity 2 with
$ H_{0}(X_{t})=H_{0}(X_{t},\ast)$ where $\ast=\{ \ast \}$ is a pointed space.
 We define an equivalent class of DNA-RNA translation processes by using the notation of a master equation for an interaction between the viral RNA and the host cell DNA by $\{x_{t},y_{t}\}=\{DNA,RNA\}$. The whole state space model of the viral replication cycle, embedded in the host cell, is denoted by $X_{t,DNA}/Y_{t,RNA} = Z_{t,GENE}=\mathbb{H}/\mathbb{H}=\mathbb{H}P^{1}$ as a moduli state space model with the definition of the genetic code as an equivalent class of the map $\alpha_{t}: X_{t,DNA}\rightarrow \{ [A],[T],[C],[G] \}\subset \mathbb{H}$, the host cell gene alphabet is defined by a hidden state space $X_{t}$ with the gene $\beta_{i}$
\begin{equation}
[A]_{DNA}:=[e^{\frac{i\pi \beta_{i}}{2}}]+[0] \mathbf{i}+[0]\mathbf{j}+[0]\mathbf{k}, \nonumber
\end{equation}

\begin{equation}
[T]_{DNA}:=[0]+[e^{\frac{-i\pi\beta_{i}}{2}}]\mathbf{i} +[0]\mathbf{j}+[0]\mathbf{k}, \nonumber
\end{equation}

\begin{equation}
[C]_{DNA}:=[0]+[0]\mathbf{i}+[e^{i\pi\beta_{i}}]\mathbf{j}+[0]\mathbf{k}, \nonumber
\end{equation}

\begin{equation}
[G]_{DNA}:=[0]+[0]\mathbf{i}+[0]\mathbf{j}+[e^{i2\pi\beta_{i}}]\mathbf{k}.
\end{equation}

In the retroviral RNA (in some virus is viral DNA, but the theory is analog with each other) of the observed state space $Y_{t,RNA}$ is a span by gene $\alpha_{i}$ with the anti-ghost field
$\Phi_{i,+}(Y_{t,RNA})$ an anti-ghost field of viral particle.
We define a pair of  ghost and  anti-ghost field genes by a middle hidden state in mRNA and ribosomal EPA state in codon and anticodon state as the ghost and anti-ghost fields in the genetic code.
Let us define a mutual genetic code as passive or dual hidden states $[s_1]^{\ast}-[s_8]^{\ast}$ and active 8 states $[s_1]-[s_8]$ for the spinor field in the genetic code by
\begin{equation}
[A]_{tRNA}:=[NU]_{mRNA}=[e^{\frac{j\pi \alpha_{i}}{2}}]+[0]\mathbf{i}+[0]\mathbf{j}+[0]\mathbf{k}, \nonumber
\end{equation}

\begin{equation}
[U]_{tRNA}:=[NA]_{mRNA}:=[0]+[e^{\frac{-j\pi\alpha_{i}}{2}}]\mathbf{i}+[0]\mathbf{j}+[0]\mathbf{k},\nonumber
\end{equation}

\begin{equation}
[C]_{tRNA}:=[NG]_{mRNA}:=[0]+[0]\mathbf{i}+ [e^{j\pi\alpha_{i}}]\mathbf{j}+[0]\mathbf{k},\nonumber
\end{equation}

\begin{equation}
[G]_{tRNA}:=[NC]_{mRNA}:=[0]+[0]\mathbf{i}+[0]\mathbf{j}+[e^{j2\pi\alpha_{i}}]\mathbf{k}.
\end{equation}

The reversed transcription process of the gene expression is defined by a moduli state space model of a coupling spinor field between the gene of a viral particle and the host cell,
\begin{equation}
 \mathbb{H}P^{1}=X_{t}/Y_{t}\ni [1,\frac{e^{2\frac{\pi}{4} n i \alpha} }{e^{2\frac{\pi}{4} m j \beta}  }   ],m,n=1,2,3,4 =[1,\frac{q|_{DNA}}{q^{\ast}|_{RNA}}]=[\frac{q|_{DNA}}{q^{\ast}|_{RNA}},1] .
\end{equation}

\begin{Definition}
Let $Sp(1)\rightarrow S^{7}\rightarrow \mathbb{H}P^{1}$ be a Hopf fibration of  $8$ states of the genetic code  $[s_1],[s_2],\cdots [s_8]\in S^{7}=T_{p}\mathcal{M}$, denoted by $[s_1]^{\ast},[s_2]^{\ast},\cdots [s_8]^{\ast} \in T_{p}^{\ast}\mathcal{M}$ states of the genetic code of the space of a viral RNA $X_{t}$ and a space of host cell DNA, $Y_{t}$.
\end{Definition}

\begin{figure}[!t]
\centering
\includegraphics[width=0.35\textwidth]{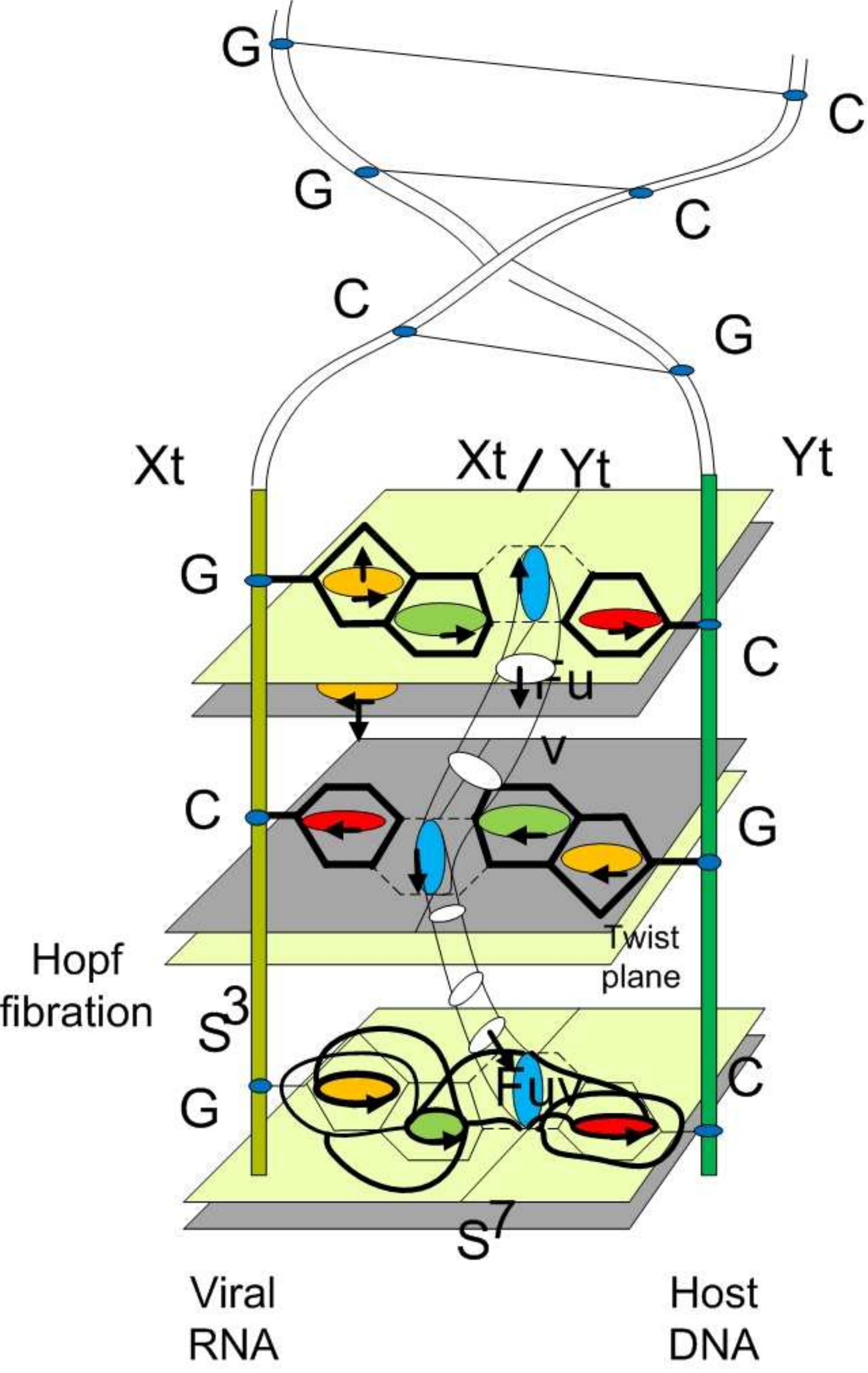}
\caption{ The picture shows the   Hopf fibration $S^{3}\rightarrow S^{7} \rightarrow \mathbb{H}P^{1}$.
The $S^{7}$ is a covering space where there is a tangent space of manifold. It is a chemical structure of right hand isomer of nucleic acid. In this figure, we show the connection between C-G.
In our model of invariant of algebraic topology, we have 2 genus (hole of carbon ring and molecule ring) of surface in G and one genus in C with the other coming from an hydrogen bond of bridge between C connected to G. We can model with principle bundle of Hopf fibration with $S^{3}$ group action of spin to base space of manifold of projective space $\mathbb{H}P^{1}$ of genetic code  which we define in previous section. }
\label{hopf}
\end{figure}

\begin{Definition}
Let $\mathcal{U}_{[A]_\alpha} \subset \mathbb{H}P^{1}$ be a chart of a local coordinate in a manifold of a genetic code over $X_{t}/Y_{t}$, where $[A]_\alpha$ is defined over the right-hand isomer genetic code $\{A,T,C,G\}$ (we use a symbol G also for U) with their dual $[A]_\alpha^{\ast}$, with the mirror symmetry of a genetic code (as an isomer on the left-hand of a nuclid acid) $\{NA,NT,NC,NG\}.$ We have a cycle and a cocycle of an orbifold as a trivialization over the tangent of the living organism manifold, so-called codon and anti-codon $\mathcal{U}_{i}\cap \mathcal{U}_{j}\cap\mathcal{U}_{k}.$
Let  $(\mathcal{M},g)$ be a living organism manifold  with $\mathcal{M}=\mathbb{H}P^{1}$ (in virus, we called a Calabi-Yau space, see Fig. \ref{hopf}) for a living organism with the Riemannian metric tensor $g_{ij}=<T_{[A]_{[A]_\alpha}}\mathcal{M},T_{[A]_\alpha}^{\ast} \mathcal{M}>$ over a tangent manifold and a cotangent manifold, $T_{[A]_\alpha}\mathcal{M}=S^{7}$ with $g_{ij}$ defined as a tensor behavior field transformation between $8\times 8=64$ states in codon as the distance in a space of the genetic code with $i=1,2,3,\dots 8 ,j=1,2,3,\cdots, 8$. The active states in genetic code are denoted as $[s_{1}],[s_{2}],\cdots
[s_{8}]$ and the hidden states inside 8 states or passive states are denoted as $[s_{1}^{\ast}], [s_{1}^{\ast}],\cdots [s_{8}^{\ast}]$.
\end{Definition}

\begin{figure}[!t]
\centering
\includegraphics[width=0.45\textwidth]{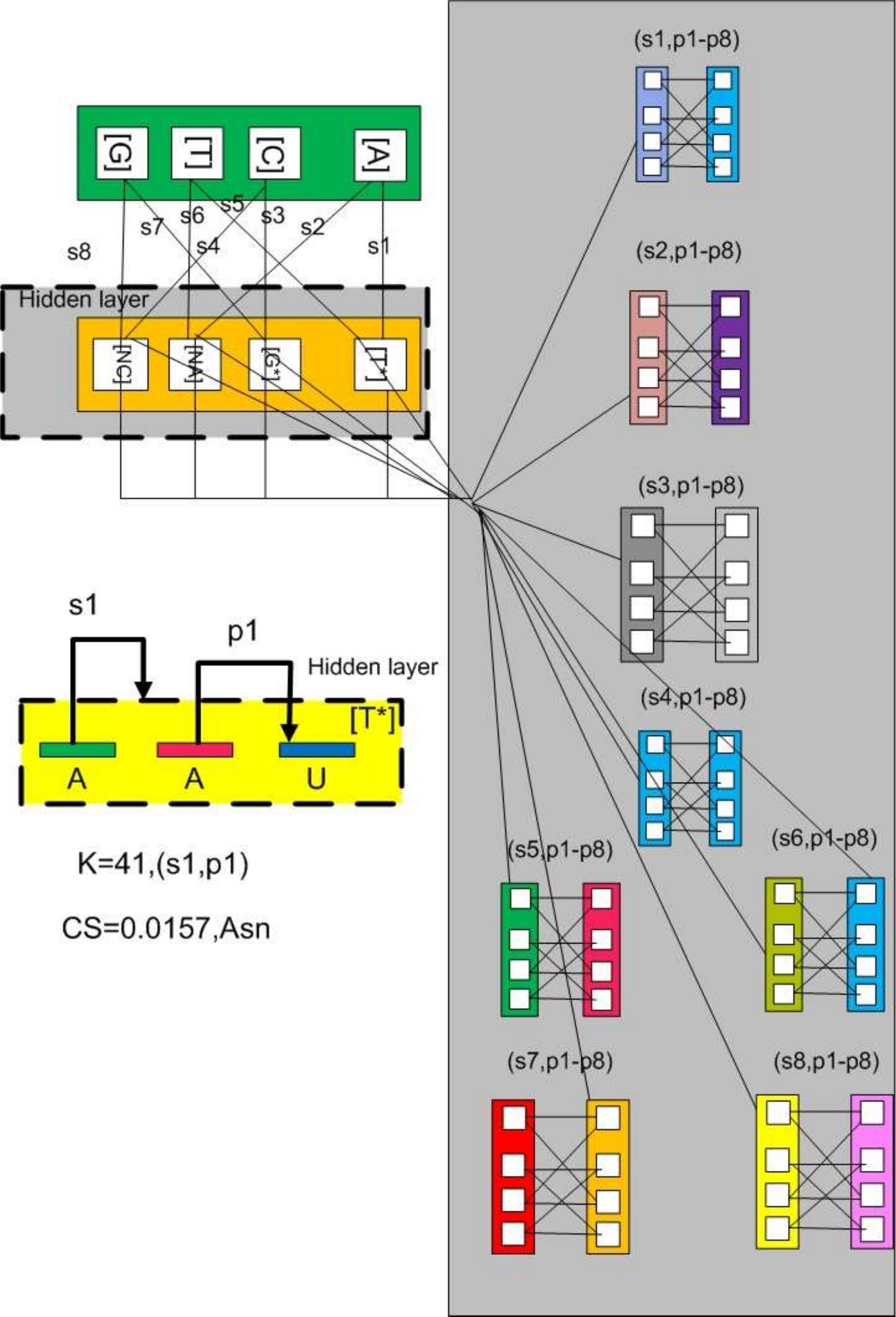}
\caption{ The picture shows  64 states of codon and anti-states of codon state.   On the upper left communication box show the first 8 states, $[s_1]-[s_8]$  of first alphabets of codon (in upper column of the small box). The down column of the box are shown hidden genetic code in codon in which does not appear in table \ref{table1}.  For the right communication box are p- states for second and third alphabet in codon. Each first alpha alphabet of $[s_{i}]$ states contain substate of p-orbital of spinor field in codon. In the left below box, we shown the example of state 41th. in codon state with predefine time series data  value of Chern-Simons current for biology.}
\label{state}

\end{figure}

\begin{Definition}

Let $[s_{i}],i=1,2\cdots 8$ be a state of a gene, we call it a genotype in the genetic code.
The state of the gene has its dual or a pair of gene $[s_{i}^{\ast}],i=1,2,3\cdots 8$ defined by a tangent of a manifold of the genetic code with the Jacobian $J=\sqrt{g_{ij}}$, where $g_{ij}=<[s_{i},[s_{j}]^{\ast}>$.
We have explicitly defined a smallest state in gene as a pair of a genetic code by a classical notation
$A-T,C-G$ with the coordinates $(A,T),(C,G)$. We define a superstate in a pair of genes as ghost and anti-ghost fields in the genetic code with the supersymmetry of a D-isomer to a L-isomer from the right-hand D-state in the light of the polarization in a nucleic acid as $[s_{i}]|_{\Phi_{i}}$ to a left-hand light L-state of an isomer of a light polarization, denoted by $[s_{i}]|_{\Phi_{i}^{+}}^{\ast}$.
\begin{equation}
[s_{1}]=([A],[T]^{\ast})\in T_{p}\mathcal{M} ,\hspace{0.5cm} [p1]=[s_{1}]^{\ast}=[s_{11}]^{\ast}=([A],[T]^{\ast})^{\ast}\in T_{p}^{\ast}\mathcal{M}  \nonumber
\end{equation}

\begin{equation}
[s_{2}]=([A],[NA]) \in T_{p}\mathcal{M} ,\hspace{0.5cm} [s_{2}]^{\ast}=  ([A],[NA])^{\ast} \in T_{p}^{\ast}\mathcal{M} \nonumber
\end{equation}

\begin{equation}
[s_{3}]=([C],[G]^{\ast}) \in T_{p}\mathcal{M} ,\hspace{0.5cm} [s_{3}]^{\ast}= ([C],[G]^{\ast})^{\ast}  \in T_{p}^{\ast}\mathcal{M} \nonumber
\end{equation}

\begin{equation}
[s_{4}]=([C],[NC])\in T_{p}\mathcal{M} ,\hspace{0.5cm} [s_{4}]^{\ast}= ([C],[NC])^{\ast} \in T_{p}^{\ast}\mathcal{M}   \nonumber
\end{equation}

\begin{equation}
[s_{5}]=([T],[T]^{\ast})\in T_{p}\mathcal{M} ,\hspace{0.5cm} [s_{5}]^{\ast}= ([T],[T]^{\ast})^{\ast} \in T_{p}^{\ast}\mathcal{M}  \nonumber
\end{equation}

\begin{equation}
[s_{6}]=([T],[NA])\in T_{p}\mathcal{M} ,\hspace{0.5cm} [s_{6}]^{\ast}=  ([T],[NA])^{\ast} \in T_{p}^{\ast}\mathcal{M}  \nonumber
\end{equation}

\begin{equation}
[s_{7}]=([G],[G]^{\ast})\in T_{p}\mathcal{M} ,\hspace{0.5cm} [s_{7}]^{\ast}= ([G],[G]^{\ast})^{\ast}\in T_{p}\mathcal{M}   \nonumber
\end{equation}

\begin{equation}
[s_{8}]=([G],[NC])\in T_{p}\mathcal{M} ,\hspace{0.5cm} [s_{8}]^{\ast}=([G],[NC])^{\ast}\in T_{p}\mathcal{M}  .
\end{equation}

The observational  states of the living organism in nature, as we know from  biochemistry, are only $[s_{1}]=([A],[T]^{\ast})$ and $[s_{3}]=([C],[G]^{\ast})$. But in the theory of supersymmetry of $S^{7}$ Hopf fibration, there exist 8 states of the ghost fields with 6 hidden states in the mirror symmetry.
In an orbifold of the living organism, in each state,   other 8 states exist which we denote as p1-p8 states. We use the notation of the Riemann tensor field for 64 states $g_{ij}=<[s_{i}],[s_{j}]^{\ast}>$ (Fig. \ref{state}).
\end{Definition}

With this definition at hands, we can define 64 states as shown in Fig. \ref{hopf}.
For example, $g_{11}=<[s_{1}],[s_{1}]^{\ast}>$, and we denote $([s_1],[p_1])$ as a pair of states in a gene.
It is a pair of the genetic code $g_{11}=<([A],[T]^{\ast}),([A],[T^{\ast}])>$. It is an analogy in codon, when we replace $[T]$ with $[U]$. We have $g_{11}=<([A],[[[T]^{\ast})]],([A],[T^{\ast}])>=<([A],([A],[T^{\ast}])> $ since the states $[T^{\ast}]$ are hidden, so we have a codon for $g_{11}$ as AAU.

\begin{Definition}
Let $\Gamma _{ij}^{k}:=[A]_{k}$ be a connection over a tangent of a manifold of $X_{t}/Y_{t}$ with $\mathcal{M}=\mathbb{H}P^{1}$ of the genetic code $k\in \{A,T,C,G\}$. We denoted $\Gamma _{\mu\nu}^{k}:=F_{\mu\nu}$ as the behavioral Yang-Mills field with its dual in Anti-de Sitter theory of supersymmetry with $\ast F_{\mu\nu}=F^{\mu\nu}$. It is a behavior of a protein folding inducing a curvature between a viral glycoprotein and a host cell receptor. This behavior of a Yang-Mills field is an interaction field between the behavior of the   virus and the host cell which can survive by a change of  curvature of the protein during the evolution. We have explicitly defined a connection by an interaction of $g_{ij}$ cycle and $g^{ij}$ cocycle of coupling between the states and the hidden states in the genetic code over the tangent of the manifold of the coupling between 2 living organisms. It is a connection in biology in the sense of an evolutional field not being in a gravitational field in a physical sense as usual,
\begin{equation}
\Gamma _{ij}^{k}=\frac{1}{2}g^{kl}(\partial_{j}g_{jk}+\partial_{i}g_{jk}-\partial_{k}g_{jk} ),
\end{equation}
where $g_{ij}=<[s_{i}],[s_{j}]^{\ast}>$.
  \end{Definition}

In an equilibrium state of evolution of an organism, we have a covariant derivative of a tensor of states with no change in the tensor field $g$,

\begin{equation}
 \bigtriangledown_{g} g=0.
\end{equation}
This situation is completely analogue in gravity \cite{wald}.

\begin{Definition}
   Let a  $F^{\mu\nu}:=(A_{\alpha})_{\mu}^{\nu}$ be a connection  on a principle bundle $Sp(1)\rightarrow S^{7}\rightarrow \mathbb{H}P^{1}$ of a genetic code over a supermanifold of a viral particle $\mathcal{A}$. It is a strength of a gauge field with an invariance under a small change of the spinor field.  The master equation is an equation for a gene structure with an excited state of the gene in the DNA according to the viral-RNA or any enzyme coming to simulate the active side along the DNA curve.
 We can use an exact sequence of the sheave cohomology  $\mathcal{O}_{X_{t}}$ with the chart over the supermanifold defined by homogeneous coordinates of $\mathbb{H}P^{1}$ for viral gene $X_{t}$ along the host cell gene $\mathcal{O}_{Y_{t}}$, while the virus attachment to the host cell is defined by the coordinate in sheave $\mathcal{O}_{X_{t}/Y_{t}}$. We apply a supersymmetry AdS theory over a Yang-Mills field of a behavioral field of the genetic code as a connection over $(A_{\alpha})\in \{[A],[T],[C],[G]\}\subset \mathbb{H}P^{1}$ with the Hopf fibration (Fig. \ref{hopf}) of viral RNA gene $F^{\mu\nu}:=(A_{\alpha})_{nu}^{\mu}=\Gamma_{\alpha\nu}^{\mu}$ with an anti self dual field over the gene of the host cell $DNA$ , $\ast F_{\mu\nu}$, we have a current of the connection between this field defined by the Chern-Pontryagin density for the interaction of behavioral fields, so-called current $J$ which varies from the curvature of docking between the behavior of the curvature over amino acid $k$ in $X_{t}$ and its dual curvature in $Y_{t}$ while docking,

\begin{equation}
  <F^{\mu\nu}\ast F_{\mu\nu}>\,,
\end{equation}
here $<->$ is an average or an expectation operator and  $k=1,2,3,\cdots$ is the number of amino acids in the genetic code.
\end{Definition}

 \begin{Definition}
Let  $R_{\nu\alpha\beta}^{\mu}$ be the curvature over a tangent space of a genetic code $(A_{\alpha})\in \{[A],[T],[C],[G]\}\subset \mathbb{H}P^{1}$. Let   $(A_{\alpha})_{\nu}^{\mu}:=\Gamma_{\alpha\nu}^{\mu}=F^{\mu\nu}$  be the connection of the  coupling between 2 alphabets of 2 organisms,  one is from DNA. The other side  is from viral RNA   (Fig. \ref{hopf} ).
The connection between genes gives 

\begin{equation}
R_{\nu\alpha \beta}^{\mu} =  \partial_{\alpha}(A_{\beta})_{\nu}^{\mu} - \partial_{\beta}(A_{\alpha})_{\nu}^{\mu}+ [ A_{\alpha}, A_{\beta} ]_{\nu}^{\mu}.
 \end{equation}
 \end{Definition}
This  is the curvature over the tangent space of  the genetic code manifold  derived from the above connection.  It is useful  also in other situations like   when t-RNA is docking with DNA in interaction between 2 D-branes of DNA and RNA.  In a   gauge field theory of DNA  and RNA genetic codes of translation process it is  the  group action of Lie-algebra one form. We have an adjoint representation of the genetic code as a translation process  over codon and anti-codon of $t-RNA$ as a new master equation for translation process in genome of coupling between 2 living organisms, in this context we mean space of virus, $X_{t}$ and space of host cell, $Y_{t}$.

In this situation, we need to define a new  value to measure the curvature in amino acids of protein structure, not in the tangent space of a genetic code. We need to define a curvature over codon and anti-codon to represent a curvature of proteins while docking to each other. The new value needs to be unique for 64 states in codon and have a meaning of curvature of genetic code with connection over manifold of genetic code also. We introduce a new definition in biology to represent a spectrum of genetic variation as curvature in protein structure while docking. We call this new quantity a \emph{Chern-Simons current for biology}.

In codon, we translate genetic code in 3 steps. The translation operator of group is given  by a behavior matrix in Lie  group, a group of supermanifold of living organism with action in  3 times. It  generates a  codon representation as an adjoint representation over gene expression. It is a  precise definition of genetic code  with parity 2  of ghost field and anti-ghost field in Chern-Simons current for the  representation of gene $A_{i}$ with current density $J^{A_{i}}=\int_{t_{1}}^{t^{n}}dJ^{A_{i}}$.

We introduce a curvature tensor for docking and anti-docking states of DNA and t-RNA at the  level of
gene expression  by borrowing the definition of Riemann curvature tensor over D-brane and anti-D-brane gauge field,
and modeling  ghost field between codon and anti-codon field with parity 2.

\begin{Definition}
  Let a  knot be a  representation of anti codon in t-RNA topological structure for amino acid $\mu$ with $J^{\mu}$ be a representation $R$ of gauge group of gene geometric translation as group action of  transcription process $G$. The genetic code is an average expectation  value of Wilson loop operator of coupling between hidden state of $x_{t}$ and $y_{t}$ twist D-brane and anti-D-brane over superspace of cell membrane. That is

\begin{equation}
W(K,R)=Tr_{R}Pexp(\oint_{K}A) ,
\end{equation}
This term  gives the asymmetric property of chiral molecule of DNA and RNA, twisted from left hand to right hand in a supersymmetry breaking as knot polynomial related to the  connection $A$.

\end{Definition}

\begin{figure}[!t]
\centering
\includegraphics[width=0.3\textwidth]{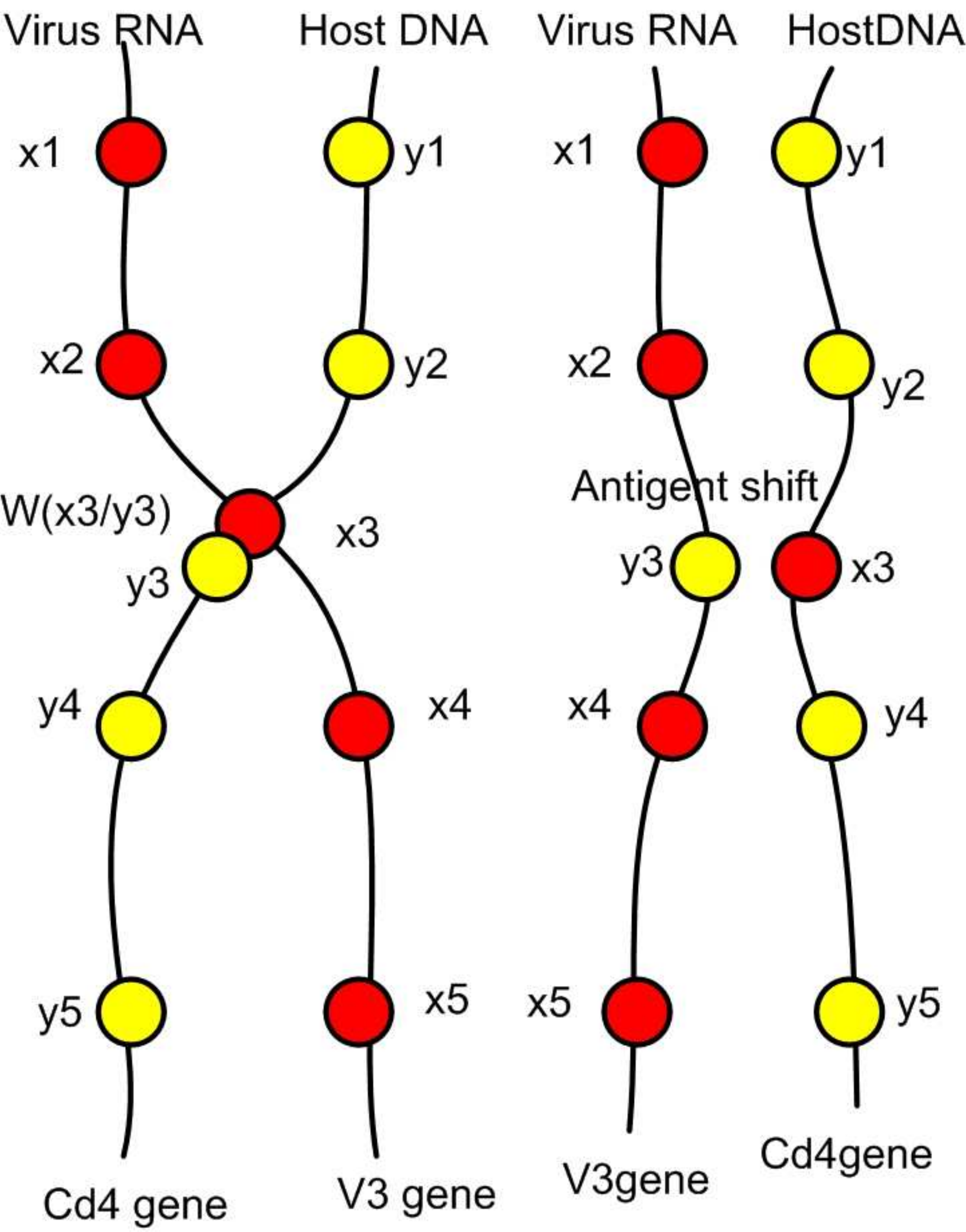}

\caption{ The picture shows the   Wilson loop of evolution in gene of Cd4 and V3 loop as twistor exchange a genetic code to $W(X_{t}/Y_{t})\rightarrow Y_{t}/X_{t}$.}
\label{shift1}
\end{figure}

\begin{figure}[!t]
\centering
\includegraphics[width=0.3\textwidth]{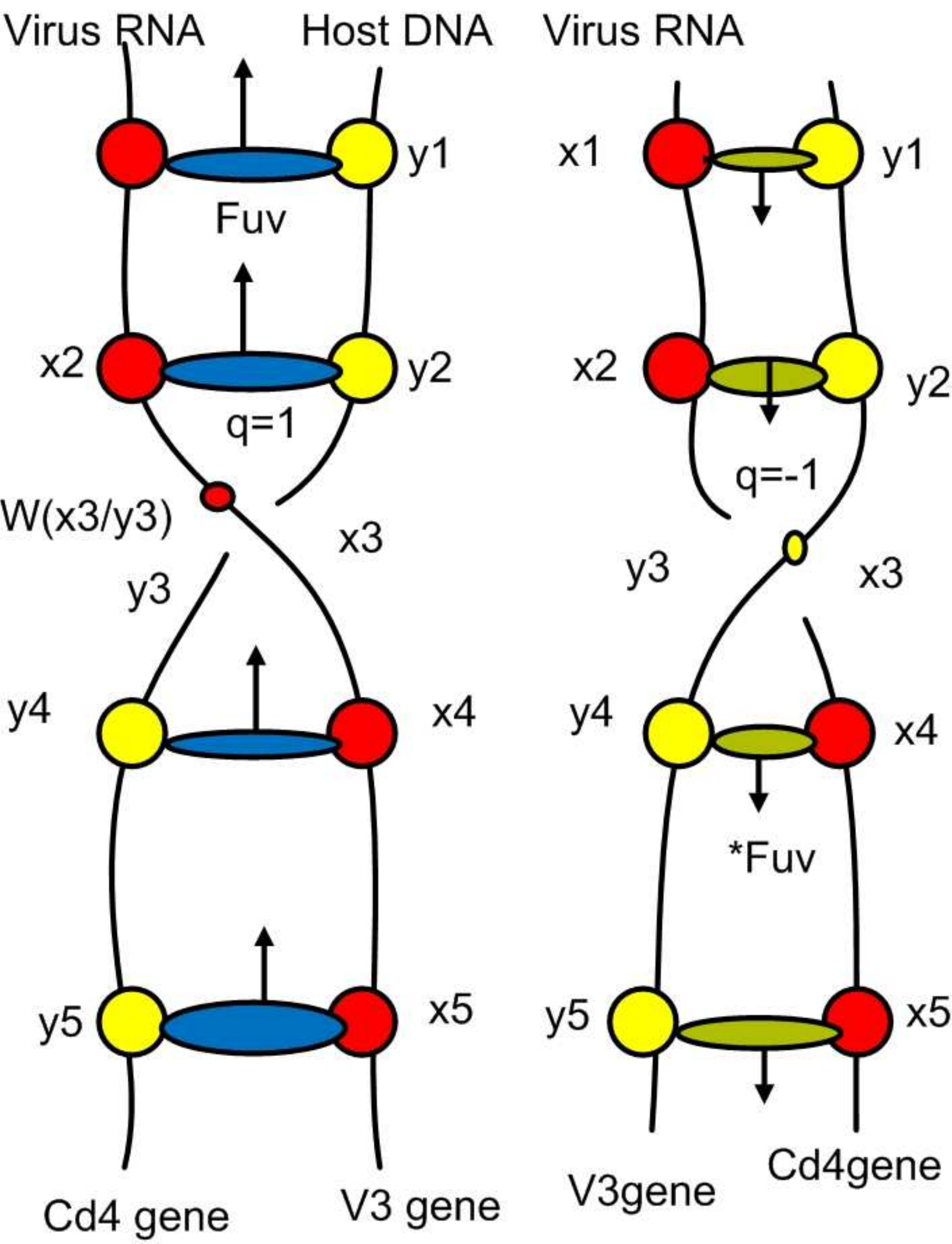}

\caption{ The picture shows   the Wilson loop of evolution in gene of Cd4 and V3 loop as twistor exchange a genetic code to $W(X_{t}/Y_{t})\rightarrow Y_{t}/X_{t}$.}
\label{shift2}
\end{figure}

In this approach, \cite{knot} we can represent the genetic code as Laurent polynomials in variable $q$
 with integer coefficients (Figs. \ref{shift1} and \ref{shift2}), that is for any knot $K$, we have 

 \begin{equation}
J(K,q)=\sum_{i=1}^{n}a_{n}q^{n}.
\end{equation}
We induce a spinor field for representation of genetic code by using the new parameter of knot $q$,
\begin{equation}
q=e^{\frac{2i\pi}{k+h}}
\end{equation}
where $h$ is the dual coexter number for group action of supersymmetry of gene expression $G$. It might be the source of evolution from the adaptive behavior derived from the  environment. For the sake of simplicity of a first study, in this work we set $h=0$ in our definition of Chern-Simons current for biology. There might  exist  a deep connection with string theory over icosahedral group $E_{8}\times E_{8}$ with $h-$ Coexter number in supersymmetry breaking of codon and anti-codon interaction.

\begin{figure}[!t]
\centering
\includegraphics[width=0.5\textwidth]{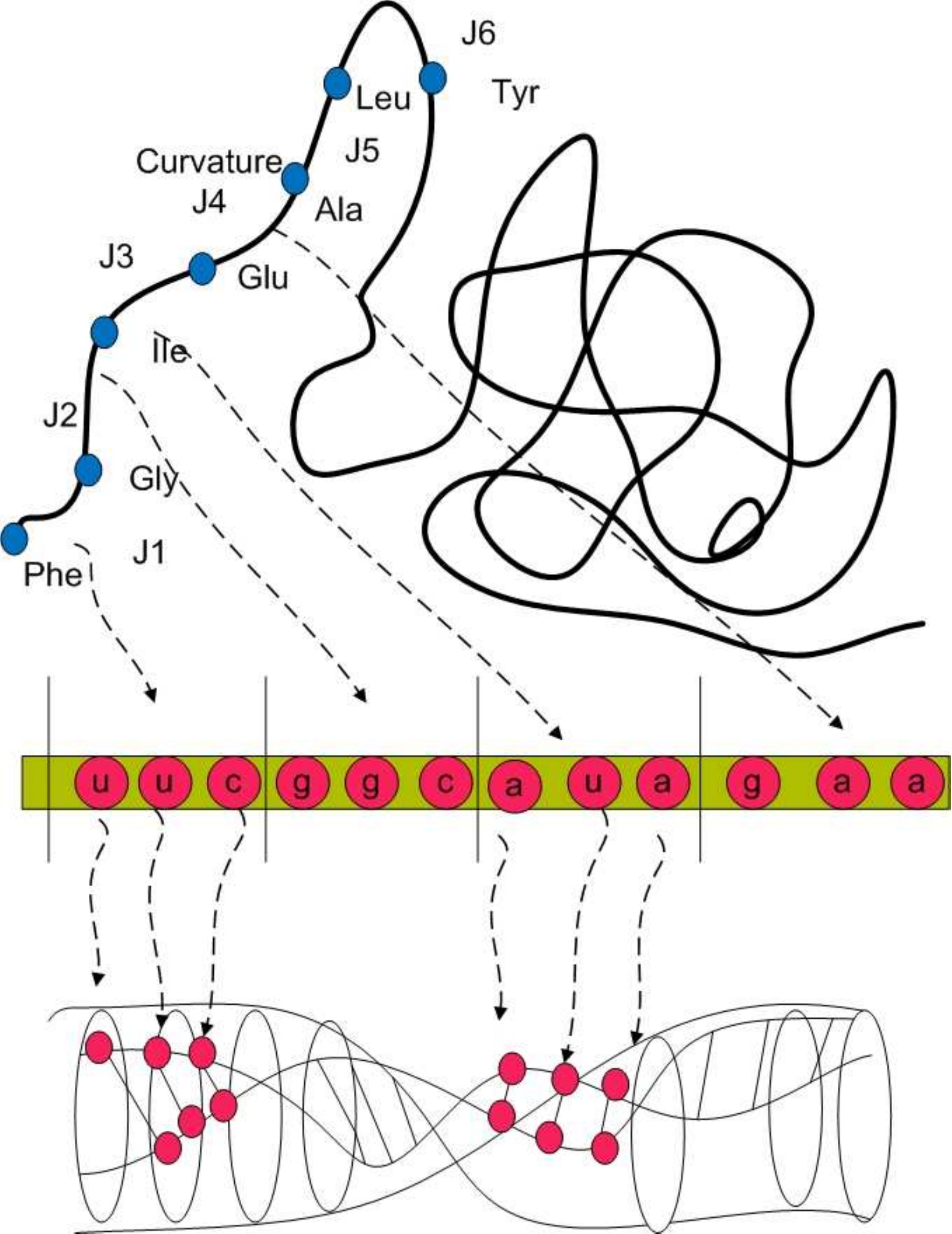}
\caption{ The picture shows   the Chern-Simons current over CD4 receptor protein as curvature  in behavior of  docking states between V3 loop and CD4. The down panel is a loop space as Hopf fibration over DNA molecule.}
\label{current}
\end{figure}

\begin{Definition}

Let   $J^{\mu}$  be a Chern-Simons current for anomaly quantum system of codons. It is a curvature of genetic code in amino acids (Fig. \ref{current}). It is also defined as the spectrum of curvature in genetic code for gene evolution detection. The meaning is a differential 3 forms in cohomology of
spin fiber  $S^{3}$ over the homotopy class $[S^{3}, X_{t}/Y_{t}]$ in the codon of t-RNA molecule.
 A path integral of gene expression is defined by the Chern-Simons theory over knots of codon and anti-codon: it  is defined by the interaction between codon and anti-codon between DNA and RNA in form of integral  $A_{i}+S_{CS}=\int \mathcal DA_{i}exp(iS_{CS})$ between genetic code in codon $A_{i}$ and genetic code in anti-codon where t is  defined by 
\begin{equation}
S_{CS}=\frac{k}{4\pi}\int_{W}Tr(A\wedge dA +\frac{2}{3}  A\wedge A\wedge A) 
\end{equation}
and
 \begin{equation}
J(q;K^{A_{i}},R_{i})= <W(K_{i},R_{i})> =<Tr_{R_{i}} P \oint_{K_{i}} A> 
= \frac{\int \mathcal DA_{i}exp(iS_{CS})  \Pi_{i}W(K_{i},R_{i})}{ \int \mathcal DA_{i}exp(iS_{CS}) }
\end{equation}
The first 2 forms  come from the kernel of cohomolgy map and the second term is a boundary map of 3 forms. It is an invariant of homotopy path over $S^{4}\times I\rightarrow X_{t}/Y_{t}$ to $S^{3}\times I \rightarrow X_{t}/Y_{t}$.
 \end{Definition}

The explicit definition of curvature over the connection of genetic code has also the new meaning of genetic sprectrum current $J^{\mu},\mu=1,2\cdots 20 $ of Chern-Simons current  for 20 amino acids. It is  generated from  the representation of Lie group over manifold of a host cell.

{\em Example:}
\begin{itemize}
\item $UUU,UUC$ for $Phe$ is define by
\begin{equation}
J^{Phe}=\epsilon^{\mu\alpha\beta\alpha}< \frac{1}{2}A_{\alpha}\partial_{\beta} A_{\gamma} +\frac{1}{3}  A_{\alpha}A_{\beta}A_{\gamma} >=\epsilon^{\mu\alpha\beta\alpha}<  \frac{1}{2} (A_{2})_{\nu}^{\mu}d(A_{2})_{\nu}^{\mu} +\frac{1}{3} (A_{2})_{\nu}^{\mu}(A_{2})_{\nu}^{\mu}(A_{2})_{\nu}^{\mu}>
\end{equation}

$dA$ is a differential form between 2 complex of genetic code variation in cohomology of $H^{3}(X_{t})$. We explicitly define the differential form of genetic code for $Phe$ by $dA_{2}=A_{2}A_{2}-A_{2}A_{3}:=UU-UC$. So,  we have $AdA=A_{2}(A_{2}A_{2}-A_{2}A_{3})=
A_{2}A_{2}A_{2}-A_{2}A_{2}A_{3}$. The minus sign here represents a linear combination of basis for codon. Therefore, we have
\begin{equation}
J^{Phe}=\epsilon^{\mu\alpha\beta\alpha}< \frac{1}{2}A_{\alpha}\partial_{\beta} A_{\gamma} +\frac{1}{3}  A_{\alpha}A_{\beta}A_{\gamma} >=\int Tr(\frac{5}{6} U\wedge U\wedge U-\frac{1}{2}U\wedge U\wedge C)=\int Tr(H^{3}(\mathcal{M})).
\end{equation}
\end{itemize}

For a translation in reversed direction of antigent shift and drift  in gene evolution theory, we can define the  group action of reversed direction of time by the $CPT$ theory for antighost field in field of time series of antibody gene. It is 
 \begin{equation}
\{x_{t},y_{t}\}=\int Tr( H^{3}(\mathcal{M}))=\sum_{i=1}^{3} g^{i}(x_{t}/y_{t})= \alpha_{t}y_{t}, g^{3}x_{t}/y_{t}\rightarrow \beta_{t}x_{t}/\alpha_{t}y_{t} \simeq [\epsilon_{t}^{\ast}]= \int TrH^{3}(\mathcal{M},g,F^{\mu\nu}).
\end{equation}

We can use  a  numerical representation for spinor field of curvature in gene expression by the Chern-Simons action defined as 

\begin{equation}
S_{CS}=\frac{k}{4\pi} \int Tr(A\wedge dA  +\frac{2}{3} A\wedge A \wedge  A)
\end{equation}
where $k=1,2,3\cdots n$   are defined by
\begin{equation}
 J^{Amino}=J^{k}\simeq \sqrt{\frac{2}{k+2}} \sin\frac{\pi}{k+2}=\int D [A] e^{iS_{CS}}.
\end{equation}
The Chern-Simons current can be derived by using a simple algorithm. The result of computation for 64 codons is shown in Table \ref{table1}.
In other words, the Chern-Simons current maps the string of genetic code into numerical values by explicit  formulas. It is  suitable for plotting   the  so called time series data directly   into the  superspace of gene expression.
We transform the alphabet string values, which cannot be computed in the classical standard definition of genetic code,  to a Chern-Simons current  of time series data of genetic code with $k=1,2,3,\cdots 64$ over spinor field with ground field of real values.  In our opinion, this approach is  suitable for intensive  computational programs   starting  from  data analysis.

\subsection{Example of Supersymmetry in  an organism}
We want now to work out an example in order to realize the above theoretical discussion.
Let us consider a virus whose   equation of state  $x_{t}$ attaches to host cell $y_{t}$ with evolution $\epsilon_{t}$:
\begin{equation}
  y_{t} =\alpha_{t} +\beta_{t} x_{t} +  \epsilon_{t}.
\end{equation}

If we consider  a   ghost  functor $\Phi_{i},\Phi_{i}^{+}:  X\rightarrow (\mathcal{A},s)\simeq [X,S^{\pm k}]$, from the above equation   we get a supermanifold

\begin{equation}
  \Phi_{i}( (y_{t} - \alpha_{t})- \beta_{t} x_{t} \simeq  \epsilon_{t}).
\end{equation}

\begin{Definition}(proposition)
The gene expression of genetic code in m-RNA, codon and in t-RNA, anti-codon (Fig.  6), can be represented  as the coefficients $\alpha_{t}$ and $\beta_{t}$ in a supermanifold model over time series.
\end{Definition}

We need to define a boundary $d: C_{4}(X_{t}/Y_{t})\rightarrow C_{3}(X_{t}/Y_{t})$ operator from $X$ to $\partial X$  which takes values in spinor field
of different angles with  curvature invariant by using 3rd cohomology group of $H^{3}(X_{t}/Y_{t})$ defined by the characteristic class $[cs_{i}]^{\ast}\in H^{3}(\mathcal{M})$ and $[cs]\in H_{3}(\mathcal{M})$. It 
 transforms by  symmetry $[cs]^{5}=0, [cs^{\ast}]^{3}=1$ of quantum virology in $E_{8}\times E_{8}$ with equation

\begin{equation}
 [cs]^{5}+ [cs^{\ast}]^{3}=1.
\end{equation}

\begin{figure}[!t]
\centering
\includegraphics[width=0.5\textwidth]{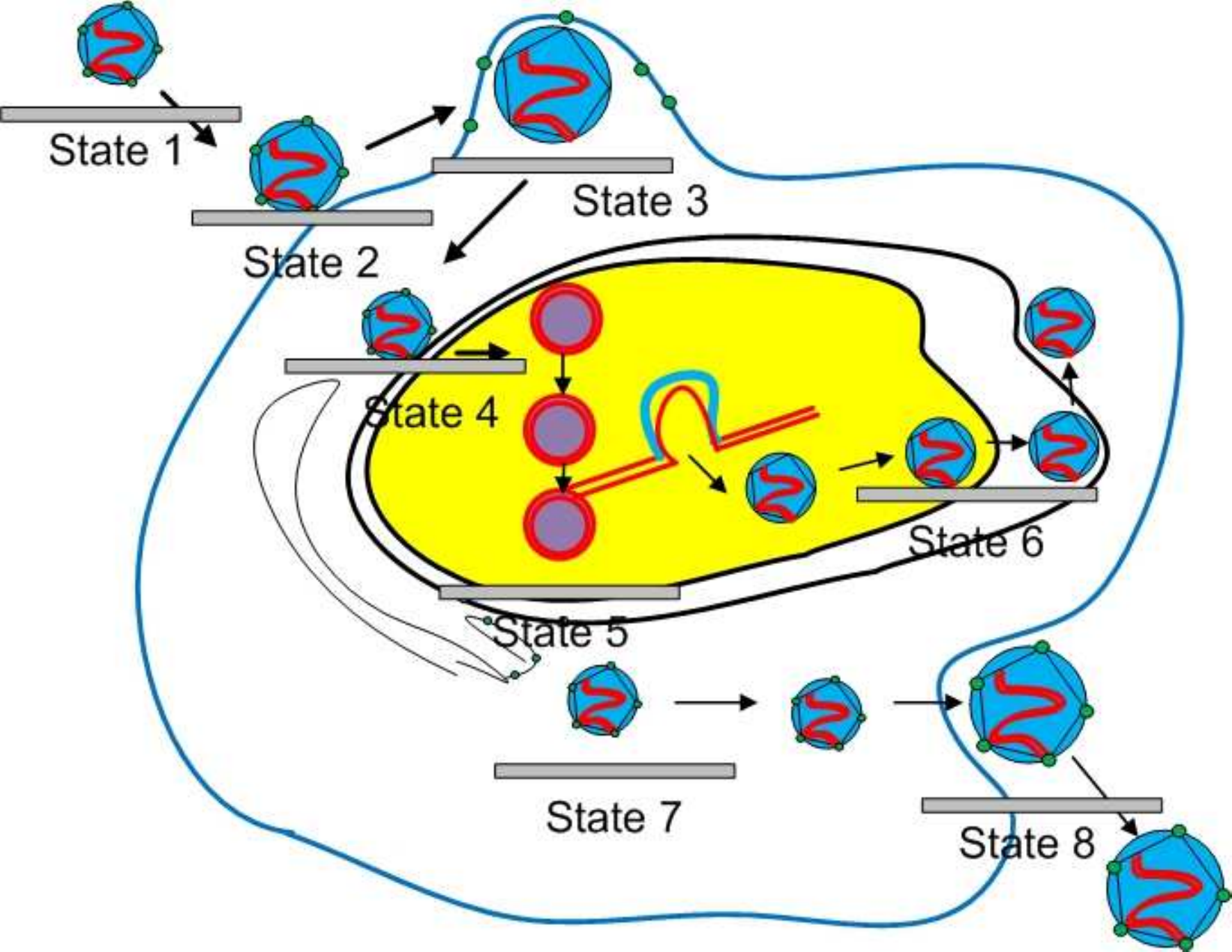}
\caption{ The picture shows  8 states of viral replication cycle as cyclic coordinate modulo 8 in viral equation of moduli state space model. It is a  long exact sequence equilibrium state in viral replication equilibrium 8 states. We denoted all 8 states of viral cocyle in host cell as parasitism behavior 8 states denoted by  $[s1]-[s8]$ in which we can be measured by using $\alpha$ cocyles basis wave function for anti-ghost field.}
\label{replication}
\end{figure}

 The anomalies  on superspace of time series data are related to the covering space of spin and spin group.
We can induce a reflection or a  time inversion, in CPT symmetry, starting  from the  intrinsic behavior of the organism $[A_{i}]$: such a transformation is  induced by the  Pauli matrices of the covering symmetry of  an organism influenced by cell cocycle.  For example, we have  $S,G1,G2, M$ ,4 states  and 4+4=8 states, of viral replication cycle (Fig. \ref{replication}) in the host cell as parasitism states. it is

\begin{figure}[!t]
\centering
\includegraphics[width=0.5\textwidth]{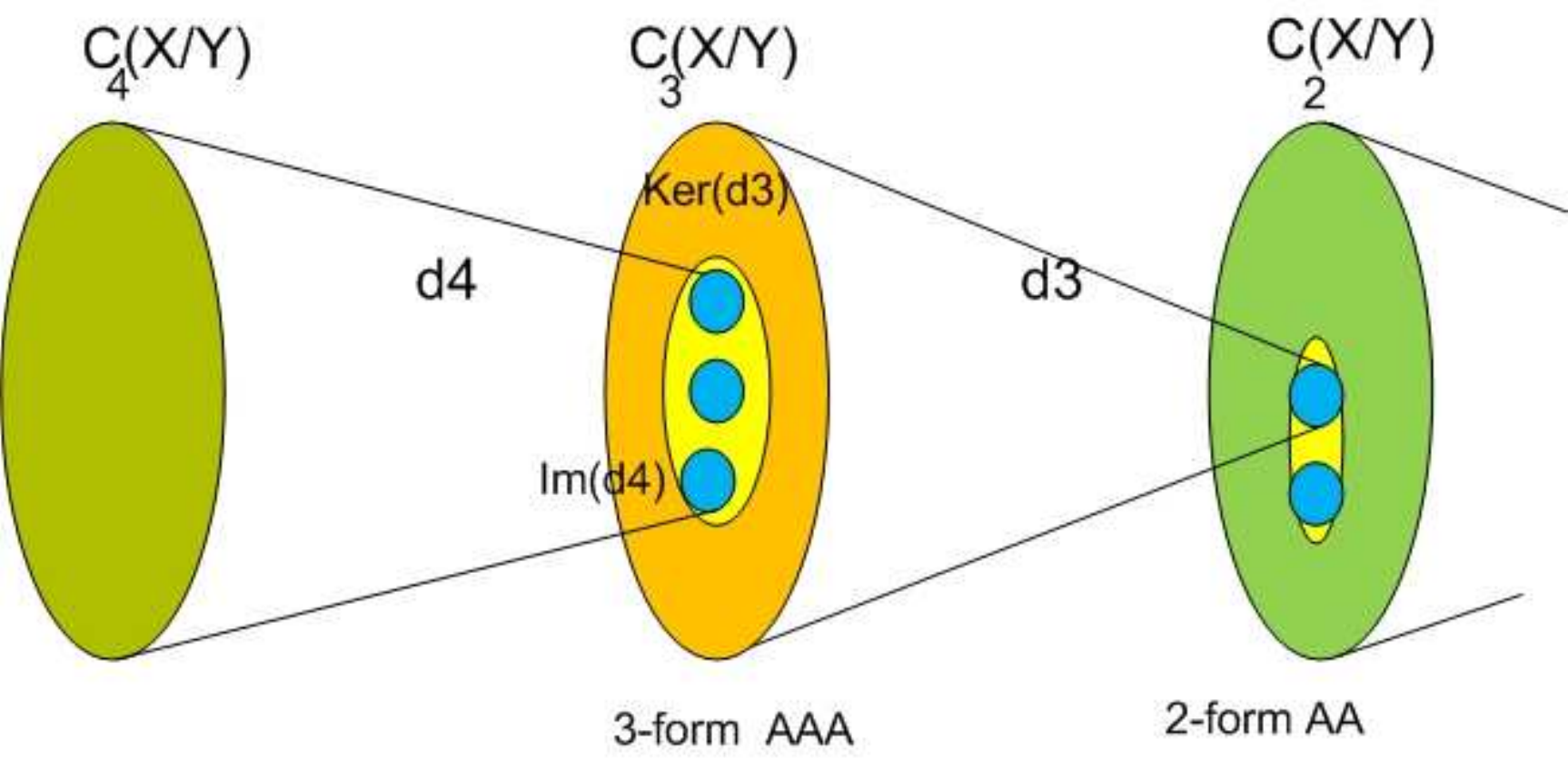}
\caption{ The picture shows the   Chern-Simons current over cohomology sequence with kernal map of 2 differntial 2 forms in genetic code $A\wedge A$ and boundary map of differential 3 forms $A\wedge A\wedge A$ in genetic code of codon .}
\label{cohomology}
\end{figure}

\begin{table}[!t]

\renewcommand{\arraystretch}{1.3}

\caption{  The table below shows  the definition of   $8\times 8=64, E_{8}\times E_{8}$ model of ghost fields and anti-ghost fields in  genetic code with predefined Chern-Simons current for codon and anticodon in real value with notation of parameter as $k$.}
\label{table1}
\centering

\begin{tabular}{||c|c|c|c|c||}\hline\hline
 &   U&  C &  A  &  G  \\ \hline\hline
 U &   0.7071 UUU Phe(k=1)&0.0534  UCU Ser(k=17) &  0.0214 UAU  Tyr(k=33)  & 0.0122 UGU  Cys(k=49)\\

&  0.5000 UUC Phe(k=2)& 0.0495 UCC Ser(k=18) & 0.0205 UAC  Tyr (k=34) &0.0118  UGC  Cys(k=50)\\

  &  0.3717 UUA Leu(k=3)& 0.0460 UCA Ser(k=19) &  0.0197 UAA  Stop (k=35)  & 0.0115 UGA  stop(k=51)\\

&  0.2887 UUG Leu(k=4)&0.0429  UCG Ser(k=20) & 0.0189 UAG  Stop(k=36)  & 0.0112 UGG  Trp(k=52)\\  \hline\hline

C& 0.2319 CUU Leu(k=5)& 0.0402 CCU Pro (k=21) &0.0182   CAU  His(k=37)  & 0.0109 CGU  Arg(k=53)\\

&0.1913 CUC Leu(k=6)& 0.0377 CCC Pro (k=22) & 0.0175 CAC  His (k=38) &  0.0106 CGC  Arg(k=54)\\

& 0.1612  CUA  Leu(k=7)& 0.0354 CCA Pro(k=23) & 0.0169 CAA  Gln(k=39)  & 0.0103 CGA  Arg(k=55)\\

& 0.1382 CUG Leu (k=8)& 0.0334 CCG Pro(k=24) & 0.0163 CAG  Gln(k=40)  & 0.0101 CGG  Arg(k=56)\\  \hline\hline


A& 0.1201 AUU Ile(k=9)& 0.0316 ACU Thr (k=25) &  0.0157 AAU  Asn (k=41) &  0.0098 AGU  Ser(k=57)\\

&  0.1057 AUC Ile(k=10)& 0.0299 ACC Thr(k=26) & 0.0152 AAC  Asn(k=42)  & 0.0096 AGC  Ser(k=58)\\

&0.0939  AUA  Ile(k=11)&0.0284  ACA Thr (k=27) &   0.0147AAA  Lys(k=43)  & 0.0093 AGA  Arg(k=59)\\

& 0.0841 AUG  Met(k=12)  & 0.0270  ACG Thr (k=28)& 0.0142 AAG  Lys(k=44)  & 0.0091 AGG  Arg(k=60)\\  \hline\hline

G &0.0759  GUU   Val(k=13)&  0.0257  GCU Ala (k=29) & 0.0138 GAU  Asp(k=45)  & 0.0089 GGU  Gly(k=61)\\

&  0.0690   GUC   Val(k=14)& 0.0245 GCC Ala (k=30)& 0.0134 GAC  Asp(k=46)  & 0.0087 GGC  Gly(k=62)\\

&  0.0630  GUA Val(k=15)&  0.0234  GCA Ala (k=31)& 0.0129 GAA  Glu(k=47)  & 0.0085 GGA  Gly(k=63)\\

&  0.0579 GUG Val(k=16)& 0.0224  GCG Ala (k=32)&  0.0126 GAG  glu(k=48)  & 0.0083 GGG  Gly(k=64)\\  \hline\hline

\end{tabular}
\end{table}

\begin{equation}
\{x_{t},y_{t}\}=ad_{y_{t}}x_{t}=g^{3}_{y_{t}}(x_{t}):=\int TrH^{3}(\mathcal{M},g).
\end{equation}

\begin{figure}[!t]
\centering
\includegraphics[width=0.5\textwidth]{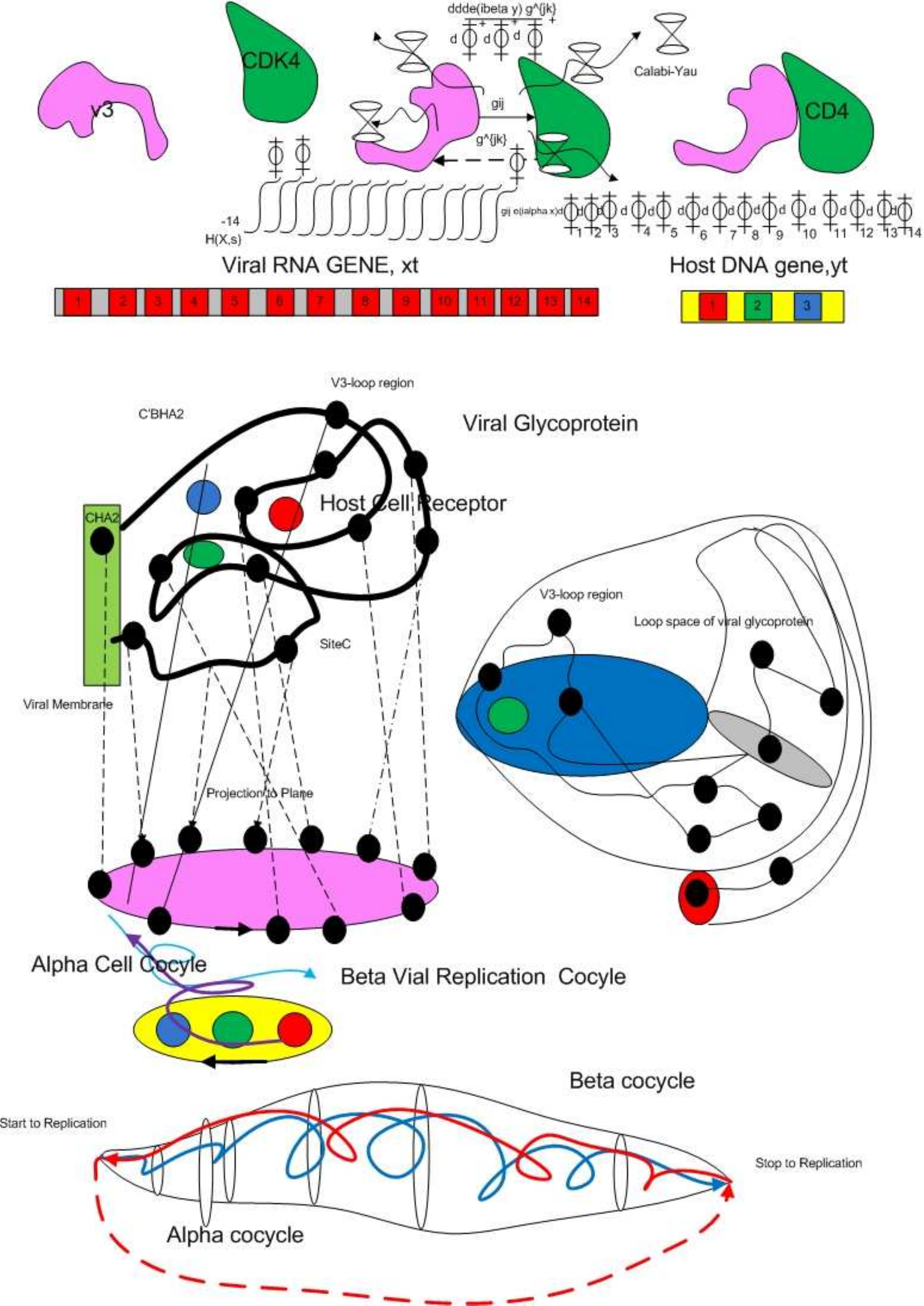}
\caption{ The picture shows the  loop space of time series data over cell membrane and docking state when a virus  attaches a  cell. The lower panel  shows the  geometry of cell cycle as the coupling between $\alpha $ and $\beta$ cocycles.}
\label{loop}
\end{figure}

For a translation in the reversed direction of an antigene shift and drift  in gene evolution theory, we can define,  by the group action of reversed direction of time in $CPT$ symmetry  for the  anti-ghost field,  a  time series of antibody gene. The $\beta$-cocyle and $\alpha$-cocoycle are coefficients of a super-regression model. The meaning of these 2 cycles is a coupling between 2 life cycles (Fig. \ref{loop}) and a coupling between the behavior of curvature while docking. It is 
 \begin{equation}
\{x_{t},y_{t}\}=\int Tr( H^{3}(\mathcal{A},s))=\sum_{i=1}^{3} g^{i}(x_{t}/y_{t})= \frac{\alpha_{t}y_{t}}{\beta_{t}x_{t}} \simeq [\epsilon_{t}^{\ast}] = \int_{H^{3}(X_{t}/Y_{t})} TrH^{3}(\mathcal{M},s).
\end{equation}
inducing an interaction between viral gene $X_{t}$ and host cell gene $Y_{t}$ with
 \begin{equation}
H_{3}(X_{t}/Y_{t})=ker(d_{3}(X_{t}/Y_{t})/Im(d_{4}(X_{t}/Y_{t})))
\end{equation}
where $H^{3}(\mathcal{M})=H_{3}(X_{t}/Y_{t})^{\ast}.$
The element of  $H^{3}(\mathcal{M})$ is  a differential 3 form of genetic code $A\wedge A\wedge A$
 modulo kernel of 2 form $A\wedge A $ (Fig. \ref{cohomology}).

\subsection{ A Unified Theory of  Codon and Anti-Codon}
Let use  use now the Chern-Simons model, coming  from the  theoretical physics,  to define a new spectrum of
 genetic code of chromosomes in  the genome of living organism including viral particles.
Typically, the Chern-Simons field is adopted as a gravitational field  and can  have also  different meanings  in differential geometry: for example it gives the connection over the principle bundle of the tangent space of a manifold or a supermanifold.

Let $E_{8}\times E_{8}$ be a space for codon and  anti-codon interaction. The number of states in classical definition of codon is 64=8$\times$ 8 states imply we have dual coexter group of codon states inside the symbolic definition.  Let $x_{t}$ be a codon state with $2^{3}=8$ states of DNA of host cell with $3$ alphabet and $y_{t}$ be another $8=2^{3}$ states of anticodon in $t-RNA$.
The 3 alphabets can be represented  as an equivalent class in $[cs]\in H^{3}(X_{t}/Y_{t}).$
\begin{figure}[!t]
\centering
\includegraphics[width=0.48\textwidth]{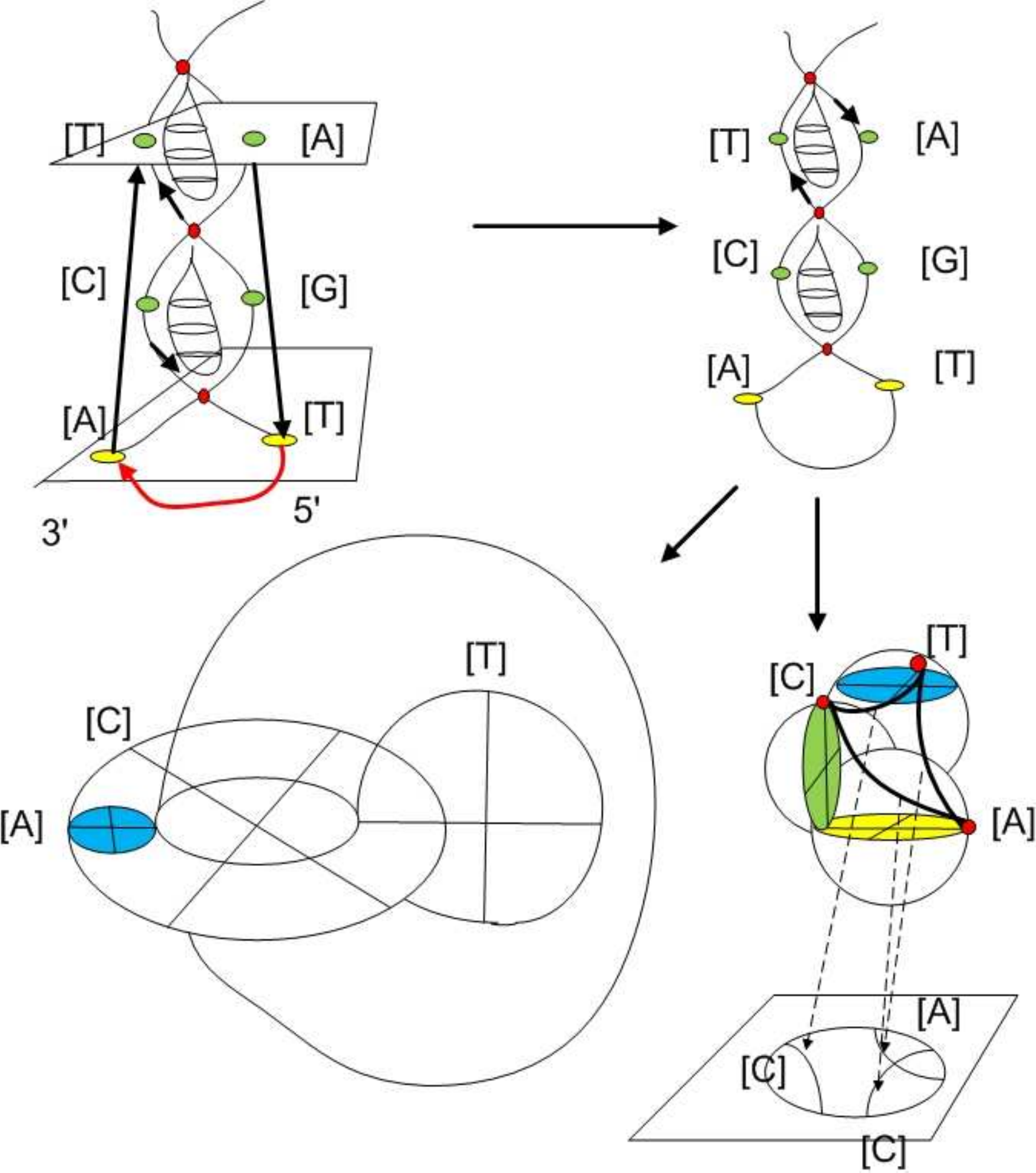}
\includegraphics[width=0.48\textwidth]{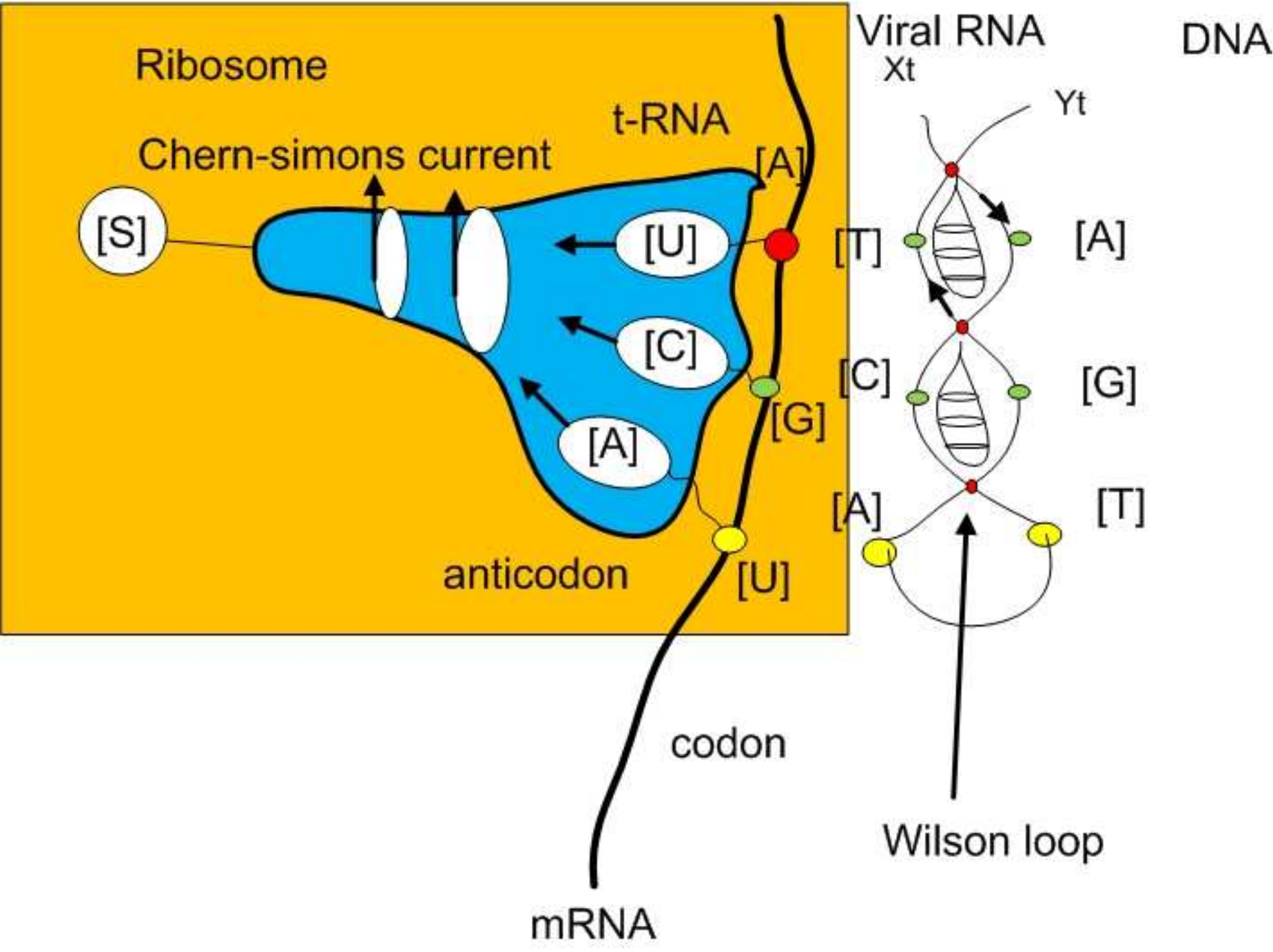}
\caption{ The picture shows the Wilson loop and the Chern-Simons current over DNA. The red  dot is a knot model if we project DNA into plane. On the left panel is the transformation of DNA into knot model with loop space  inside. There are many possibilities  to the classify space of DNA into disjoint sum of unit spheres or
M\"obius strips with disjoint sum of spheres. It  depend on chosen mathematical model behind the real situation of DNA module in active  chiral states. We can use knot model and spinor field on $S^{3}$ to define a 3 alphabet code by using disjoint sum over unit sphere. The Poincare disk model for time series data is a  projection of 4 dimensional sphere of quanternion field as quaternionic projective plane. On the right panel, there  is a picture of codon ghost field and anti-codon  field. In t-RNA over ribosome, we induce a Yang-Mills behavior field with coupling between 3 alphabets. We called this a connection of 3 codon and anti-codon curvatures. It is in our new definition,  namely the Chern-Simons current for biology.}

\label{trna}
\end{figure}

In viral genome replication,  generally a single-stranded positive-sense RNA virus with a DNA intermediate,  for example a  retrovirus,  are used in cloning technology, and they have been  successfully used  in gene delivery systems for gene therapy.
 Viruses copy their genomes directly to DNA again  and RNA  (Fig. \ref{trna}). Their single strand RNA genome are directly transformed  reversed   to mRNA for translation process with host ribosome and t-RNA. We use the notation of Chern-Simons current of tRNA and ribosome  for anti-codon and codon for RNA virus. The parasitism is going on at this step, when a virus uses host $t-RNA$ and ribosome for the synthesis of its capsid proteins.
\begin{Definition}

 Let $X_{t}$ be superspace of virus RNA gene and $Y_{t}$ be superspace of host cell DNA gene. We have an exact sequence over tangent of manifold $(\mathcal{M},g,F^{\mu\nu})$ by the sheave sequence

\begin{equation}
0\rightarrow \mathcal{O}_{X_{t}(RNA)} \longrightarrow  \mathcal{O}_{Y_{t}(DNA)}  \longrightarrow \mathcal{O}_{X_{t}/Y_{t}(RNA)} \longrightarrow 0
\end{equation}

\end{Definition}

This  chain sequence is the basis to define a new BV-cohomology for biological system and Chern-Simons current for biology in this work (Fig. \ref{exact}).
We take a $H_{0}(-):TOP\rightarrow GROUP$ to the chain sequence above and 
induce an exact sequence of sheave cohomology  with local coordinate defined by the distances in genetic code  over codon.  We have an induced infinite sequence of biological cohomology 

\begin{equation}
0 \rightarrow \cdots\rightarrow H_{0}(X_{t}(DNA)) \longrightarrow  H_{0}(Y_{t}(mRNA))  \longrightarrow H_{0} (X_{t}/Y_{t}(tRNA))\cdots\longrightarrow H_{3}(Protein) \rightarrow \cdots.
\end{equation}

with

\begin{equation}
\cdots\leftarrow H^{0}(X_{t}(DNA)) \longleftarrow  H^{0}(Y_{t}(mRNA))  \longleftarrow H^{0} (X_{t}/Y_{t}(tRNA))\cdots     \longleftarrow H^{2}( X_{t}/Y_{t}(tRNA))  \longleftarrow H^{3}(Protein) \leftarrow \cdots .
\end{equation}

This cohomology sequence  can be considered a  grand unified model in biology.  
\subsection{Anomaly in Codon and Anti-Codon ghost field}
In this section, we introduce the Chern-Simons current of the connection $(A_{\alpha=1,2,3,4})_{\nu}^{\mu}$ to connect the genetic code into the gene sequence. We want to  give a precise definition for time series data of gene expression of codon as   curvature over a connection.

\begin{figure}[!t]
\centering
\includegraphics[width=0.28\textwidth]{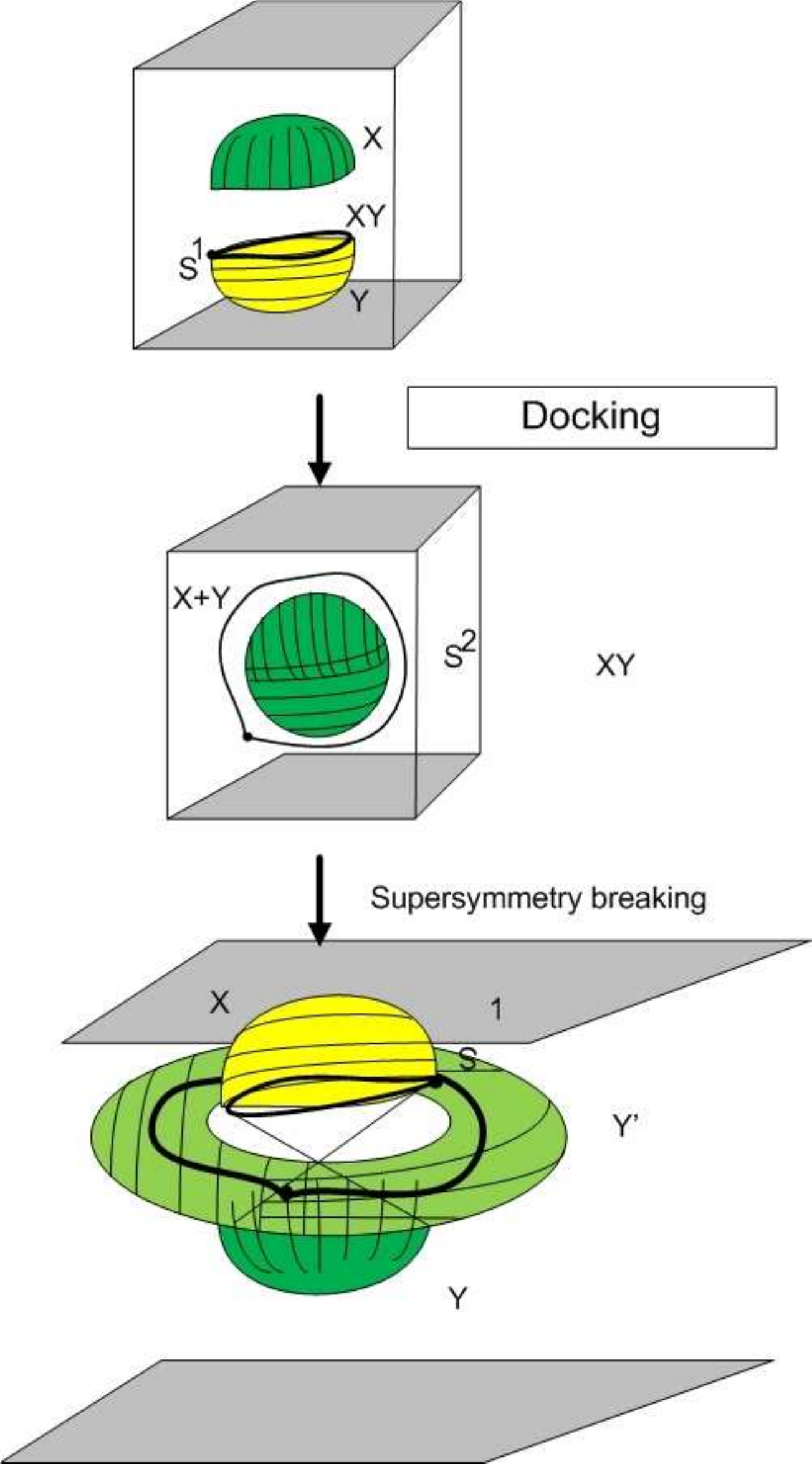}
\includegraphics[width=0.48\textwidth]{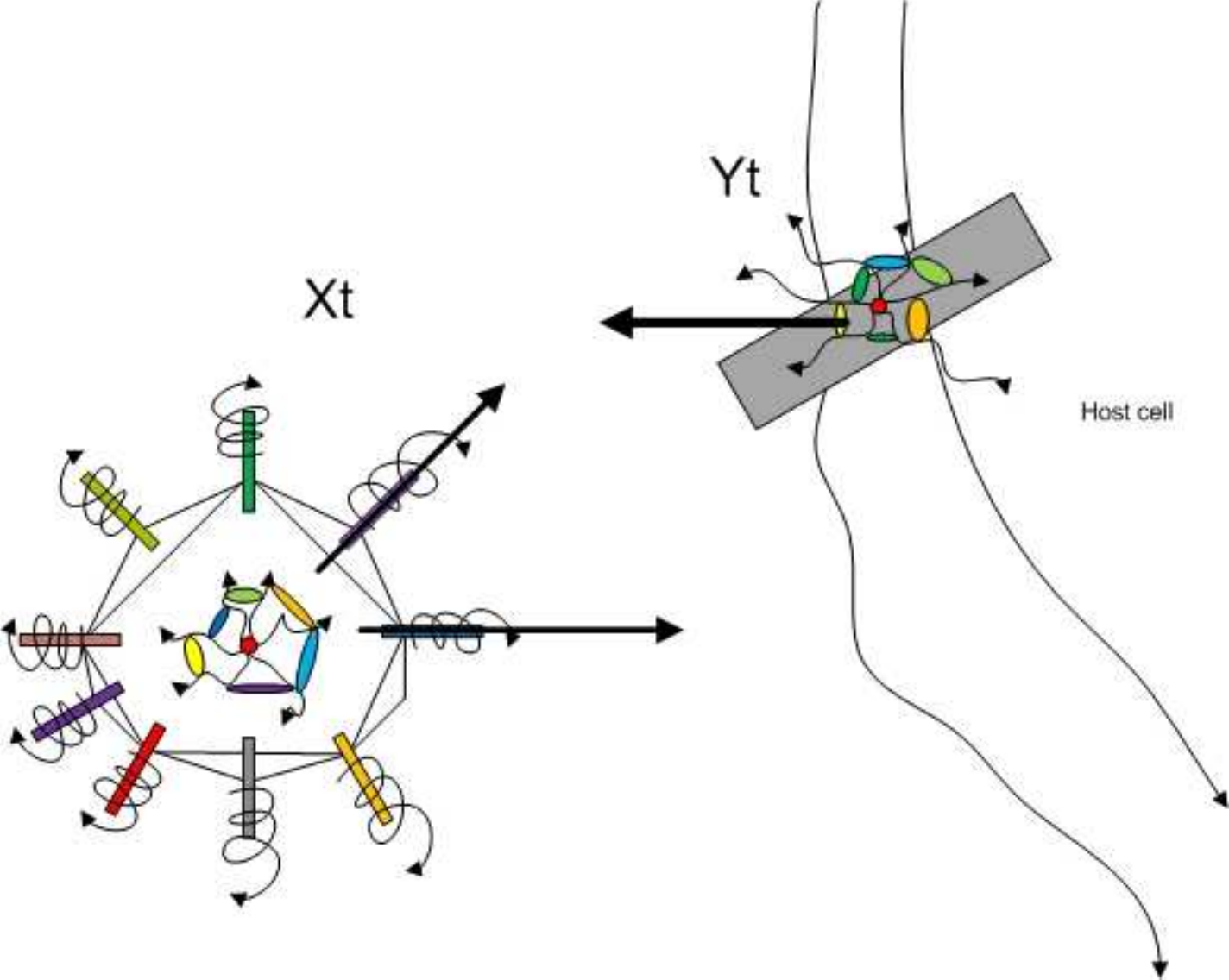}
\caption{ The picture shows the Mayer-Vietoris sequence of 2 biological objects attached to each other and the 
exchange of their supersymmetry while separated. On the right drawing picture, there  is a model of  Calabi-Yau surface of D-brane and anti-D-brane of protein-protein interaction between V3 loop of HIV virus and CD4 of host cell recepter as the cohomology sequence.}
\label{exact}
\end{figure}

Therefore, let us  introduce a spinor field from the  equation of viral particle attached  to the  host cell by using the behavior matrix of the Yang-Mills  field between ghost field of the gene expression in viral particle and anti-ghost field of host cell. We use the fact that second cohomology group of topological space is associated with the definition of spinor field and the Mayer-Vietoris sequence (Fig. \ref{exact}). We have a new master equation for  HIV viral V3 glycoprotein attached to the host  cell CD4 receptor protein. This is  an exact sequence of second cohomology group of gene expression and,  for a complex of chemical reaction of hydrogen bond in interaction between 2 living organisms, parasitism and host cell interaction of states. This behavior can be defined in the  framework of supersymmetry breaking along cell membrane while virus attaches  the cell in Wilson loop model of knots.  The  link  of biological moduli state space is $Z_{t}=X_{t}/Y_{t}\stackrel{Wilson loop}{\longrightarrow} Y_{t}/X_{t}$.

\begin{figure}[!t]
\centering
\includegraphics[width=0.48\textwidth]{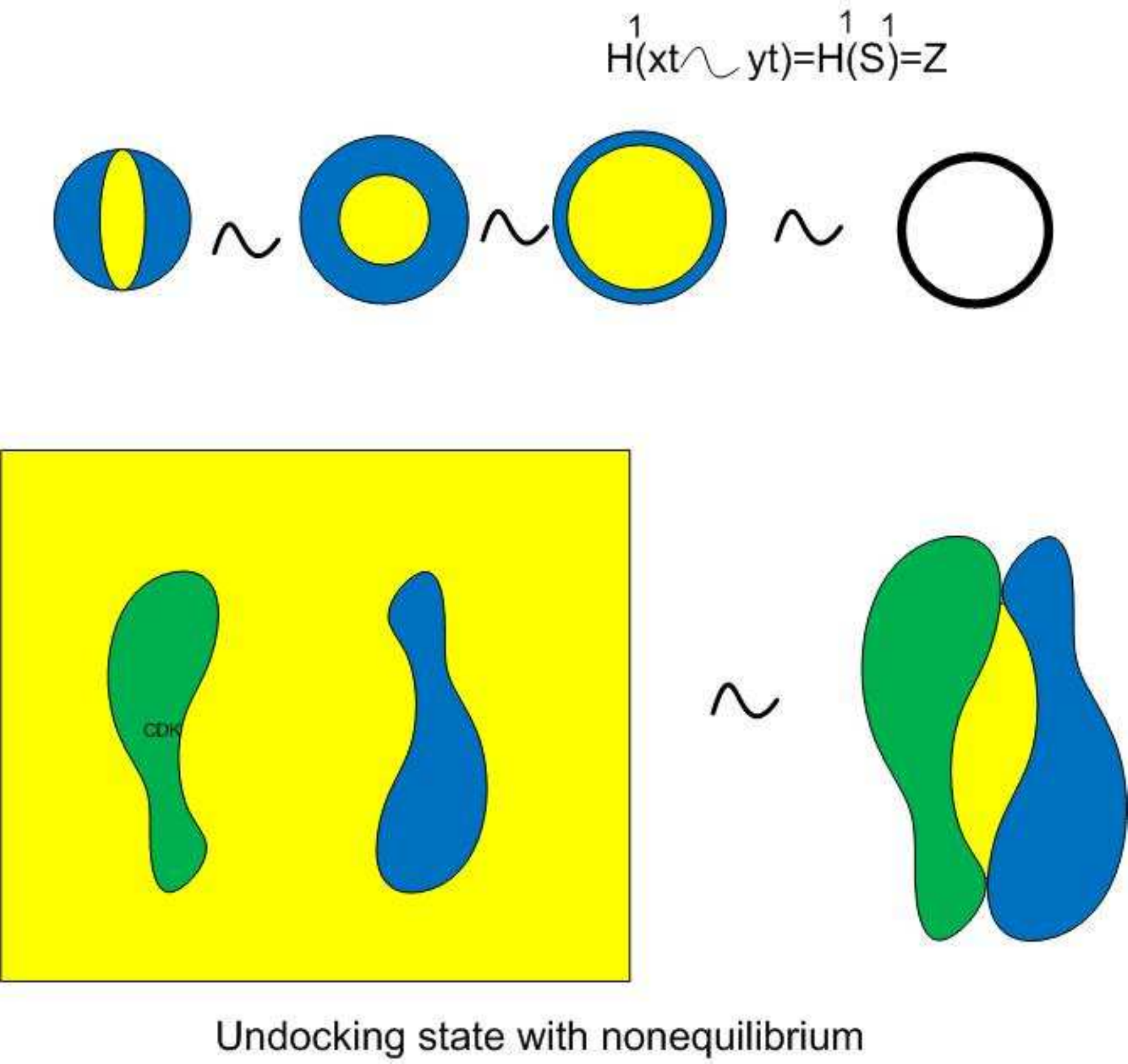}
\includegraphics[width=0.48\textwidth]{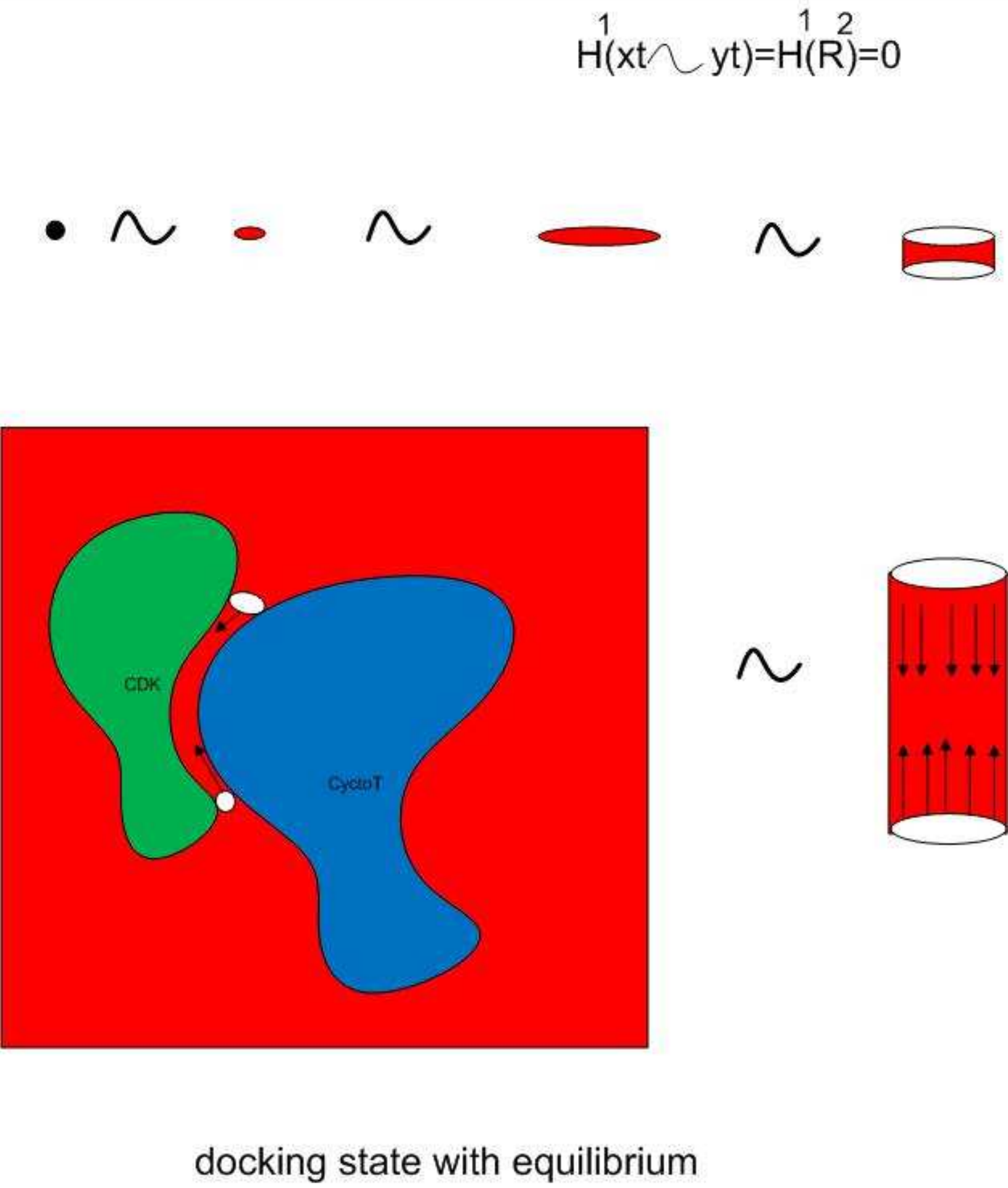}
\caption{ The picture shows the   duality in plane while the virus attaches  the surface of cell membrane with the same curvature and with different curvature. On the left side, it is  demonstrated the homotopy equivalent state of protein-protein interaction with different curvature while docking. The remaining area  between docking curvature is not zero. This area can induce a contractible to new 14-extradimensions by using the homotopy equivalent map. When we take the cohomology group to that surface,  the computation of cohomology is not zero. This fact is implying a  nonequilibrium state of docking between V3 loop HIV virus and host cell receptor CD4 in difference curvature. On the right hand situation of docking state, there  is an equilibrium state with the same curvature. The rest of the area  between docking can be contractible to a point and the cohomology group is zero. }
\label{dock}
 \end{figure}

Let $x_{t}$ be  a glycoprotein  of viral particle and $y_{t}$ be the anti-body host cell (Fig. \ref{dock}). The protein-protein interaction in organism generates an induced ghost field of cell cocyle $y_{t}-\alpha_{t}\simeq \beta_{t}x_{t}\simeq \epsilon_{t}$ where $\epsilon_{t}\in H^{3}(X_{t}/Y_{t})$ as superstatistical un-normal superdistribution in form of super-Lagragian over ghost field of viral RNA  $\Phi(x_{t})$ and anti-ghost field of host DNA $\Phi^{+}(y_{t})$ with its parity $ p(\Phi(x_{t}))+p(\Phi^{+}(y_{t}))=-1. $

Let  an equivalent class of DNA of organism by a pair of chromosome from host cell $x_{t}$ and from virus  $y_{t}$  be a Poisson structure in G-theory for viral replication
\begin{equation}
 ad_{\Phi_{i}^{+}(y_{t})}\Phi_{j}(x_{t})=\{\Phi_{i}(x_{t}),\Phi_{j}^{+}(y_{t})\}=\left\{\int Sdt,\int Sdt  \right\} = J^{k}(t)
\end{equation}
where $J^{k}(t)$ is  a master equation for gene expression between 2 genes of $\Phi_{i}(x_{t})$ gene and $\Phi_{j}^{+}(y_{t})$ gene.
It is a genotype in pair of ghost field $\Phi_{i}(x_{t})$ of gene and anti-ghost field of gene $\Phi_{i}^{+}(y_{t})$ in dominant states. The recessive state of gene expression is denoted by $\phi_{i}(x_{t})$ and $\phi^{+}_{i}(y_{t})$. A pair of ghost field and anti-ghost field with parity 2 is a spinor field of genotype of a supermanifold of pairs of living organism $(\mathcal{M},F^{\mu\nu})$.

  The equation of glycoprotein attached to the  CD4 T-host cell is given  by

\begin{equation}
 adj\left[   \begin{array}{cccccccccc}
   D\Phi_{1}(x_{t}) &0&0&0&0&0&0&0&0&0\\
0&D\Phi_{2}(x_{t})&0&0&0&0&0&0&0&0\\
0& 0&\cdots&0&0&0&0&0&0&0\\
0&0&0&0&0&0&D\Phi_{14}(x_{t})&0&0&0\\
0&0&0&0&0&0&0&d\Phi_{1}^{+}(y_{t})&0&0\\
0&0&0&0&0&0&0&0&d\Phi_{2}^{+}(y_{t})&0\\
0&0&0&0&0&0&0&0&0&  d\Phi_{3}^{+}(y_{t})\\
 \end{array}  \right ]  \nonumber
\end{equation}

\begin{equation}
=adj_{y_{t}}x_{t}=\{y_{t},x_{t}\}=0.
\end{equation}

  We have
 14 bases for ghost fields,  3 bases for anti-ghost fields (for details on definition of basis ordering, see Table \ref{table3})

\begin{equation}
D\Phi_{k}(x_{k}),k=1,2,\cdots 14 = \frac{d}{dS^{-1}}\int_{S^{-1}}^{\Phi_{k}} g_{kk}e^{-j2\pi\beta_{k}(x_{k})} d\beta_{k} := \frac{d}{d\beta_{k}}\int_{S^{-1}}^{\Phi_{k}}g_{kk} e^{-2\pi j\beta_{i}(x_{k})} d\beta_{k} \nonumber
\end{equation}
and
\begin{equation}
d\Phi_{p}^{+}(y_{t}),p=1,2,3=\frac{d}{dS^{1}} g^{pp}\Phi_{p}^{+}(y_{t})=\frac{g^{pp}de^{2\pi i\alpha_{p}(y_{p})}}{d\alpha_{p}}\nonumber
\end{equation}
with $i=j=\sqrt{-1}$ used for defined 2(14-11)=3+3=6 hidden states derived  by the coupling between these 2 complex numbers and  interaction of Mayer-Vietoris sequence between ghost  field and anti-ghost field by using the BV-cohomology  $  H^{3}(Y_{t})/H^{-14}(X_{t})\simeq  H^{11}(Y_{t}/X_{t})=[[\alpha]:Y_{t}/X_{t}\rightarrow S^{11} ] \ni [\epsilon_{t}^{\ast}]$ with homotopy class $[\alpha]$ deformed between ghost field and  anti-ghost field in evolutional spinor field of cell biology in 11 dimensions. From modern researches on unified theory in 11-dimensions, it might be possible to modify the Nahm equation for 3 forms of codon and to  study the induced Lie algebras of manifold tangent space  of living host cell  in 3-anti-ghost fields in CD4 structure of host cell receptor.


We have a change of state and a dual state in genotype measured by the covariant derivative of connection $\bigtriangledown_{g_{ij}}g^{jk}=0$.  When viral particle   inflects or contacts the host cell with zero curvature between the  gene expression over connection of genetic code,  we can use the Yang-Mills  field $*F^{\nu\mu}$ of viral glycoprotein and $F_{\mu\nu}$ of host cell receptor protein. We obtain  a modified  Seiberg-Witten   docking equation over connection of genetic code $[A_{i}]$ for gene expression  over the curvature of docking, that is 
\begin{equation}
D^{[A]}\Phi_{i}(x_{t})=0\nonumber
\end{equation}
and in equilibrium of docking, the interaction of behavior of gene between $[A]$ in  virus and $[A]^{\ast}$ in  host cell are anti-self dual in mirror symmetry theory, that is 
\begin{equation}
F_{[A]}^{\mu\nu}=\ast F_{\mu\nu}^{[A]^{\ast}}.
\end{equation}
where $D^{[A]}$ is a Dirac operator over the connection $[A]$. When it is not in contact or docking to between anti-gene in ghost field $\Phi_{i}(x_{t})$ and anti-ghost field and $[A]^{\ast}$, it  is a dual genetic code in L-isomer.  The difference of Chern-Simon current of curvature of proteins  will not be zeros and the plot of the gene spectrum  will not be on the same line.

\subsection{Algebraic Construction of Trash DNA}

As discussed above, a central dogma is an exact sequence of cohomology. It is composed by a chain sequence and a co-chain sequence of 3
 types of complex over the fiber  of superspace of time series data of genetic code as an open set in sheaf cohomology.
\begin{itemize}
\item $D-brane(C_{\cdot}(X_{t})^{DNA,R}):=C_{\cdot}(\mathcal{O}_{D};X_{t})$ be a pre-sheaf of coordinates of living organism in DNA sequence on an open strand. It is a covery of sites in the  Grothendieck topology over sheaf cohomology.
\item $D-brane(C_{\cdot}(Y_{t})^{RNA,R}):=C_{\cdot}(\mathcal{O}_{R};X_{t})$
\item $D-brane(C_{\cdot}(X_{t}/Y_{t})^{Pro,R}):=C_{\cdot}(\mathcal{O}_{<D,R>};X_{t}/Y_{t}) :=C_{\cdot}(\mathcal{O}_{ P};X_{t}/Y_{t}) $
\item $anti-D-brane(C^{\cdot}(X_{t}^{\ast})^{DNA,L}):=C^{\cdot}(\mathcal{O}_{D^{\ast}};X_{t}^{\ast})$
\item $anti-D-brane(C^{\cdot}(Y_{t}^{\ast})^{RNA,L}):=C^{\cdot}(\mathcal{O}_{R^{\ast}};Y_{t}^{\ast})$
\item $anti-D-brane(C^{\cdot}(X_{t}^{\ast}/Y_{t}^{\ast})^{Pro,L}):=C^{\cdot}(\mathcal{O}_{<D,R>^{\ast}};X_{t}^{\ast}/Y_{t}^{\ast})  :=C^{\cdot}(\mathcal{O}_{P^{\ast}}; X_{t}^{\ast}/Y_{t}^{\ast} ).$
\end{itemize}
We have an equivalent class of transition states in a central dogma functor and twistor between these 3 complexes in both left and right symmetry 
and also adjoint functor $ C_{\cdot}(X_{t})^{DNA,L,R}\rightarrow C_{\cdot}(Y_{t})^{RNA,L,R} \rightarrow C_{\cdot}(X_{t}/Y_{t})^{Pro,L,R}. $

\begin{Definition}
A state of gene expression in DNA replication as a function of time series of spinor fields in genetic code is independent  of the chosen  coordinates as 8 hidden states from  DNA to other chain of DNA when cell divisions are defined by the based pairing relationship over site and sieve of the sheaf cohomology in the  Grothendieck topology,
\begin{equation}
[s_{1}]:\mathcal{O}_{D,[A],X_{t}}\rightarrow \mathcal{O}_{D,[T^{\ast}],X_{t}^{\ast}}, [s_{3}]:\mathcal{O}_{D,[C],X_{t}}\rightarrow \mathcal{O}_{D,[G^{\ast}],X_{t}^{\ast}}.
\end{equation}
Let $\mathcal{O}_{ X_{t}}$ be the sheaf of an organism  superspace.
We have an equivalent class of coupling states between spinor fields in the 2 side of D-brane and anti-D-brane of DNA based pairing
 with $[A],[C]\in D-brane(C_{\cdot}(X_{t})_{DNA}^{R}),[T]^{\ast},[G]^{\ast}\in anti-D-brane(C^{\cdot}(X_{t})_{DNA}^{R})$ where $R$ is a right handed supersymmetry of DNA molecule as D-brane.
 These 2 states are explicit states  of gene transfer expression to warp  state between  active protein in the central dogma 
during the replication process
\begin{equation}
 d[s_{1}],d[s_{3}]:D-brane(C_{\cdot}(X_{t})^{DNA,R})\rightarrow anti-D-brane(C_{\cdot}(X_{t})^{DNA,R}),d[s_{1}]\mapsto [s_{3}^{\ast}],d[s_{3}]\mapsto [s_{1}^{\ast}].
\end{equation}
It is a wrap state between DNA based pairs in the  replication process in the central dogma in the context of sheaf cohomology theory for biology.
A state of gene expression (geneon state) in the DNA transcription process from  DNA to  RNA in central dogma is defined by the coordinates over the state of  based pairing relationship as
\begin{equation}
[s_{2}]:\mathcal{O}_{[A]}\rightarrow \mathcal{O}_{[NA]}, [s_{4}]:\mathcal{O}_{[C]}\rightarrow \mathcal{O}_{[NC]}.
\end{equation}
where $[NA]_{X_{t}}=[T]_{Y_{t}}.$

A state of gene expression in DNA is a self dual transcription (transposon state) from  DNA to  DNA  in passive state of inactive gene in trash DNA area of 98\%. A central dogma is defined by based pairing relationship defined as
\begin{equation}
[s_{5}]:\mathcal{O}_{[T]}\rightarrow \mathcal{O}_{[T^{\ast}]}, [s_{7}]:\mathcal{O}_{[G]}\rightarrow \mathcal{O}_{[G^{\ast}]}.
\end{equation}
where $[NA]_{X_{t}}=[T]_{Y_{t}}.$

 Let $X_{t}$ be a superspace of time series data of genetic code in DNA in mirror symmetry of reversed transcription of retransposon of HIV viral gene with host organism and $Y_{t}$ be a superspace 
of  time series data of retrotransposon map $D^{\ast}$ in reversed direction   in host organism into viral $R^{\ast}$.

A state of gene expression in DNA reversed transcription (retrotransposon state) from  $DNA^{\ast}:=D$ to a  nonactive state (passive state of left symmetry) $RNA^{\ast}=R^{\ast}$ in central dogma is defined by based pairing relationship defined as
\begin{equation}
[s_{6}]:\mathcal{O}_{[T]}\rightarrow \mathcal{O}_{[NA](NA=T_{L})}, [s_{8}]:\mathcal{O}_{[G]}\rightarrow \mathcal{O}_{[NC]([NC=G_{L}])}.
\end{equation}
 \end{Definition}
They are site and sieve in the Grothendieck topology of sheaf cohomology. The classical central dogma in biology extends to  a new K-theory of central dogma  independent of coordinates of the chosen gene in the gene expression. We have a universal objection in the category of  gene objects among all species with site and sieve over the Grothendieck topology in form of adjoint functor of codon defined by triplets of left and right adjoint functor in the left translation and right translation over extended central dogma functor map on  supermanifold fiber of living organism, 
\begin{itemize}
\item $s_{1},s_{3}:\mathcal{O}_{D}\rightarrow \mathcal{O}_{D}$(DNA replication in active state of geneon) with mirror symmetry of dual state in trash  DNA $s_{5},s_{7}:\mathcal{O}_{D^{\ast}}\rightarrow \mathcal{O}_{D^{\ast}}$ (passive state of non-geneon, transposon, self dual state replication).
\item $s_{2}.s_{4}:\mathcal{O}_{D}\rightarrow \mathcal{O}_{R}$ (transcription) with mirror symmetry of dual state in trash as retransposon DNA$ s_{6},s_{8}:\mathcal{O}_{R^{\ast}}\rightarrow \mathcal{O}_{D^{\ast}}$.

\item Ribozyme $ \mathcal{O}_{R}\rightarrow \mathcal{O}_{D}$  with mirror symmetry of dual state in trash DNA, unknown non-coding RNA hidden state$ \mathcal{O}_{D^{\ast}}\rightarrow \mathcal{O}_{R^{\ast}}$.
\item siRNA $ \mathcal{O}_{R}\rightarrow \mathcal{O}_{R}$ with mirror symmetry of dual state in trash DNA, unknown un-coding RNA hidden state $\mathcal{O}_{R^{\ast}}\rightarrow \mathcal{O}_{R^{\ast}}$.
\end{itemize}
All the above states  are  open states (single strand of DNA and RNA in both direction) so they are exited state of genetic code.
So they need dual pairs to be a ground state with a closed spin shell. We will define a wave function of states over all 4 types of gene by using differential 2 forms of closed state in the following  Section. The transition will be defined by using a modified Dirac operator  with left and right transitions over 4 types of ground states to exited open states which we define and compute in the last Section.

\begin{Theorem}

There exists a super-codon hidden state associated with the left and right adjoint translation over unoriented   tangent of the supermanifold of a living organism  which reacts as hidden repeated translation and transcription process in the trash area of DNA.
\end{Theorem}

{\bf Proof.}\\
We have equivalent class of transition state in a central dogma functor and twistor between these 3 complexes in both left and right symmetry 
and also adjoint functor $ C_{\cdot}(X_{t})^{DNA,L,R}\rightarrow C_{\cdot}(Y_{t})^{RNA,L,R} \rightarrow C_{\cdot}(X_{t}/Y_{t})^{Pro,L,R}. $
All 8 gene-s-orbitals for $s1-s8$ have hidden 8 p-orbitals of below chain of ribozyme and RNAi with their dual states.
All these states generate a codon table with 64 codon and anti-codon with $8\times 8=64$ codon states.
The extra of 4 chain in central dogma is a protein as geneon and anti-protein while docking in passive hidden state as anti-geneon. We define a pair $\mathcal{O}_{<D,R>}\rightarrow \mathcal{O}_{P}:=\mathcal{O}_{D}\rightarrow \mathcal{O}_{R} \rightarrow \mathcal{O}_{P}  $, a protein output from the translation process and $\mathcal{O}_{<D^{\ast},R^{\ast}>}\rightarrow \mathcal{O}_{P^{\ast}}$ a protein for control error and also reacts as a switch open and closed state of gene expression in reversed transcription, $\mathcal{O}_{P}\rightarrow \mathcal{O}_{<D^{\ast},R^{\ast}>}$ a complex of ribosome and protein in HIV reversed transcription. 

The source of protein layer, as docking and undocking states,  shows a tensor correlation layer in the  behavior of the gene expression in both active states of $2\%$ area and in hidden transition states of control expression of active states induced by a spinor field in the left supersymmetry of  $98\%$ area  of inactive  undocking states of non equilibrium of repeated inactive hidden states of spinor field of gene. It is  a supersymmetric property of central dogma of all possibilities of transcription and reversed transcription from 
DNA $(D)$ to RNA $(R)$ and to protein $(P)$ as a $(8+4)$ chain of short exact sequences in sheaf cohomology. The extra canonical layer $\mathcal{O}_{P}$ and $\mathcal{O}_{P^{\ast}}$ in central dogma is induced from the group action between $D$ and $R$ in left and right translation. It is a network of gene from both DNA and RNA  having a  tensor correlation to each other as a tensor field from gene expression $\mathcal{O}_{<D,R>}:=\mathcal{O}_{D\otimes R}\simeq \mathcal{O}_{P}$ as free product induced from superspace D and R ($P^{\ast}$ induced from $\mathcal{O}_{<D^{\ast},R^{\ast}>}:=\mathcal{O}_{P^{\ast}}$ in left supersymmetry).

We can define a new induced   coupling between left and right supersymmetry  field of central dogma  in pair  of $(D,R)$. 
We say that two functors with left and right supersymmetry in biology are defied by right translation of DNA and RNA in central 
dogma and left revered transcription of  DNA and RNA in extend central dogma. We use a Lie superalgebra with adjoint functor as the Poisson  brackets for the quantization of states between hidden transitions state jumping into hidden geneon state in transposon and retrotransponson. So we have an adjoint representation of supercodon in left representation from reversed translation. This alphabet starts, for example, with the codon $AUG$ and can be  adjoint transformed to $(GUA)_{R}$, $(AUG)_{L}$ and $(GUA)_{R}$ in the left symmetry.  Eeach central dogma chain will duplicate 4 canonical states of active and passive canonical states of protein layer in left and right adjoint translation over codon.   A superanti-codon in right translation  is induced by the  interaction  of equilibrium and not equilibrium state of docking  and undocking states in the  feedback loop of ghost  and anti-ghost fields in the sheaf of supermanifold   of living organism by $ \mathcal{O}_{P}: \mathcal{O}_{DNA}\rightarrow \mathcal{O}_{RNA} $ a  right translation of genetic code, $ \mathcal{O}_{P^{\ast}}: \mathcal{O}_{RNA^{\ast}}\rightarrow \mathcal{O}_{DNA^{\ast}}$.  A left translation of reversed transcription of retrotransposon
 is adjoint if it forms an adjunction control feed back loop of adaptive behavior in the evolution among all living organisms
 induced from the interaction of docking and non-docking states between passive and active protein representation layer as
 parallel transport in spinor field in Kolmogorov space of genetic code $ \mathcal{O}_{
P}  \dashv  \mathcal{O}_{P^{\ast}} $. It is a  hidden control force  for the  interaction between left and right translation as source of transponson in repeated inactive gene of the  transition state between the coupling state of DNA and RNA in transcription process.
 
This means that they are equipped with natural transformations of $\beta-$cycle and $\alpha-$ co-cycle of trivialization  state
 over the fiber of sheaf in supermanifold of all organisms by the universal object as site and sieve in Groethdieck topology
 $ \beta:1_{\varphi_{transposon}} \mathcal{O}_{DNA^{\ast}} \rightarrow   \mathcal{O}_{P^{\ast} } \circ   \mathcal{O}_{P } $
 and $ \alpha :\mathcal{O}_{P } \circ   \mathcal{O}_{P^{\ast} } \rightarrow 1_{\varphi_{retrotransposon}} \mathcal{O}_{RNA^{\ast}} $.
We notice, from the proof, that the retrotansposon is independent of the chosen coordinate of nonfunctional RNA,
  satisfying the triangle identities of supercodon and super-anti-codon, that is the compositions 
$ \mathcal{O}_{P^{\ast} } \stackrel{\mathcal{O}_{P^{\ast} }\beta}{\rightarrow}   \mathcal{O}_{P^{\ast} } 
\mathcal{O}_{P } \mathcal{O}_{P^{\ast} }    \stackrel{   \beta \mathcal{O}_{P^{\ast} }}{\rightarrow } \mathcal{O}_{P^{\ast} } $,
where $[A]_{amino}\in \mathcal{O}_{p}$ is an observed state of amino acid in left supersymmetry of supercodon 3 alphabets with other 2 hidden states from feedback loop of transposon hidden protein layers from left translation of reversed transcription. For the right translation in right supersymmetry, we have an adjoint representation,
$  \mathcal{O}_{P }\stackrel{\alpha \mathcal{O}_{P } }{\rightarrow}    \mathcal{O}_{P } \mathcal{O}_{P^{\ast} } \mathcal{O}_{P }    \stackrel{  \mathcal{O}_{P } \alpha}{ \rightarrow } \mathcal{O}_{P} $
that are identities. The left or right adjoint of any functor, if it exists, is unique up to a unique isomorphism.

We have totally 4 chain sequences with extra $8\times 4=32 $ states of central dogma cohomolgy  as follows,

\begin{itemize}
\item $\mathcal{O}_{<D,R>}\rightarrow \mathcal{O}_{P}$ direct translation of protein (8 hidden state induced from canonical form of 
8 possible chain complexes $\mathcal{O}_{<D,R>}$  which have been defined above).
\item $\mathcal{O}_{P^{\ast}}\rightarrow \mathcal{O}_{<D,R>}$ indirect translation of DNA and RNA polymerase to open and closed gene expression of 8 hidden states induced from canonical form of 
8 possible chain complexes $\mathcal{O}_{<D,R>}$  which we have defined above.
\item $\mathcal{O}_{P^{\ast}}\rightarrow \mathcal{O}_{<D^{\ast},R^{\ast}>}$, indirect revered translation of $DNA^{\ast}$ host cell and $RNA^{\ast}$, viral retransposon   forming a ribonucleocomplex in reversed transcription process to open and close gene expression of 8 hidden states induced from canonical form of 
8 possible chain complexs $<D,R>$  which have been defined above.
\item $\mathcal{O}_{<D^{\ast},R^{\ast}>}\rightarrow \mathcal{O}_{P}$, viral capsid protein reversed translation of 8 hidden state induced from canonical form of 
8 possible chain complexes over the sheaf $\mathcal{O}_{<D,R>}$  which  have been defined above. 
\end{itemize}

We already gave  an adjoint representation in $(8+3)$ dimensions with left and right supersymmetry of codon as the quantization state of Poisson bracket in adjoint function as the group action in  left and right translation with adjoint representation  of genetic code,   called supersymmetry-codon state.\\
{\bf  End of the proof. }

\vspace{5.mm}

Specifically, $C_{\cdot}(\mathcal{O}_{ D,R,P})$ is a chain complex of triplet states of central dogma with passive and active layer of protein translation. These extra 32 codons are called supercodon and superanti-codon. It comes from supersymmetry property of central dogma as defined above. They are defined for each pairs of doublet states of central dogma $C_{\cdot}(\mathcal{O}_{D,R})$. One can  induce 
4 active layers over active protein $P$ with $C_{\cdot}(\mathcal{O}_{D,R})\rightarrow  C_{\cdot}(\mathcal{O}_{P})$  and also dual state of passive feedback loop 
of protein layer in dual state $C_{\cdot}(\mathcal{O}_{P^{\ast}})\rightarrow C_{\cdot}(\mathcal{O}_{D,R}).$ This can be the source of Chern-Simons 3 form with the definition of differential 3 forms over a triplet  state of genetic code as the Chern-Simons current for biology. Totally we have 
$64+32=96$ states of supercodon and anti-codon complete code in supersymmetry theory with a ratio of expression in trash
 DNA in $98\%$  area.

From this construction, we have 20 amino acids because of the  adjoint representation  amplified  by the ratio  $64/32$, where the hidden $32$ states are  a partition function of hidden states of central dogma. The other interesting output is a number of human chromosomes where pair states reflect adjoint representation of left and right supercodon with the lack of last pairs of chromosome due to the break down into small piece of nonfunctional RNA. It  acts as a  passive hidden docking protein state for complementary  sex chromosome in supersymmetry as a power of regeneration with equal sex in poputation of human. It is a hidden passive layer of sheaf cohomology in biology to fullfill equilibrium state of docking. The equilibrium needs $8\times 3=24$ (3 triplet states reflect also 24 types of codons  represented by  20 amino acid with 3 stop and one in start with common in amino acid $AUG$) chromosomes in human genome.

\begin{figure}[!t]
 \centering
\includegraphics[width=0.5\textwidth]{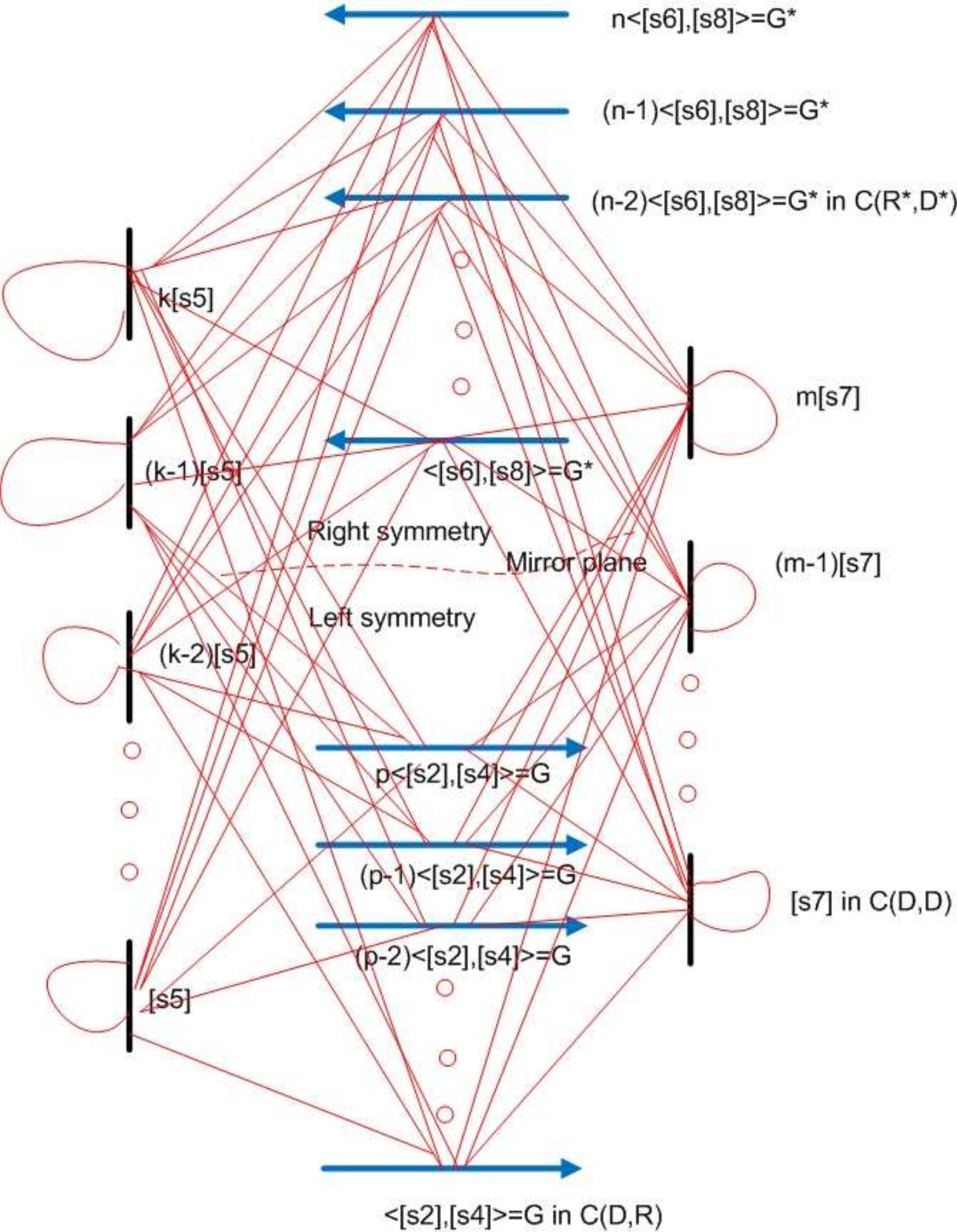}
  \caption{In the picture,  we show transition state in trash  DNA between transposon, retrotransposon and active gene in the moduli-state  space   Diophantine equation. The transitions have active and passive layer. The active layer is in the active genes while the passive layer is in the area of trash DNA. The transition between states and hidden states of  these 2 layers satisfies the  moduli -state space Diophantine equation. The modulo makes the gene expression into circle and in fiber of the transition state.\label{fig_moduli} }
\end{figure}
 
Therefore, we have a relationship between equivalent class of hidden states (see Fig. \ref{fig_moduli} for details) and states as transition in trash DNA defined as

\begin{equation}
[s_{6}] \equiv [s_{8}] \hspace{0.5cm}  mod \hspace{0.5cm} \varphi^{transposon},\hspace{0.5cm} [s_{5}] \equiv [s_{7}]\hspace{0.5cm}  mod \hspace{0.5cm} \varphi^{retrotransposon}
\end{equation}

\begin{equation}
[s_{1}] \equiv [s_{3}]\hspace{0.5cm}  mod \hspace{0.5cm} \varphi^{anti-geneon}\hspace{0.5cm} [s_{2}] \equiv [s_{4}]\hspace{0.5cm}  mod \hspace{0.5cm} \varphi^{geneon}.
\end{equation}

We have $\varphi^{retrotransposon}=G^{\ast}, (G^{\ast})^{\ast}=G$ to be a geneon, so we have
$k[s_{5}]+m[s_{7}]=n[G^{\ast}],  \exists k,m, d \in \mathbb{Z} , (k,m)=d,d|n. $
This equation can be recursive if $k=m-1$ for example. It is a transition function between state and hidden state as explained in the diagram of active and passive state of gene expression moduli state space equation. A detailed discussion of this point can be found in \cite{ssm}. 
For the trash area of DNA, we have a new definition of the system of moduli state space.

\begin{Definition}
The equation of transposon and retrotransposon in trash DNA is a system of moduli state space  with cycle and 
co-cycle mode as coefficients for the evolution field in the Dionphatine equation,
\end{Definition}

In total,  we have  32 more extra  states of  noncoding RNA, i.e. tRNA, sRNA, and other unknown non-function RNA where  transition state from intron, telomere and repeated $(CA)_{n}$ code for extra dynamic codon-anticodon of genetic code come from the  supersymmetry  and cohomology theory for biology.

\vspace{3.mm}

{\bf Remark: } The calculation of average $98\%$ of trash DNA and $2\%$ of active side of gene expression in living organism, especially in the human genome, can be obtained from the ratio of possible active state of geneon and all inactive state
 of non-geneon state in cohomology of the above central dogma. We have only 2 states for transcription of well known 
$s_{2},s_{4}:D\rightarrow R\rightarrow P$ over $64+32=96$ hidden states with canonical form of hidden state. 
Therefore $2/96=0.020833\simeq 0.02=2\%$. The rest  $98\%$ is induced from the coupling states  of  hidden states of cohomology in central dogma.
 \begin{Definition}
A sheaf cohomology of replication cycle of retransposon is defined as a short exact  sequence in loop as follows,
\begin{equation}
0_{D-brane}\rightarrow \mathcal{O}_{X_{t}} \rightarrow \mathcal{O}_{Y_{t}} \rightarrow \mathcal{O}_{X_{t}/Y_{t}} \rightarrow 0_{anti-D-brane}  \rightarrow \mathcal{O}_{Y_{t}^{\ast}/X_{t}^{\ast}}\rightarrow \mathcal{O}_{X_{t}^{\ast}} \rightarrow \mathcal{O}_{Y_{t}^{\ast}}  \rightarrow 0_{D-brane}
\end{equation}
\end{Definition}
Starting from these definitions and algebraic constructions, we can compute all the hidden state in codon considering a a Chern-Simons 3-form for DNA.

\begin{figure}[!t]
 \centering
\includegraphics[width=0.38\textwidth]{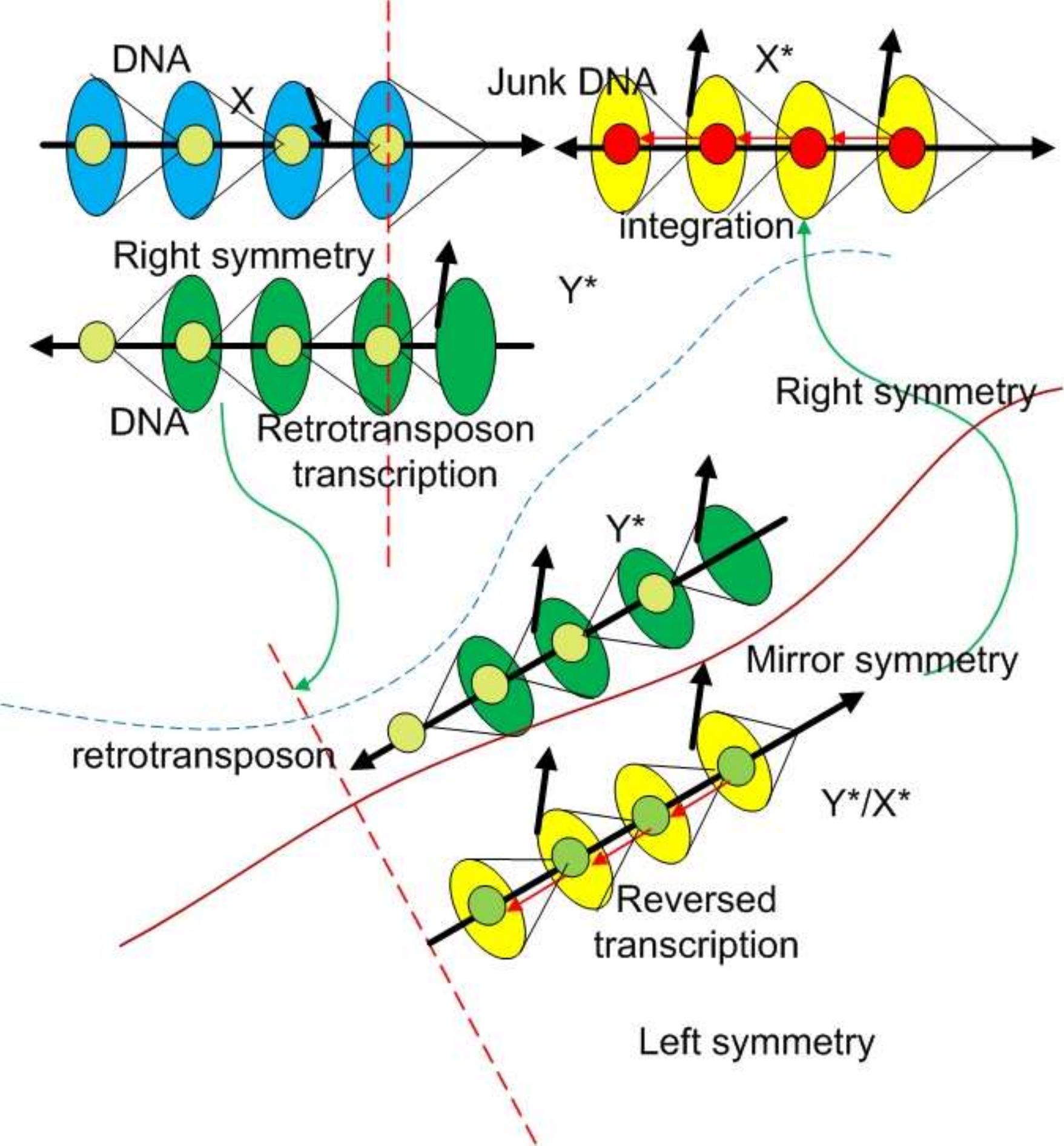}
\includegraphics[width=0.28\textwidth]{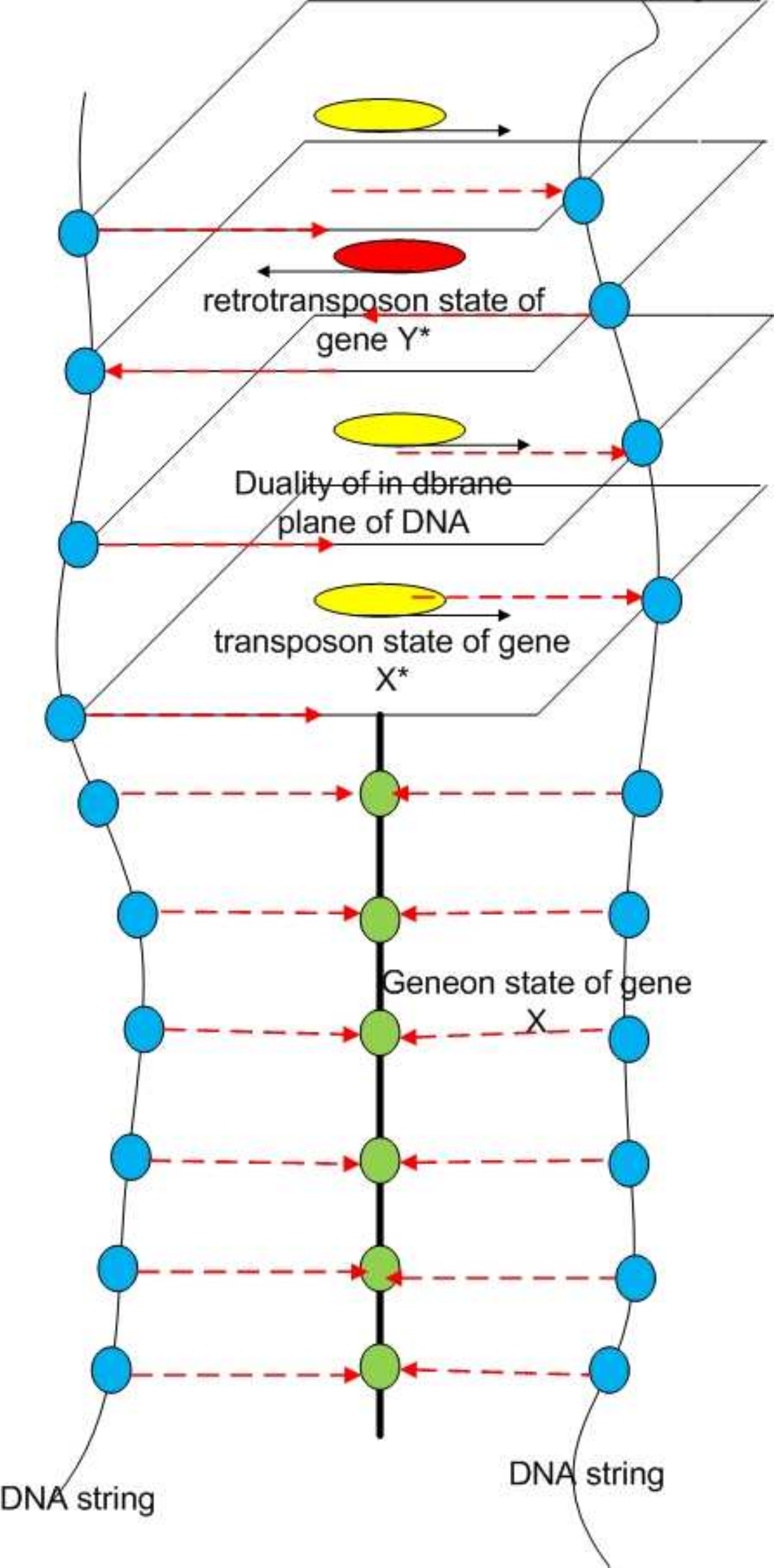}\\
\includegraphics[width=0.45\textwidth]{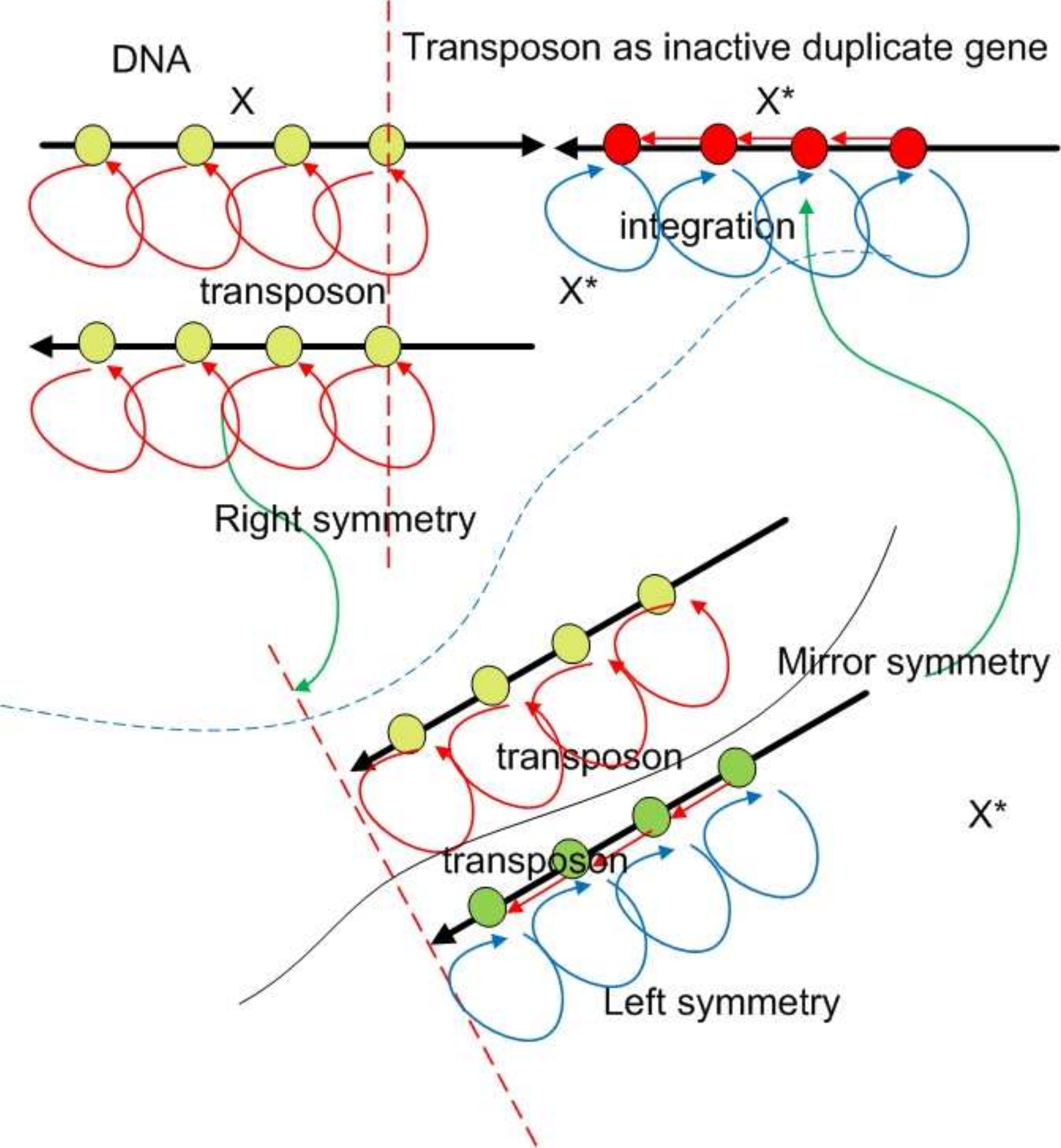}
  \caption{In the above drawing on the right, the picture shows the  cohomology sequence of central dogma of DNA in the  2\% active area or the  trash DNA with retrotransposon hidden state as duplication geneon loop space orbital in hidden plane of D-brane and anti-D-brane in DNA molecule with repeated pattern. For active area of geneon, the shape of cohomology sequence  of DNA space is contractible to points and lines.  This area is rich of active states of transcription gene. In the above picture on the left, the  mechanism of reversed transcription of retrotransposon is showed.
This process is an example of HIV infection and induces an obstruction component by turning a  hidden direction of supersymmetry of active gene to an inactive direct intervening  process of control gene expression in trash DNA area.
 Therefore the cohomology sequence of docking state is non-zero with these induced  obstruction components and cannot go to zero.  The  immunosystem induces a loop of control that destroys the  central dogma developing  an  infinite loop process with a  immunodefieciency syndrome. The reasons why the gene is not active is that  organic molecules can  work in living organism on right symmetry. Retransposon turns gene to insert back into left and makes  them inactive in repeated patterns. On below drawing, we show the mechanism of loop space with spinor field  in time series of genetic code. It is a self dual map
 of genetic code to itself in loop for holding memory in gene as in active part of trash DNA. The flip-flop model 
of open and closed gene into loop can be explained by inserting and deleting gene by transposon. This mechanism turns DNA into a knot model where  a twistor appear as a modified Wilson loop for biology.   \label{fig5}  }
 \end{figure}

\section{Computation of all hidden states of codon with Chern-Simons 3-forms in DNA}

We  use differential  2 forms induced from the equilibrium $d^{2}=0$ over 8 states and hidden states to  generate codons with 64 basis in exact sequence. Then we use triplet states over presentation layer of protein $P$ to generate the Chern-Simons 3 form over the canonical form of exact sequence of triplet state $P$ with its dual $P^{\ast}.$ The computation is as followings.
For a chain central dogma of codon,  we induce 64 states from differential 2 forms of spinor field over short exact cohomology
 over 8 states. Let $C_{\cdot}(s_{1},s_{2},\cdots, s_{8}):=C_{\cdot}(\mathcal{O}_{<D,R>})\coprod C_{\cdot}(\mathcal{O}_{ <D,D>}) \coprod C_{\cdot}(\mathcal{O}_{<R,D>}) \coprod C_{\cdot}(\mathcal{O}_{<R,R>})  :=C_{\cdot}(D,R)$ which induces $n-states, C_{n}(s_{1},s_{2},s_{3},\cdots ,s_{8};\mathcal{O}_{X_{t}} )$ and a dual state be co-chain $C^{n}(s_{1}^{\ast},s_{2}^{\ast},\cdots, s_{8}^{\ast};\mathcal{O}_{X_{t}^{\ast}}):=C^{\cdot}(\mathcal{O}_{<D^{\ast},R^{\ast}>})\coprod 
C^{\cdot}(\mathcal{O}_{<D^{\ast},D^{\ast}>}) \coprod C^{\cdot}(\mathcal{O}_{<R^{\ast},D^{\ast}>}) \coprod 
C^{\cdot}(\mathcal{O}_{<R^{\ast},R^{\ast}>})  :=C^{\cdot}(D^{\ast},R^{\ast};\mathcal{O}_{X_{t}^{\ast}})$ . Let the state $s_{i}$ be  $\varphi_{n}(s_{i})=: g_{1}\rightarrow g_{2},\rightarrow g_{3} \cdots \rightarrow g_{n} $
  a wave function of gene expression in any direction of  transcription process in the active area and 
let the state $s_{i}^{\ast}$ be  $\varphi_{n}(s_{i}^{\ast})=: g_{1}^{\ast}\leftarrow g_{2}^{\ast},\leftarrow g_{3}^{\ast} \cdots \leftarrow g_{n}^{\ast} $
 be a wave function of dual state of gene expression in reversed direction of  transcription process in the  inactive area
 of trash DNA. From our definition of supersymmetry in central dogma,  we use the 
 notation of differential one form $d_{i}\varphi_{n}$ at the position $i$ of the state $s_{i}$, coming from the hidden 8 states, 
in such a way that we have  homogeneous coordinates over the tangent of the unoriented supermanifold of $\mathbb{H}P^{1}$ with 1 at the position of state $s_{i}$ to induce the chart as coordinate in the trivialization for the translation of the coordinate of the state as a gene expression.  

\begin{equation}
\cdots \rightarrow C_{n}(s_{1},s_{2},s_{3},\cdots ,s_{8};\mathcal{O}_{X_{t}}) \stackrel{\partial_{[s_i]}}{ \rightarrow}  C_{n+1}(s_{1},s_{2},s_{3},s_{i-1}\cdots ,1_{i},s_{i+1} ,s_{8};\mathcal{O}_{X_{t}}) \rightarrow \cdots 
\end{equation}
and
\begin{equation}
\cdots \rightarrow C^{n}(s_{1}^{\ast},s_{2}^{\ast},s_{3}^{\ast},\cdots ,s_{8}^{\ast};\mathcal{O}_{X_{t}}) \stackrel{d_{[s_i]^{\ast}}}{ \rightarrow}  C^{n-1}(s_{1}^{\ast},s_{2}^{\ast},s_{3}^{\ast},s_{i-1}^{\ast}\cdots ,1_{i},s_{i+1=7}^{\ast} ,s_{8}^{\ast};\mathcal{O}_{X_{t}})\cdots \rightarrow 
\end{equation}

We have an open set $[s_{i}]\in U_{i}$ with co-cycle $g_{ij} \in U_{i}\cap U_{j}, g(x)=gx$ for the  transcription process and $g(x)=xg$ for the reversed transcription over the supermanifold of the living organism. We use the fact that, from differential geometry, the wave function of geneon, transposon, retrotransposon, and anti-geneon are independent of the chosen coordinate as function of the gene. When the cohomology sequence of the central dogma is a short exact sequence of predefined superspace in gene with property of sheaf cohomology, it   induces an  infinite sequence of cohomolgies with equilibrium states in nature.
  
The differences between living organism and non-living organism is the life energy over brane in nonactive DNA with octomer as memory  in loop space in time series data with D-brane-interaction as induced by the Chern-Simons current with a coherent state like a superconductor in Josephson tunnel effect  between non-stationary part of active geneon and inactive part of retrotransposon. The  transposon switches the  nonstationary Chern-Simons current tunnel through a stationary path in the region of repeated gene in trash area of 98\%. The algebraic definition of the life energy of a living organism is a ghost field in physics when the living organism is living and when it dies:  it will wrap and transform to anti-ghost field with hidden  state over  fifth-dimensions and wrap back to the retrotransposon ground field in a new born organism in the region of trash DNA with no memory. It  can be transmitted between the 4th and 5th dimension since the extradimension is a non-oriented supermanifold of living organism.

\begin{Definition}
Every  protein, DNA and RNA in a  living organism has its  ghost field to turn their gene spinor  state to open (activate) or close down while docking  and undocking state as a decision field for  gene expression in feedback loop. We define two main types of  hidden spinor fields:  a   ghost field $\Phi_{i}$ for activating a gene or a geneon for the docking operator  (minimum state in superspace of  time series data) and an anti-ghost field $\Phi_{i}^{+}$ for  activating an anti-geneon for undocking state  (maximum state in superspace of time series data).  We have

\begin{equation}
\Phi: ( \mathcal{O}_{\mathcal{A}},s) \rightarrow  \mathbb{Z}/2=\{[s2],[s4]\}(:=\{1,-1\})
\end{equation}
where $\mathcal{O}_{\mathcal{A}}$ is a sheaf  collection of all fiber states over tangent of  supermanifold of  living organism with an endowed  master equation $s$ as a hidden ground spinor field of states, $[s_{4}]:=<[s_{6}],[s_{8}]>$
 since $(-1=((i)(i)))_{anti-D-brane}$ and  $[s_{2}]=<[s_{5}],[s_{7}]>$ sicne $(1=((i)(-i)))_{D-brane}$.
\end{Definition}

When a protein docking state is in equilibrium, the genetic code will exist as different 2 forms over spinor field of second cohomology group of coupling between 2 states as basis for genetic code in form of wave function of gene in equilibrium state with 64 codon. This 64 state  comes from  the  exact cohomology sequence (see Fig. \ref{fig5} for detail of  cohomolgy space of DNA and its mechanism to generate obstruction components to make homology with non-zero element and make it in non-equilibrium state with retransposon) with induced 2 forms $d^{2}=0$. Let now define all well known
 active and inactive state of gene expression as followings,
 
\begin{Definition}
Let $W_{\mu}(z)=\frac{1}{z}$ be a twistor of telomere or from DNA knot state induced from the genetic recombination or transposon, etc. The wave function is a  superstate of superprobability distribution that satisfies  a  Seiberg-Witten equation \cite{seiberg} with modified Dirac operator for biology $D_{\pm}\varphi_{i}=0$ in equilibrium state. We can define

\begin{equation}
| \varphi_{[A_{i},s_{i},\mu]}^{geneon}>=\oint_{H^{2}(X;D,R)}\Pi_{\mu}W_{\mu} \frac{1}{([s^{\ast}]-[s_{2}])
 ([s^{\ast}]-[s_{4}])}d[s_{2}]\wedge d[s_{4}]
\end{equation}
where $H^{2}(X;D,R) $ is a second cohomology group induced from the central dogma of state  $ C_{\ast}(D,R)$ transcription chain;

 \begin{equation}
| \varphi_{[A_{i},s_{i},\mu]}^{anti-geneon }>=\oint_{H^{2}(X;D,D)}\Pi_{\mu}W_{\mu} \frac{1}{([s^{\ast}]-[s_{1}])([s^{\ast}]-[s_{3}])}d[s_{1}]\wedge d[s_{3}]
\end{equation}
where $H^{2}(X;D,D) $ is a second cohomology group induced from the central dogma of state  $C_{\ast}(D,D)$ transcription chain.
We adopted an  inversed Dirac operator projection from the pole in the histone protein. We have a wave function of transposon,
 $  |\varphi_{[A_{i},s_{i}^{\ast},\mu]}^{transponson}>=<\varphi_{[A_{i},s_{i}^{\ast},\mu]}^{transponson}|D^{\ast}$ with the property that
 $< \varphi_{[A_{i},s_{i}^{\ast},\mu]}^{transponson}|D^{\pm}{D^{\pm}}^{\ast}| \varphi_{[A_{i},s_{i}^{\ast},\mu]}^{transponson}>=1$
\begin{equation}
 |\varphi_{[A_{i},s_{i}^{\ast},\mu]}^{transponson}>=\oint_{H^{2}(X;D^{\ast},D^{\ast})}\Pi_{\mu}W_{\mu} \frac{1}{([s^{\ast}]-[s_{5}])([s^{\ast}]-[s_{7}])}d[s_{5}]\wedge d[s_{7}]
\end{equation}
where $H^{2}(X;D^{\ast},D^{\ast}) $ is a second cohomology group induced from the central dogma of state  $C_{\ast}(D^{\ast},D^{\ast})$ for self adjoint component of  the involution map of the insert transcription $g^{\ast}(s_{i})=s_{i}^{\ast}$ or the delete transcription (${g^{\ast}}^{-1}(s_{i}^{\ast})=s_{i}$) chain.

\begin{equation}
| \varphi_{[A_{i},s_{i}^{\ast},\mu]}^{retrotransponson}>=\oint_{H^{2}(X;D^{\ast},D^{\ast})}\Pi_{\mu}W_{\mu} \frac{1}{([s^{\ast}]-[s_{6}])([s^{\ast}]-[s_{8}])}d[s_{6}]\wedge d[s_{8}]
\end{equation}
where $H^{2}(X;R^{\ast},D^{\ast}) $ is a second cohomology group induced from the central dogma of state  $C_{\ast}(R^{\ast},D^{\ast})$ for self adjoint component of   involution map of insert
 $g^{\ast}(s_{i})=s_{i}^{\ast}$ or delete transcription (${g^{\ast}}^{-1}(s_{i}^{\ast})=s_{i}$) chain.
\end{Definition}
The definition of wave function holds hidden information of the gene expression. The information can be taken off only when we solve the Seiberg-Witten equation  with the new definition of Dirac operator according to the  Chern-Simons current  over the cohomology sequence of living organism.
  
We induce a state in the gene of central dogma without transitive layer of protein $P$. One has a  cycle and a co-cyle $\beta $ of the superspace of  living organism $X_{t}$ by defining the Jacobian over the fiber of the supermanifold $g_{ij}=\frac{\partial s_{i}}{\partial s_{j}}.$  

The source of gravitational field, the  Chern-Simons current in codon, is induced from the connection over the parallel transport of co-cycle $g^{ij}$ and cycle $g_{ij}$ in strict analogy with the definition of connection $\Gamma_{ij}^{k} :=[A_{k}]$ in General Relativity theory where there is a  related  definition of the Riemann curvature  $R^{\mu}_{\nu\alpha\beta}$ (see the discussion in Section III). 

One can  define a  3 form over the alphabet code  by another Chern-Simons 3 form over the alphabet of genetic code representing a  protein layer as triplet state which induces more 32 states from differential 3 forms over hidden differential 2 forms of the states of geneon.
Specifically, we have the Chern-Simons current for biology over the supermanifold of living organism $\mathcal{M}$ defined as

\begin{equation}
  J^{\mu=k}=\int tr H^{3}(\mathcal{M}):=\frac{k}{4\pi}\int_{H^{3}(D,R,P)}  A\partial A+\frac{2}{3}A\wedge A\wedge A. \end{equation}

\begin{Definition}
A modified Dirac operator for left chiral geneon basis $x\in X$ is defined by turning the  mirror symmetry 
of D-brane $x\in X$ to anti-D-brane with retransposon dual basis $x^{\ast}\in X^{\ast} $with reversed time scale of transcription $dt^{\ast}.$ We have a Wigner ray  $Wigner<x,x>_{R}:=<x,x>_{R}:= \{D^{+}x,D^{+}x\}= Ad_{D^{+}}D^{+}x={D_{+},D^{+}}x={D_{+}x,D^{+}x}$ for the right translation with exponential map $Ad_{D^{+}}D^{\pm}=e^{D^{\pm}}, Ad_{D^{-}}D^{\pm}=\ln(D^{\pm})$. The projection of the Riemann sphere of the histone to the hidden state of geneon and also the dual back to geneon   represents the  hidden state in the extradimension perpendicular to the  D-brane.
The direction of imaginary number is perpendicular to the D-brane for wrapping  up and down between interaction of D-brane of protein interaction of ghost field $\Phi^{\pm}$. 
 We have 
 \begin{equation}
Ad_{D^{+}}D^{+}\Phi^{-}(x)=\Phi^{+}(x^{\ast}),ad_{D^{+}}D^{+}\Phi^{-}(y)=\Phi^{+}(y^{\ast}),ad_{D^{-}}D^{-}\Phi^{+}(x^{\ast})=\Phi^{-}(x),
 \end{equation}
 \begin{equation}
Ad_{D^{-}}D^{-}\Phi^{+}(y^{\ast})=\Phi^{-}(y),ad_{D^{+}}D^{+}\Phi^{-}(y^{\ast})=0,ad_{D^{+}}D^{+}\Phi^{-}(x^{\ast})=0,Ad_{D^{-}}D^{-}\Phi^{+}(x)=0.,ad_{D^{-}}D^{-}\Phi^{+}(y)=0.
 \end{equation}
The definition of the Dirac operator is used to explain the gene expression with the algebraic operation of docking and undocking states   as in   quantum mechanics in  the framework of supersymmetry.

\end{Definition}

Let $W_{k}$ be a twistor in  telomere or in transposon inserted with a twist state in the  gene expression equation. We have a solution for repeated obstruction component which is 
\begin{equation}
Z=\Pi_{k}W_{k}(Y_{t}^{\ast}/X_{t}^{\ast})e^{\frac{\beta_{i}}{\alpha} D^{\pm}[s_{i}]_{X_{t}/Y_{t}}}=\Pi_{k}W_{k}(Y_{t}^{\ast}/X_{t}^{\ast})e^{\oint_{H^{2}(X_{t}^{\ast}/Y_{t}^{\ast})}\frac{\Pi_{j}(y_{t}^{\ast}-[s_{j}])^{\alpha_{j}}}{\Pi_{i}(x_{t^{\ast}}-[s_{i}^{\ast}])^{\beta_{i}}}dx_{t^{\ast}}\wedge dy_{t}^{\ast}}.
\end{equation}
This is a source in  loop space $\pi_{1}(X_{t},[s_{i}])$ with spinor field in time series data. Transcription and translation can be in reversed direction of dual time analogy with transposon and retrotansposon obsruction component in the
area of trash DNA.

\begin{equation}
0 \rightarrow     \mathbb{Z}/2 \rightarrow  \mathcal{O}_{X_{t}} \rightarrow 
\mathcal{O}_{Y_{t}} \rightarrow \mathcal{O}_{Y_{t}/X_{t}}\rightarrow 0 
\end{equation}
This sequence induces an  infinite sequence 

\begin{equation}
0 \rightarrow     \mathbb{Z}/2 \rightarrow  \mathcal{O}_{X_{t}} \rightarrow \mathcal{O}_{Y_{t}} \rightarrow 
\mathcal{O}_{Y_{t}/X_{t}}\rightarrow  H^{1}(\mathbb{Z}/2;\mathbb{Z}/2) \rightarrow  H^{1}(\mathcal{O}_{X_{t}};\mathbb{Z}/2) \rightarrow  H^{1}(\mathcal{O}_{Y_{t}};\mathbb{Z}/2) \rightarrow H^{1}(\mathcal{O}_{Y_{t}/X_{t}};\mathbb{Z}/2)\rightarrow \nonumber
\end{equation}

\begin{equation}
 \rightarrow  H^{2}(\mathbb{Z}/2;\mathbb{Z}/2) \rightarrow  H^{2}(\mathcal{O}_{X_{t}};\mathbb{Z}/2) \rightarrow  
H^{2}(\mathcal{O}_{Y_{t}};\mathbb{Z}/2) \rightarrow H^{2}(\mathcal{O}_{Y_{t}/X_{t}};\mathbb{Z}/2)\rightarrow \nonumber
\end{equation}

\begin{equation}
 \rightarrow  H^{3}(\mathbb{Z}/2;\mathbb{Z}/2) \rightarrow  H^{3}(\mathcal{O}_{X_{t}};\mathbb{Z}/2) \rightarrow 
 H^{3}(\mathcal{O}_{Y_{t}};\mathbb{Z}/2) \rightarrow H^{3}(\mathcal{O}_{Y_{t}/X_{t}};\mathbb{Z}/2)\rightarrow \cdots
\end{equation}
It is a source of feedback control loop of gene in adjoint map of ghost field and anti-ghost field in docking state of protein $\Phi_{i}^{\pm}$. Here  $[s_{4}] ,[s6]$  stand for the retransposon in the  time series data as ground field and $<[s_{5}],[s_{7}]>$ are  the hidden transition state to  the  ground field in the  inactive area of transposon trash DNA. The master equation is defined by Poisson bracket of the super-Lagragian of ghost  and anti-ghost fields in protein presentation layer
\begin{equation}
 s:=\left \{ \int Sdt,-   \right \}=0
\end{equation}
where $S$ is the  least action of super-Lagragian of ghost field in superspace of time series data.
Th BV-cohomology group for the pullback functor of   adjoint representation of supersymmetry in codon and anti-codon is  

\begin{equation}
 \cdots \longrightarrow H^{-7}(\mathcal{O}_{\mathcal{A}},s) 
  \longrightarrow H^{-8}(\mathcal{O}_{\mathcal{A}},s) 
\longrightarrow  \longrightarrow H^{-9}( \mathcal{O}_{\mathcal{A}},s) 
\cdots
\longrightarrow H^{-14}( \mathcal{O}_{\mathcal{A}},s) 
\end{equation}

\begin{equation}
 \rightarrow    \cdots \rightarrow  \nonumber
 \end{equation}
\begin{equation}
 \rightarrow  H^{11}(\mathbb{Z}/2;\mathbb{Z}/2) \rightarrow  H^{11}(\mathcal{O}_{X_{t}};\mathbb{Z}/2) \rightarrow  H^{11}(\mathcal{O}_{Y_{t}};\mathbb{Z}/2) \rightarrow H^{11}(\mathcal{O}_{Y_{t}/X_{t}};\mathbb{Z}/2)\rightarrow \cdots  
\end{equation}
In the hidden layer of superspace of time series data , one can induce a infinite long exact sequence of BV-cohomology of negative dimension. The cohomolgy is used to explain the gene expression by the sheaf resolution from the local section of the geneon wave function with respect to the global section  of immunosystem and to the global physiology of cell.

\begin{equation}
( \mathcal{O}_{\mathcal{A}},s)\rightarrow H^{-1}( \mathcal{O}_{\mathcal{A}},s) \rightarrow H^{-2}( \mathcal{O}_{\mathcal{A}},s)\longrightarrow \cdots \rightarrow H^{-14}( \mathcal{O}_{\mathcal{A}},s)\longrightarrow \cdots .  
\end{equation}
 When it  is in  equilibrium docking, we have $H^{-k}( \mathcal{O}_{A},s)=0$  for all $k>0$.  This means that  we cannot  observe ghost  and anti-ghost fields if docking system  is in equilibrium. There will exists a hidden negative dimension area of trash DNA, active  only when the state of undocking in protein system appear. The immunosystem will induce a feedback loop 
of retransposon and transponson by the cohomology in negative dimension in feedback path to cohomology in positive dimension to  find a new equilibrium point of docking until it  holds; then the negative path is zeros again recursively. The main expression of gene can be written in a super-regression $[\alpha] \sim [\beta]\in H^{2}(\mathcal{O}_{\mathcal{A}};\mathcal{O}_{X/Y})$, so one can  induce an evolutional field in trash area as a hidden field to control the second round of recursive form of induced long infinite cohomology on the sphere, that is  
\begin{equation}
\varphi_{Y}^{retransposon}=[\alpha]  +[\beta] \varphi_{X}^{geneon}+[\epsilon_{t}]_{X/Y} 
\end{equation}
The copy process of loop space of retrotransposon is a characteristic class of cohomology in negative dimensions induced by  fthe  master equation with adjoint functor as a cohomology functor in this model.
 
The extended Chern-Simons current for trash DNA is possible thanks to the  Coexter number $h$ in the Laurant series of poles of states $[si]^{\ast}$:  from the  icosahedral group $E_{8}$,  we get $h=30$.  Therefore 
we can extend this approach to represent the genetic code as Laurent polynomials in the variable $q$
 with integer coefficients, that is for trash area with knot $K$ over sheaf of DNA, that is 
\begin{equation}
J(K ,q)=\sum_{i=1}^{n}a_{n}q^{n}:=\oint_{H^{n}(\mathcal{O}_{\mathcal{X}})}W_{D^{+}} \frac{1}{(\Pi_{i}([s]-[s_{i}])}d[]s.
\end{equation}
where $W_{D^{-}}([s]-[s_{i}])=\frac{1}{[s]-[s_{i}]}$ and $W_{D^{+}}(\frac{1}{z})=z,W_{D^{-}}(z)=(\frac{1}{z})$, 
gives a  modified Wilson loop operator as a M\"obius map for knot state in DNA transpospon and retrotransposon according to this new definition of modified Dirac operator. This operator sends a pole to the principle line bundle over the fiber space of the DNA molecule from histone octomer protein. In the Cauchy integral, it is a Laurent series of poles with a singularity in the Riemann sphere with a M\"obius map of duality in complex plane.
We induce a spinor field for the representation of the genetic code by using a new parameter of knot  $q$ for retrosponson state $[s_{6}^{\ast}]$. It is 
\begin{equation}
[q]=e^{\frac{2i\pi}{k+h}}  = [s_{6}]_{[T]}  =e^{-\frac{\pi  i\beta_{i}}{2 }}\,.
\end{equation}
where $h=30$ is the dual Coexter number for the retronposon of HIV icosahedral   group action of supersymmetry of gene expression $G$; $\beta_{i}=1,2,3,\cdots ,N$ is a partition function of repeated state of retrotransposon in integers of 
orbit over the fiber space of the gene.  When HIV attaches, we use the   icosahedral group $E_{8}\times E_{8}$ with $h=30$ Coexter number in the supersymmetry breaking of codon and anti-codon interaction for trash area.

In the case of retransposon of HIV $[s_{6}]=([T],[NA])$, we get the relationship 
$\frac{2i\pi}{k+h}=-\frac{\pi  i  \beta_{i} }{2}$
so, in this case, we have repeated $N$ states of the Chern-Simons current $J^{\mu=k}$ in the hidden direction   with $k=-30 +\frac{2}{\beta_{i}} $ where $\beta =1,2,3,\cdots, N $ because
$k=\frac{2}{\beta_{i}}-30.$ The Chern-Simons current induced from trash DNA is inactive in hidden state since it is a complex number with repeated exited state for transition state of retransposon, that is

\begin{equation}
 J^{k=retrotransposon}=J^{k=\beta_{i}}= -i\sqrt{\frac{2}{28-\frac{2}{\beta_{i}}}} \sin\frac{\pi}{28-\frac{2}{\beta_{i}} } .
\end{equation}
where $\beta_{i} =1,2,3,\cdots,N $ so we have repeated $N$ hidden Chern-Simons currents in this area that cannot be observed as  real values. 

The above considerations  can be  adopted for some concrete realizations.  Below, we will take into account the specific examples of   some genetic codes, in particular a viral gene, and develop an empirical analysis of the Chern-Simons current for  the time series.

\section{Methodology}
Let us now  use  a  numerical representation for spinor field of gene expression by using the Chern-Simons  current map string of genetic code to measure the curvature over the amino acid sequence of a gene. It is suitable for plotting the  genetic code into mathematical object that are  time series data in the superspace of gene expression.
We transform the alphabet string value, in which cannot be computed  the classical standard definition of genetic code,  to Chern-Simons current  of time series data of genetic code $J^{\mu}$ with $\mu=1,2,3,\cdots 64$ over spinor field with ground field of real values  which is suitable for massive  computations  of data.

The empirical analysis is allowed  by using the  Chern-Simons current representation for the spectrum of genetic code  of viral capsid proteins. We implement the algorithm for  the new representation of spinor field for superspace of viral genetic variation by using a novel algorithm of analysis of nonstationary time series.   We use the $\ICHAIN(1)$ transformation. The algorithm can be find in \cite{Kolmogorov, cohomo7} to detect the spinor field in time series data of genetic code.The main  tool is the so called $\ICHAIN(n)$ in time series prediction with time series of superspace of genetic code for real application with viral gene expression. We compare the result with traditional  bioinformation string matching the representation with our new methodology of time series expression as the  main result of our work.

We prepare empirical   6 time series data  downloaded  from GenBank (see Appendix A for details). The target of our time series data analysis is to reproduce the  Chern-Simons current with CD4 and V3 of HIV virus from 3 types of  time series data of host cell. We transform alphabet into Chern-Simons current and perform $\ICHAIN(1)-\ICHAIN(6)$ transformation. After we get a result of time series data, we perform the  analysis of Hilbert transformation  to be a closed look at behavior of instantaneous frequency of 3 genes of CD4 compared with V3 pattern. The main methodology is the algorithm to find gene evolutional tensor network. The algorithm consists of  2 steps: first we compute tensor correlations of 6 imfs of CD4 time series data and other 6 imf from V3 time series data. The time series of target of data analysis of gene tensor network evolution is CD4 time series (Appendix A)   of 78 amino acids. For the V3 time series data, we choose first 78 time series data for empirical test with the same length of CD4 time series data. We compute the average tensor correlation matrix with size of (36x36)x(78x78) (the number  36  comes from the number of imf1-imf6 of CD4 and imf1-imf6 of V3 loop) with slice window of length 5.
The algorithm to construct the tensor network is a planar graph algorithm \cite{cohomo7}. The network is  built and visualized by using algorithm of tensor correlation over gene sequence tensor network.
The edge of network shows how much genes are correlated with each other. The edge of gene tensor network is a sequence of animo acids or genetic codes ordered by the number of alphabet. A gene tensor network corresponds to one hidden state in genetic code. We study in CD4 genes with 36  coupling states  between 6 states of CD6 genes and 6 states of 6 imfs of 6 imfs  of V3 genes. Result of computation are discussed in next section.

\begin{table}[!t]
\renewcommand{\arraystretch}{1.3}
\caption{  The table below shows  the definition of first 14-ghost fields   in $E_{8}\times E_{8}$ model of artificial HIV viral v3 loop and cd4 as a supermanifold structure. The details of the  definition and picture of all components below can be found  in \cite{cd4_drawing}.}
\label{table3}
\centering
 \begin{tabular}{|c|c|c|c|c|c|} \hline \hline
 Ghost field $\Phi_{i}$  &  site name &  moduli state space variable &Anti-Ghost field $\Phi_{i}^{+}$  &  site name &  moduli state space variable \\ \hline \hline
  $\Phi_{1}$& $ C  $& $x_{t}$  &$\Phi_{1}^{+}$ & CD4 receptor& $y_{t}$     \\ \hline

$\Phi_{2}$& $N   $& $x_{t}$ & $\Phi_{2}^{+}$& CXCR4 coreceptor& $y_{t}$     \\ \hline

$\Phi_{3}$& $\alpha_{1}  $& $x_{t}$ & $\Phi_{3}^{+}$& CCR5 coreceptor& $y_{t}$      \\ \hline

$\Phi_{4}$& $L_{A} $& $x_{t}$          \\ \hline

$\Phi_{5}$& $ \alpha_{5}    $& $x_{t}$          \\ \hline

$\Phi_{6}$& $L_{C} $& $x_{t}$             \\ \hline

$\Phi_{7}$&  $ \alpha_{3}    $ & $x_{t}$      \\ \hline

$\Phi_{8}$&   $L_{E} $ & $x_{t}$            \\ \hline

$\Phi_{9}$&    V5& $x_{t}$          \\ \hline
$\Phi_{10}$& V4& $x_{t}$          \\ \hline

$\Phi_{11}$& V3& $x_{t}$          \\ \hline

$\Phi_{12}$& V1/V2  & $x_{t}$    \\ \hline

$\Phi_{13}$& Bridging Sheet& $x_{t}$    \\ \hline

$\Phi_{14}$& $\beta$ Sheet & $x_{t}$    \\ \hline
  \hline\hline
	 \end{tabular}
\end{table}

\section{Results  of Time Series on Viral Gene}

\begin{figure}[!t]
\centering
\includegraphics[width=0.75\textwidth]{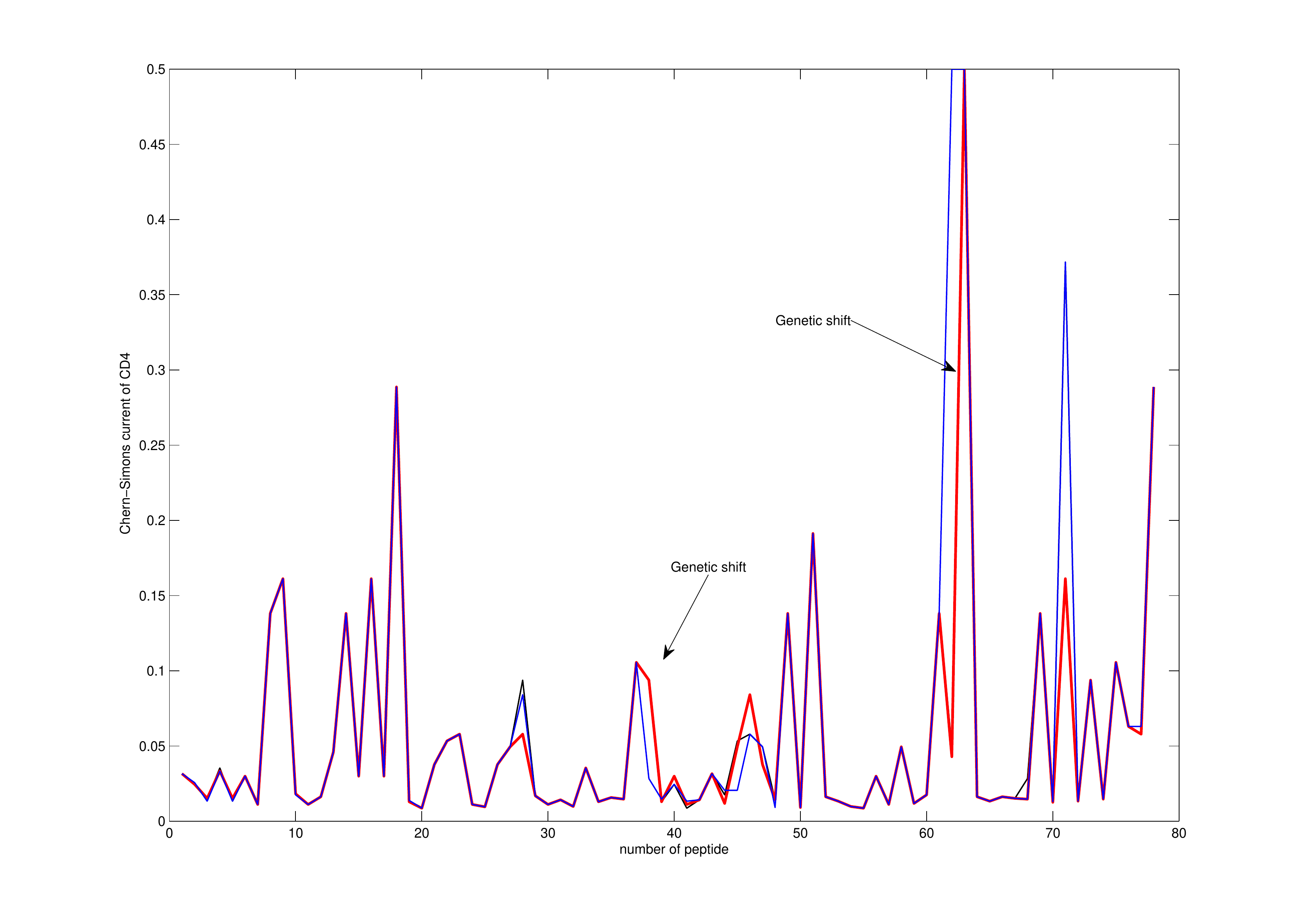}
\caption{In this figure, we plot the Chern-Simons current over 78 kinds of peptides. We can notice the genetic shift over CD4 on the plot while the graph is not in the same position. The grey line is the plot of the CD4 gene in the species Oryctolagus cuniculus algirus. The red line is the plot of the CD4 gene in Sylvilagus bachmani and the blue plot is the plot of the CD4 gene in Lepus saxatilis. The details of all the genetic codes of the plot can be found in Appendix A.}
\label{fig13}\end{figure}

\begin{figure}[!t]
\centering
\includegraphics[width=0.75\textwidth]{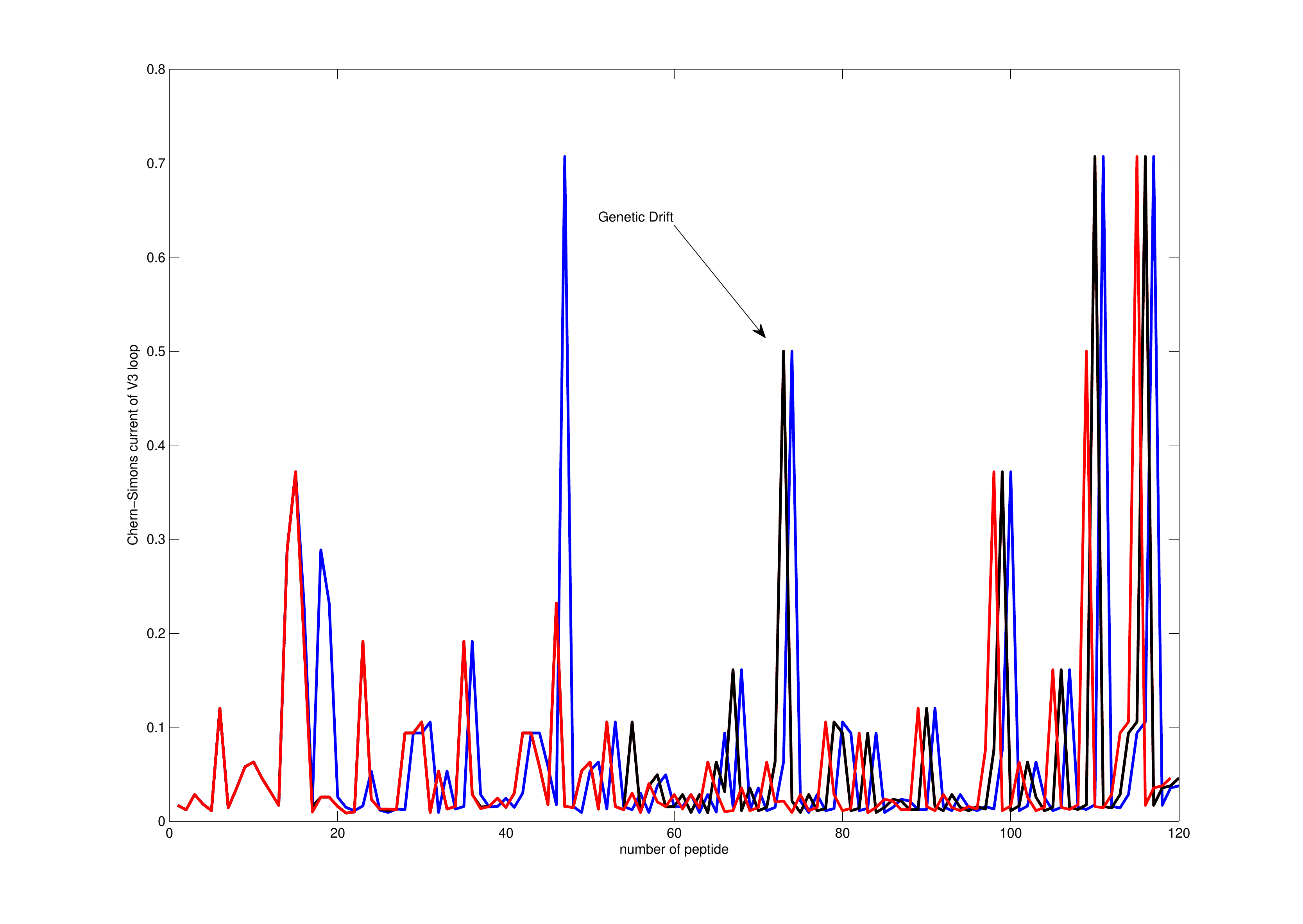}
 \caption{The picture shows the Chern-Simons current of the V3 loop region of HIV with 3 different genetic codes. The red line is the v3 loop region of HIV-1 isolated HN46-clP from Viet Nam envelope glycoprotein gene, partial cds. The isolation is on 05-MAR-2013. The black line is HIV-1 isolated HN46-clQ  and the blue line is HN46-clR from Viet Nam envelope glycoprotein gene, partial cds. The details of all the genetic codes and peptides are  shown in Appendix A.}
\label{fig9}
 \end{figure}

\begin{figure}[!t]
\centering
\includegraphics[width=0.48\textwidth]{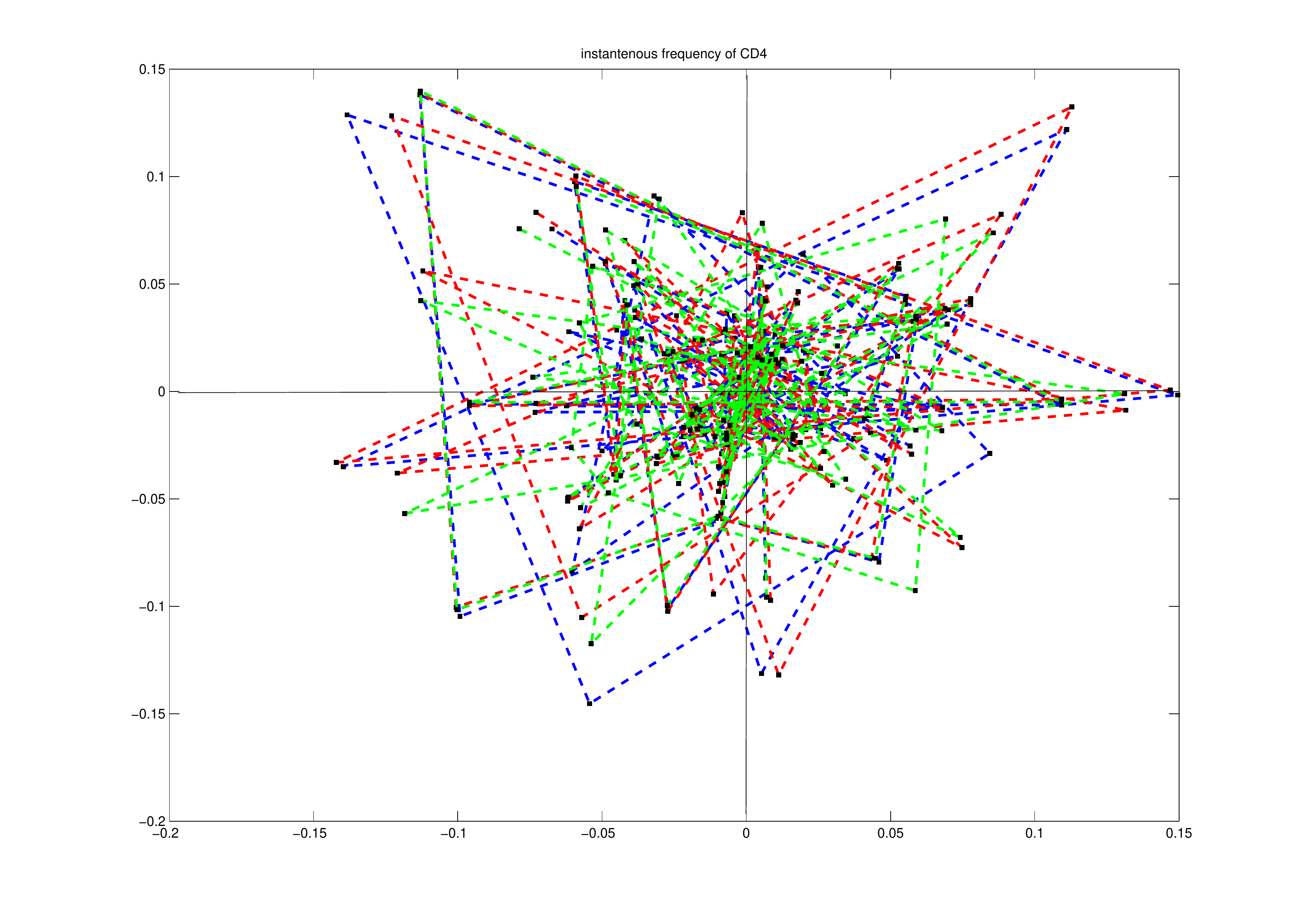}
\includegraphics[width=0.48\textwidth]{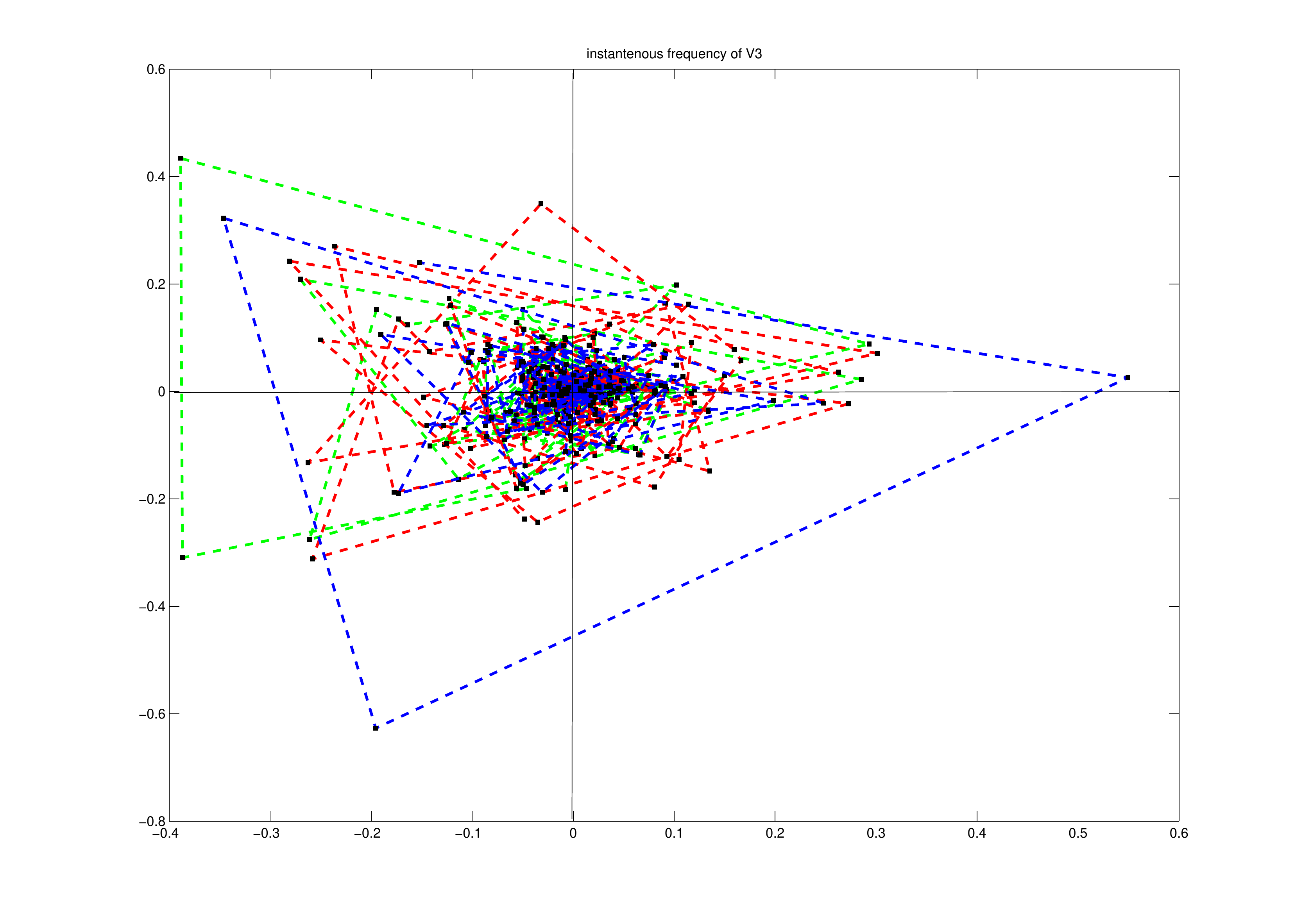}
\caption{On the left panel, we show the Hilbert transformation  of   $\ICHAIN(1) $ of CD4 to the complex plane.
On the right  panel, we show the Hilbert transformation  of   $\ICHAIN(1) $ of the V3 loop region of HIV to the complex plane.}
\label{fig18} \end{figure}

\begin{figure}[!t]
\centering
\includegraphics[width=0.48\textwidth]{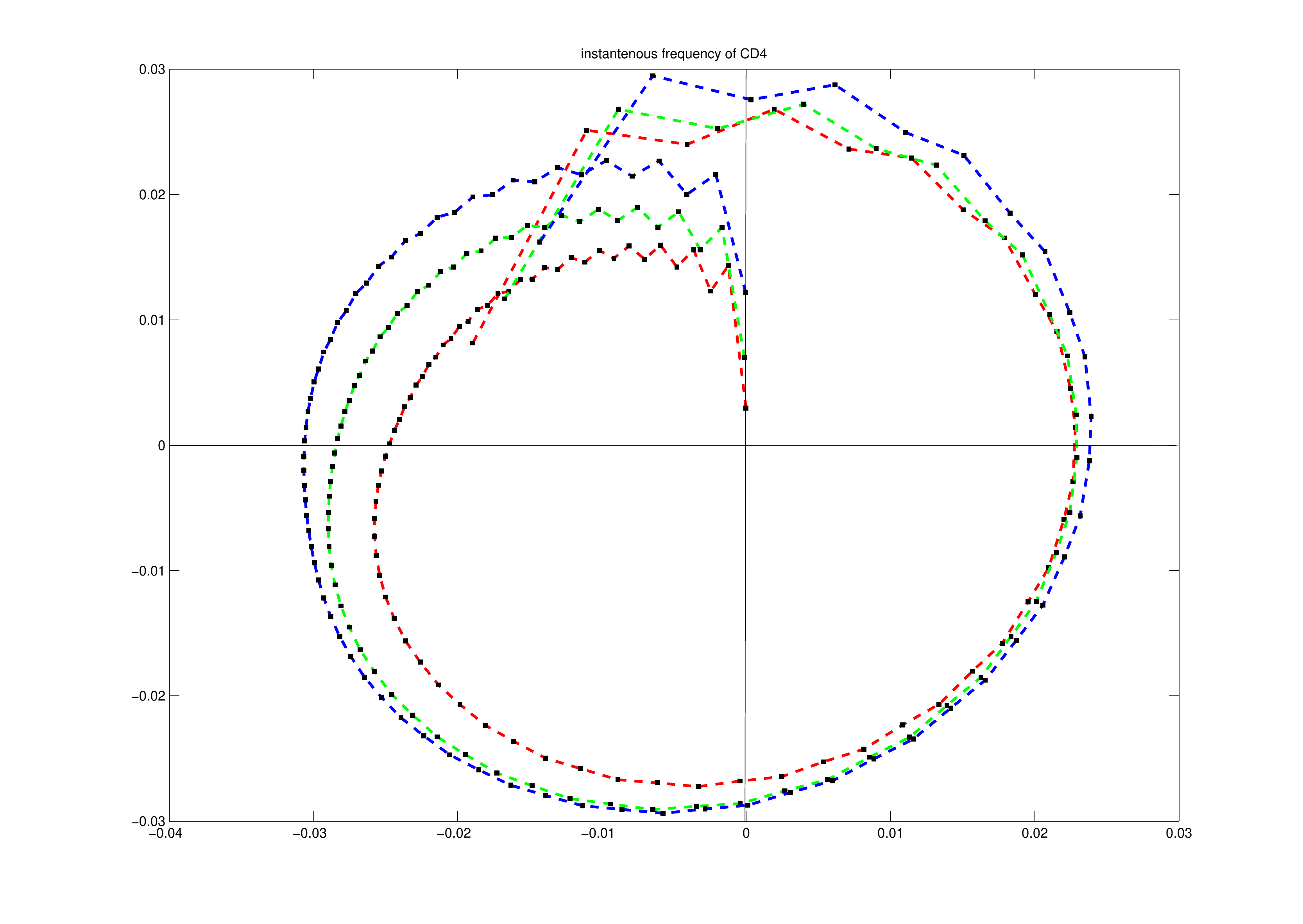}
\includegraphics[width=0.48\textwidth]{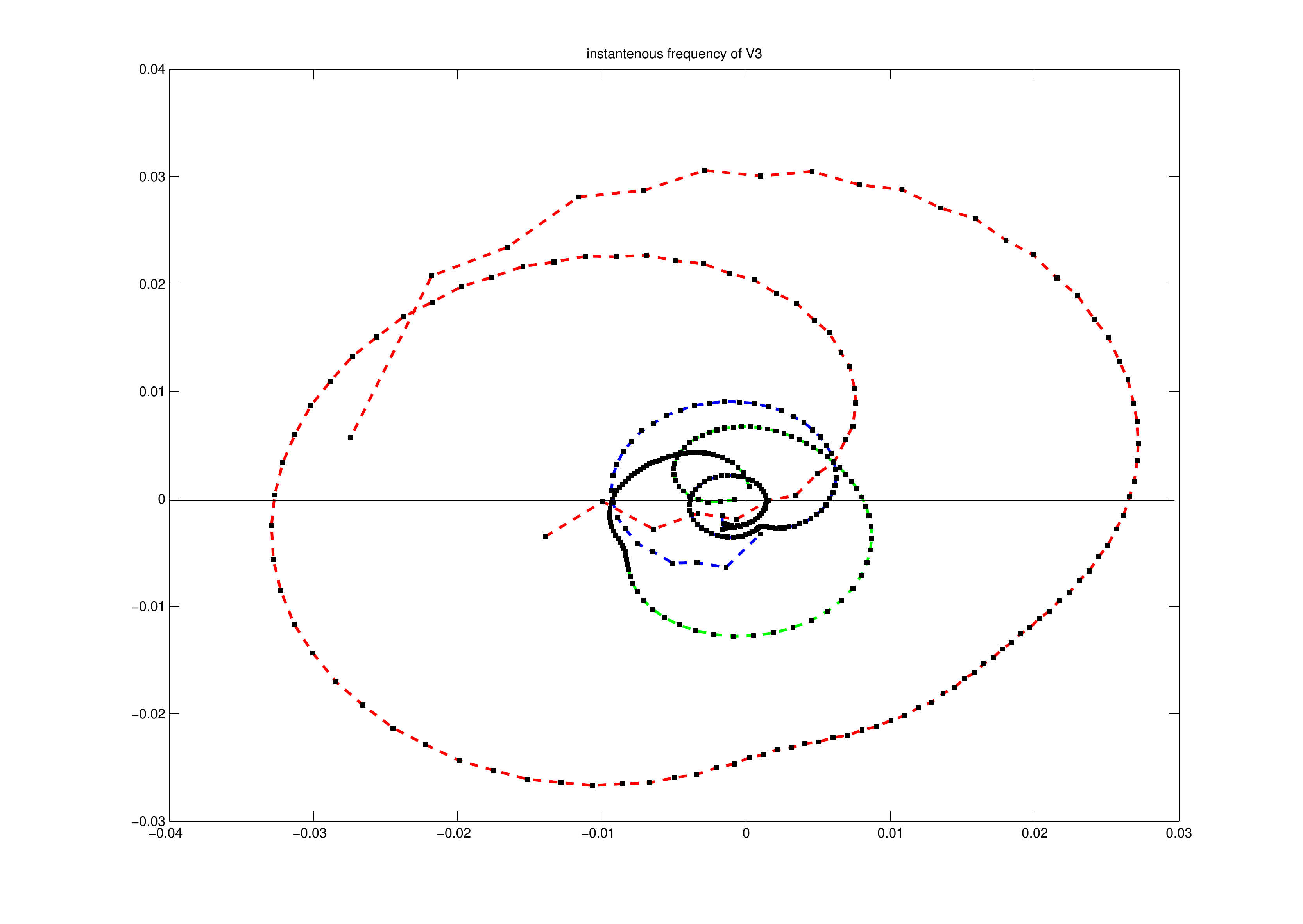}

\caption{On the left panel, we show a Hilbert transformation  of $\ICHAIN(6)$ of CD4 to the complex plane.
The trajectory of the plot can be interpreted as an $\alpha$-cocycle of the CD4 host cell $x_{t}$ in our new definition.
On the right panel, we show the Hilbert transformation of $\ICHAIN(6)$ of V3 loop region of HIV to the complex plane. The trajectory of the plot can be interpreted as a $\beta-$cocycle of V3 of HIV virus $y_{t}$ in our new definition.}
\label{fig19} \end{figure}

 We obtained  the genetic shift in CD4 (Fig. \ref{fig13}) and the genetic drift in V3 (Fig. \ref{fig9}) of HIV virus, a new representation of gene, by using a spinor field in a time series data. We found that in our representation, the time series of the gene structure is non-stationary and used the modern methodology to detect the genetic shift and the genetic drift over the non-stationary hidden states in the gene structure demonstrated by the Hilbert transformation  of $\ICHAIN(1)$ (Fig. \ref{fig18}).
 The $\beta$-cocycle is calculated by using a Hilbert transformation of  $\ICHAIN(6)$   $V3$  and  the $\alpha$-cocycle is calculated by using a Hilbert transformation of  $\ICHAIN(6)$   $CD4$. The result of the calculations can be found in Fig. \ref{fig19}.

\begin{figure}[!t]
\centering
\includegraphics[width=0.75\textwidth]{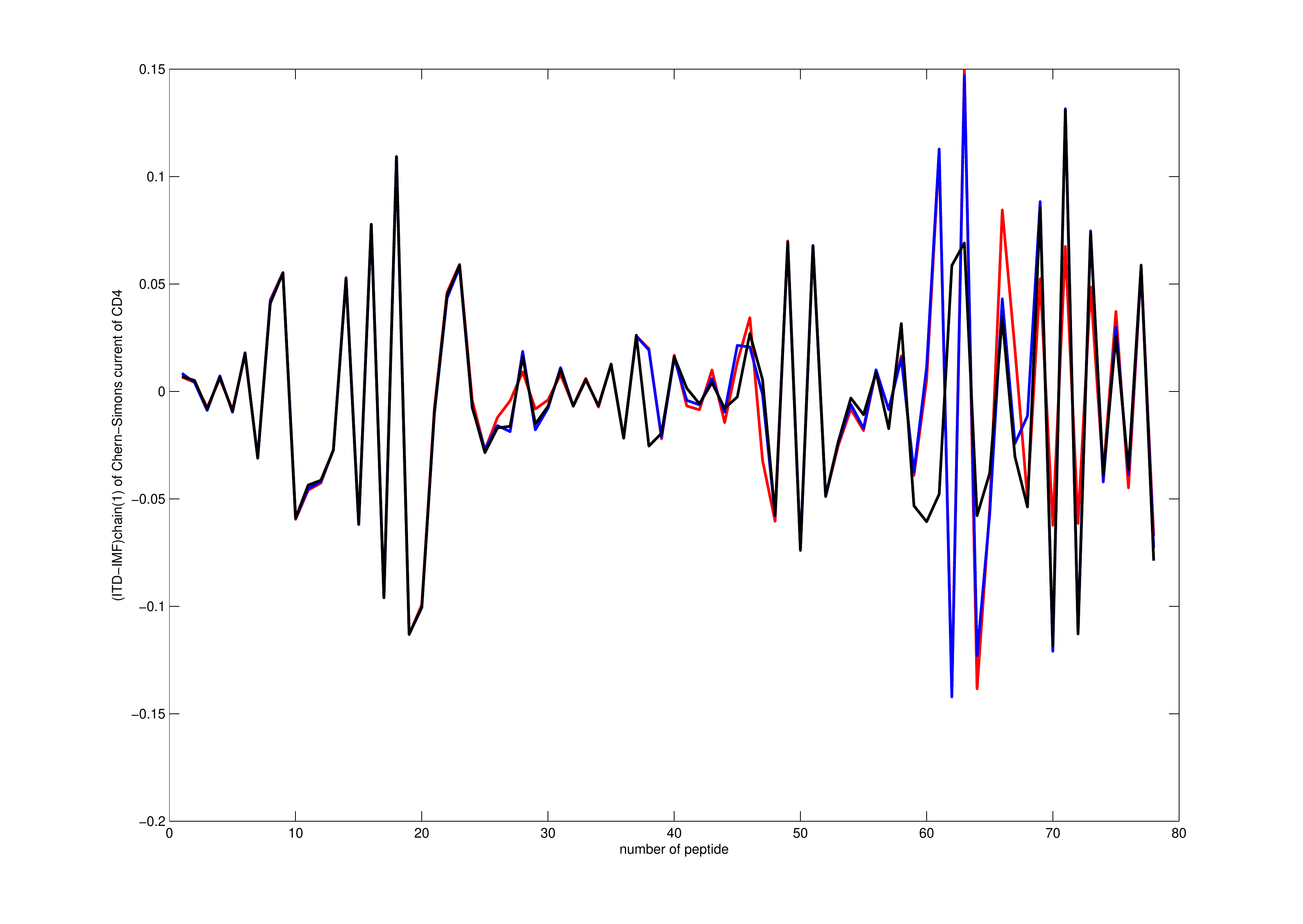}
\caption{ The picture     is    the plot of   $\ICHAIN(1)$ of the Chern-Simons current of  CD4 genes.  The grey line is the plot of the CD4 gene in the species Oryctolagus cuniculus algirus. The red line is the plot of the CD4 gene in Sylvilagus bachmani and the blue plot is the plot of the CD4 gene in Lepus saxatilis. The details of all the genetic codes of the plot can be found in Appendix A.
 }
\label{fig14} \end{figure}

\begin{figure}[!t]
\centering
\includegraphics[width=0.75\textwidth]{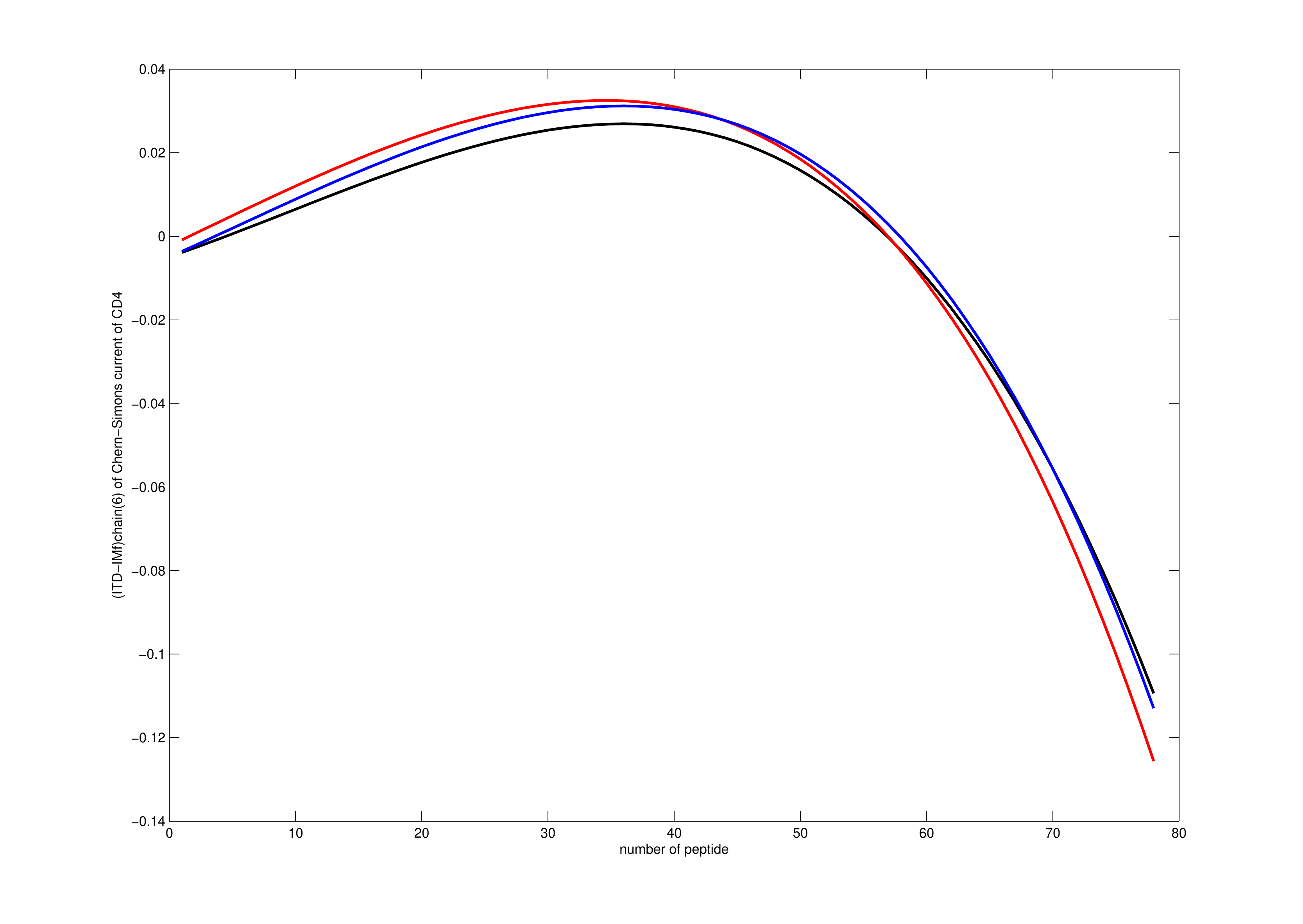}
\caption{The picture  is the plot of $\ICHAIN(6)$ of the Chern-Simons current of the CD4 genes.
 The grey line is the plot of the CD4 gene in the species Oryctolagus cuniculus algirus. The red line is the plot of the CD4 gene in Sylvilagus bachmani and the blue plot is the plot of the CD4 gene in Lepus saxatilis. The details of all the genetic codes of the plot can be found in Appendix A.
}
\label{fig15} \end{figure}

\begin{figure}[!t]
\centering
\includegraphics[width=0.75\textwidth]{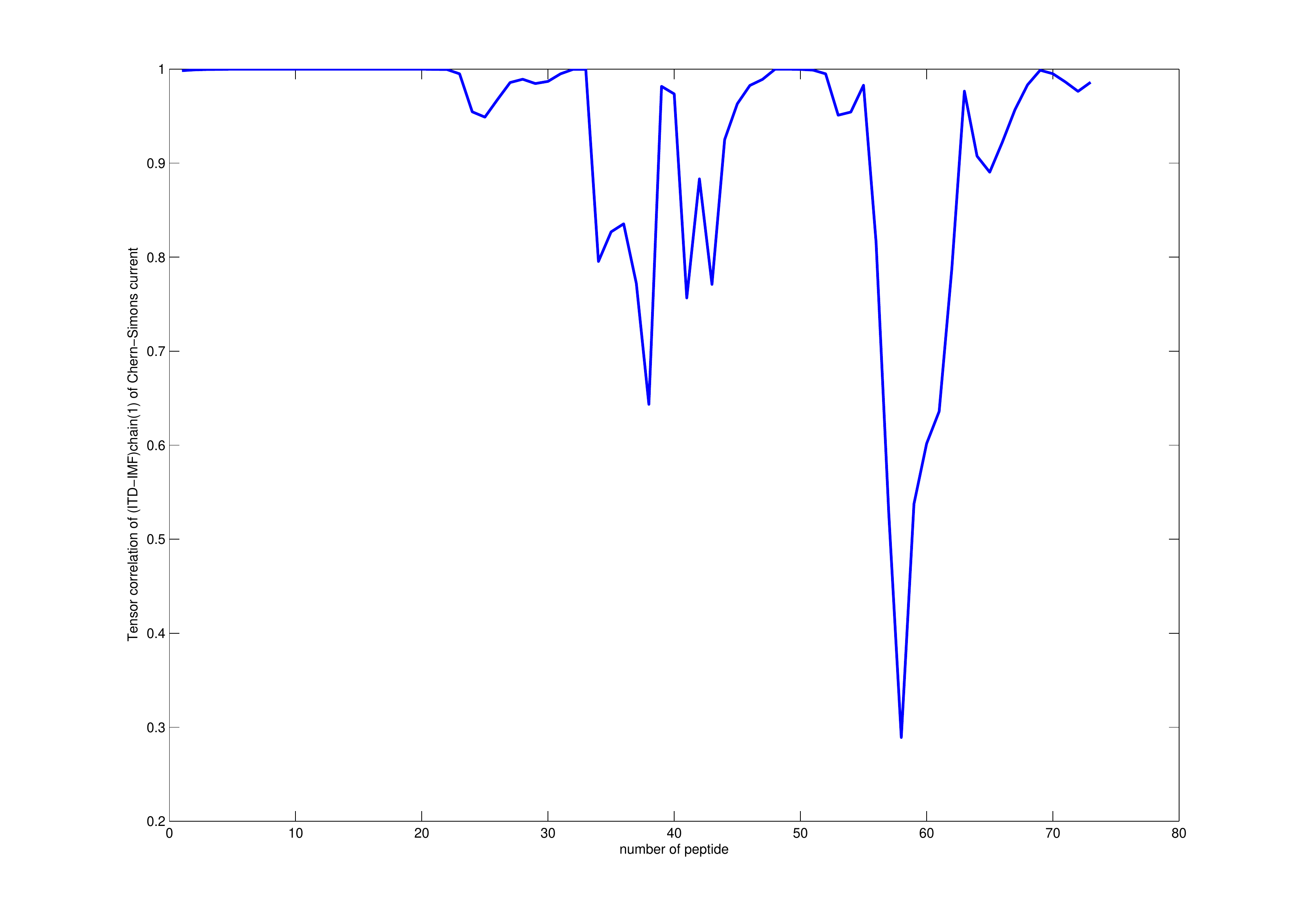}
\caption{The picture shows the average tensor correlation of $\ICHAIN(1)$ of the Chern-Simons current of the CD4 genes in rabbit 3 species.  In these plots, we noticed the evolution site for the genetic drift in the CD4 genes at the minimum position of the tensor correlation since all the 3 genes do not correlate to each other. }
\label{fig16} \end{figure}

\begin{figure}[!t]
\centering
\includegraphics[width=0.75\textwidth]{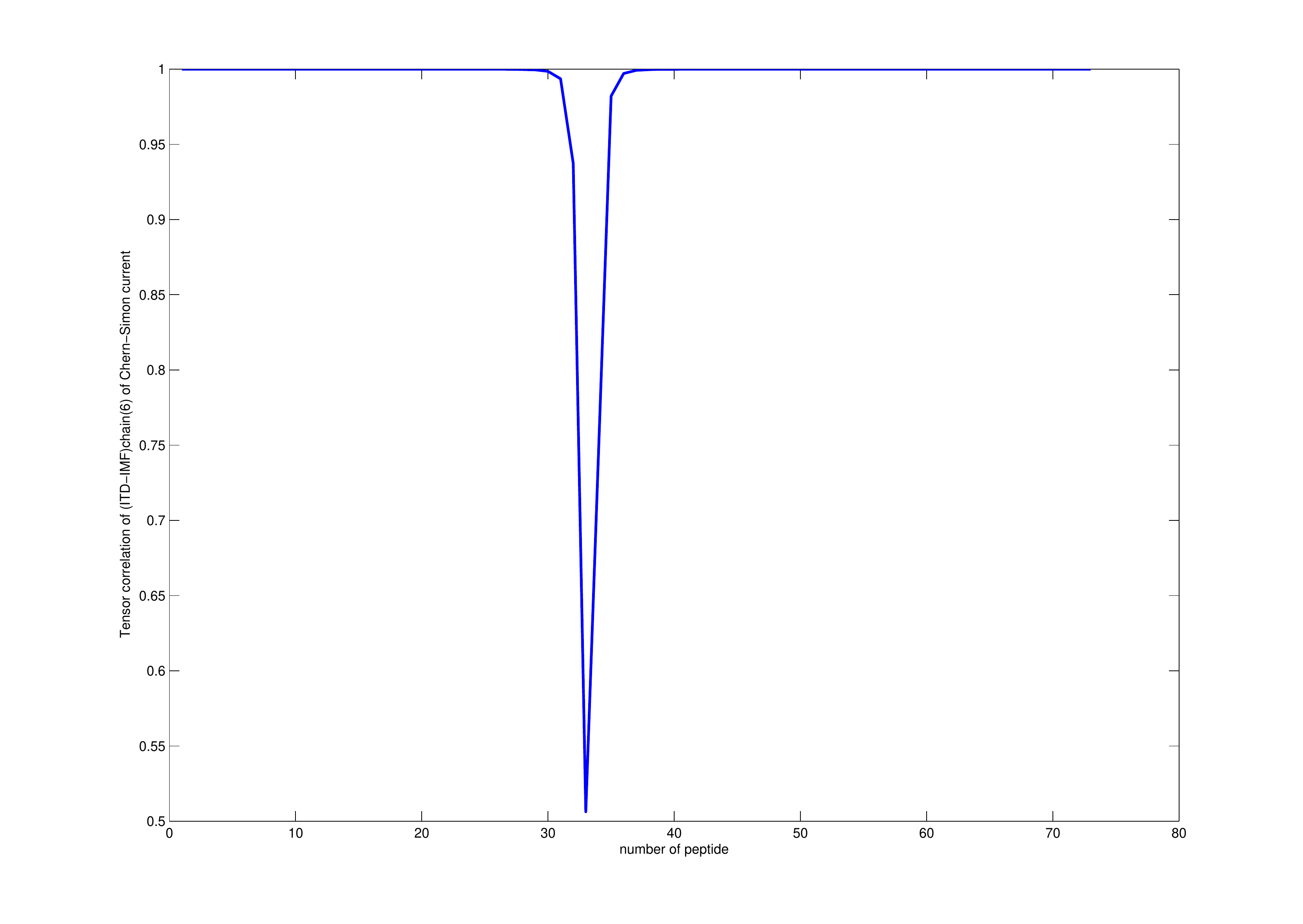}
\caption{ The picture shows   the average tensor correlation of $\ICHAIN(6)$ of the Chern-Simons current of the CD4 genes in the rabbit 3 species. In these plots, we notice the evolution site for the genetic drift in the CD4 genes at the minimum position of the tensor correlation since all the 3 genes are not correlating  each other. }
\label{fig17} \end{figure}

We provide an empirical analysis of the Chern-Simons current. We found that 6 moduli state spaces of $\ICHAIN(1)-\ICHAIN(6)$ for the V3 loop region over glycoprotein of HIV virus with the first 120 gene sequences of 360 time series of genetic codes of the antigene drift in the V3 loop region of the HIV glycoprotein and the host cell of CD4 from 3 species of rabbits.
  The results of the antigene shift in CD4 gene are calculated by using $\ICHAIN(1)-\ICHAIN(6)$.
 One of the results of  $\ICHAIN(1)$ can be found in Fig. \ref{fig14}
and $\ICHAIN(6)$ can be  found in Fig. \ref{fig15}.
We obtained the result of  the calculation of the antigene shift in the CD4 region of the HIV host cell of 3 species of rabbits with 3 samples by using a tensor correlation.
The genetic shift can be found in the position of peptide number 38-40 from 3 samples. Our result is correct with the tensor correlation of $\ICHAIN(1)$ (Fig. \ref{fig16}) and  $\ICHAIN(6)$ (Fig. \ref{fig17})

\begin{figure}[!t]
\centering
\includegraphics[width=0.75\textwidth]{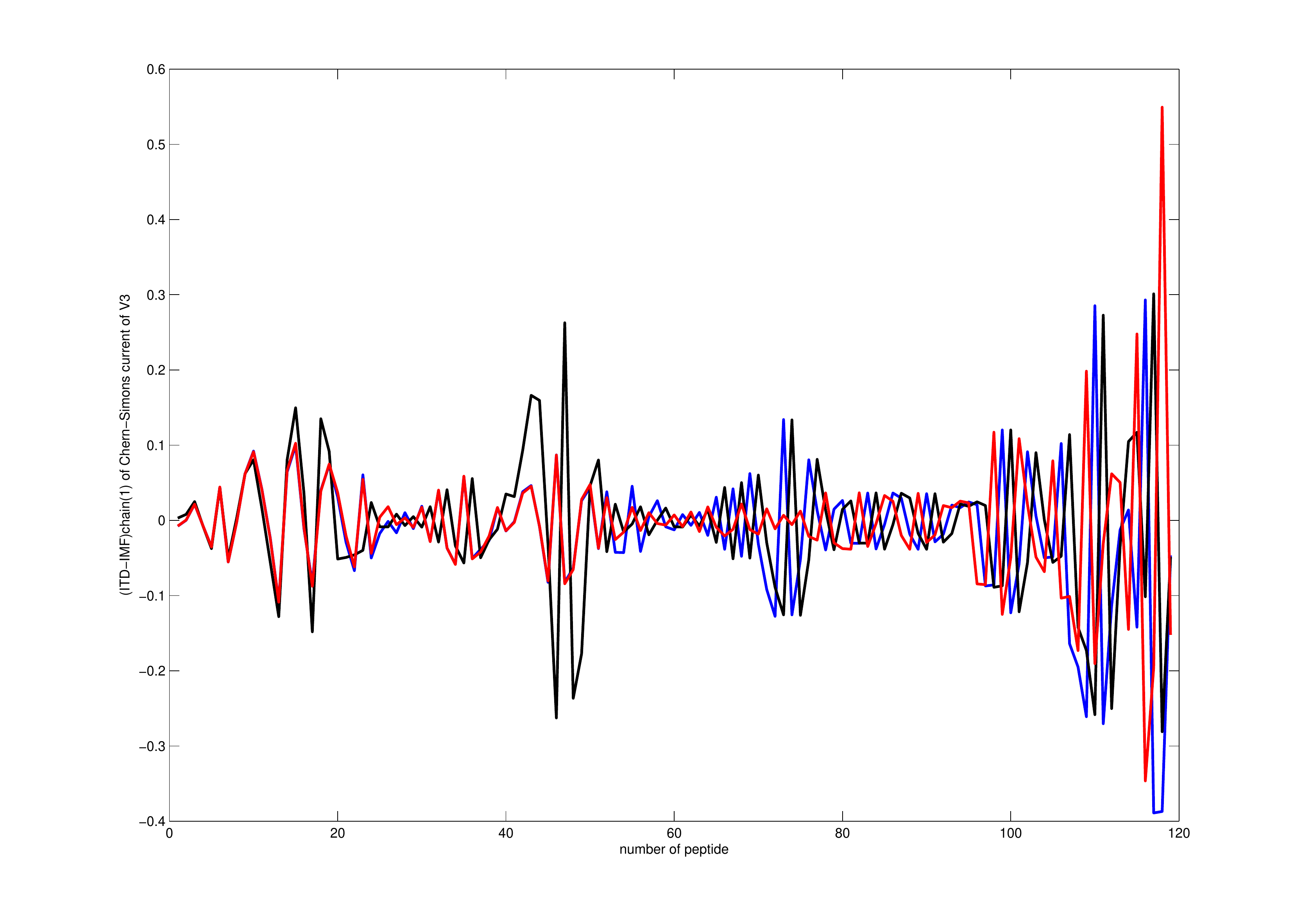}
 \caption{The picture is a plot of the Chern-Simons current of $\ICHAIN(1)$ of the V3 loop region of HIV with 3 different genetic codes.
 The red line is the V3 loop region of HIV-1 isolated HN46-clP from the Viet Nam envelope glycoprotein gene, partial cds. The isolation is on 05-MAR-2013. The black line is HIV-1 isolated HN46-clQ and the blue line is HN46-clR from the Viet Nam envelope glycoprotein gene, partial cds. The details of all the genetic codes and number of peptides is shown in Appendix A.
   }
\label{fig10}
\end{figure}

\begin{figure}[!t]
\centering
\includegraphics[width=0.75\textwidth]{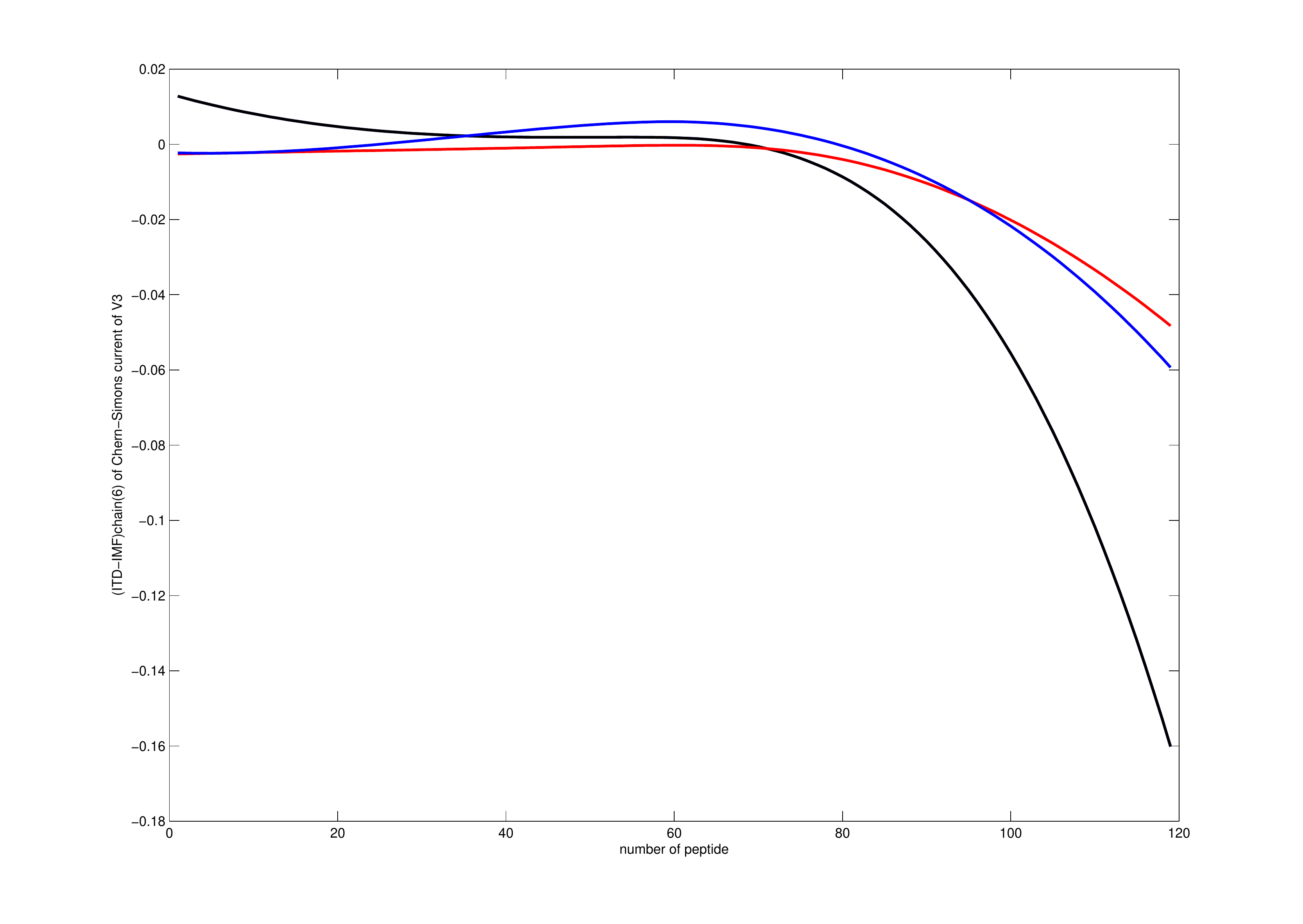}
\caption{The picture is a plot of the Chern-Simons current of $\ICHAIN(6)$ of the V3 loop region of HIV with 3 different genetic codes.
 The red line is a v3 loop region of HIV-1 isolated HN46-clP from the Viet Nam envelope glycoprotein gene, partial cds. The isolation is on 05-MAR-2013. The black line is HIV-1 isolated HN46-clQ  and the blue line is HN46-clR from the Viet Nam envelope glycoprotein gene, partial cds. The details of all the genetic codes and the number of peptides is shown in Appendix A. }
\label{fig11}\end{figure}

\begin{figure}[!t]
\centering
\includegraphics[width=0.75\textwidth]{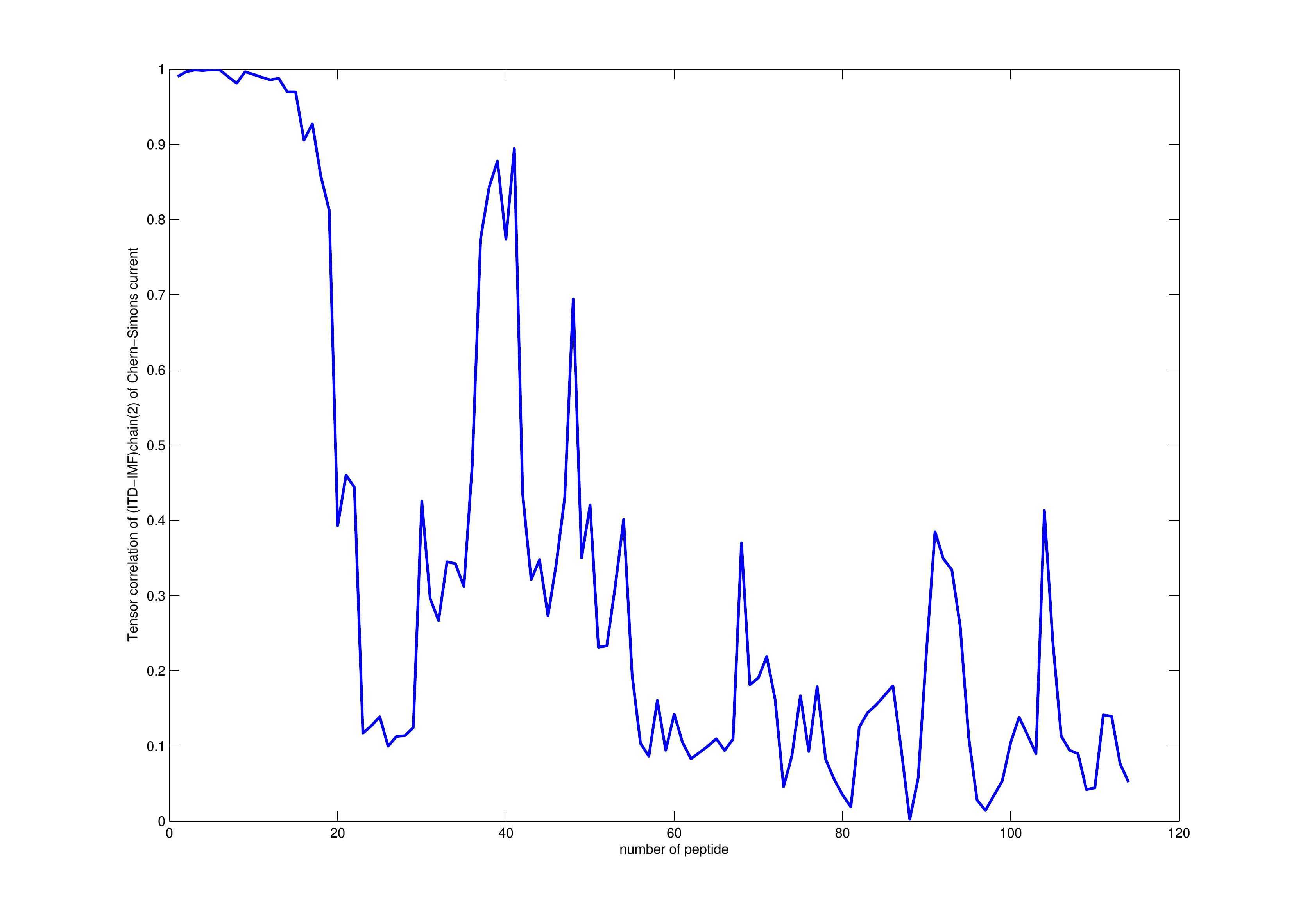}
\caption{The picture  is a plot of a tensor correlation of the Chern-Simons current of $\ICHAIN(1)$ of the V3 loop region of HIV with 3 different genetic codes.
  }
\label{fig11_1} \end{figure}

\begin{figure}[!t]
\centering
\includegraphics[width=0.75\textwidth]{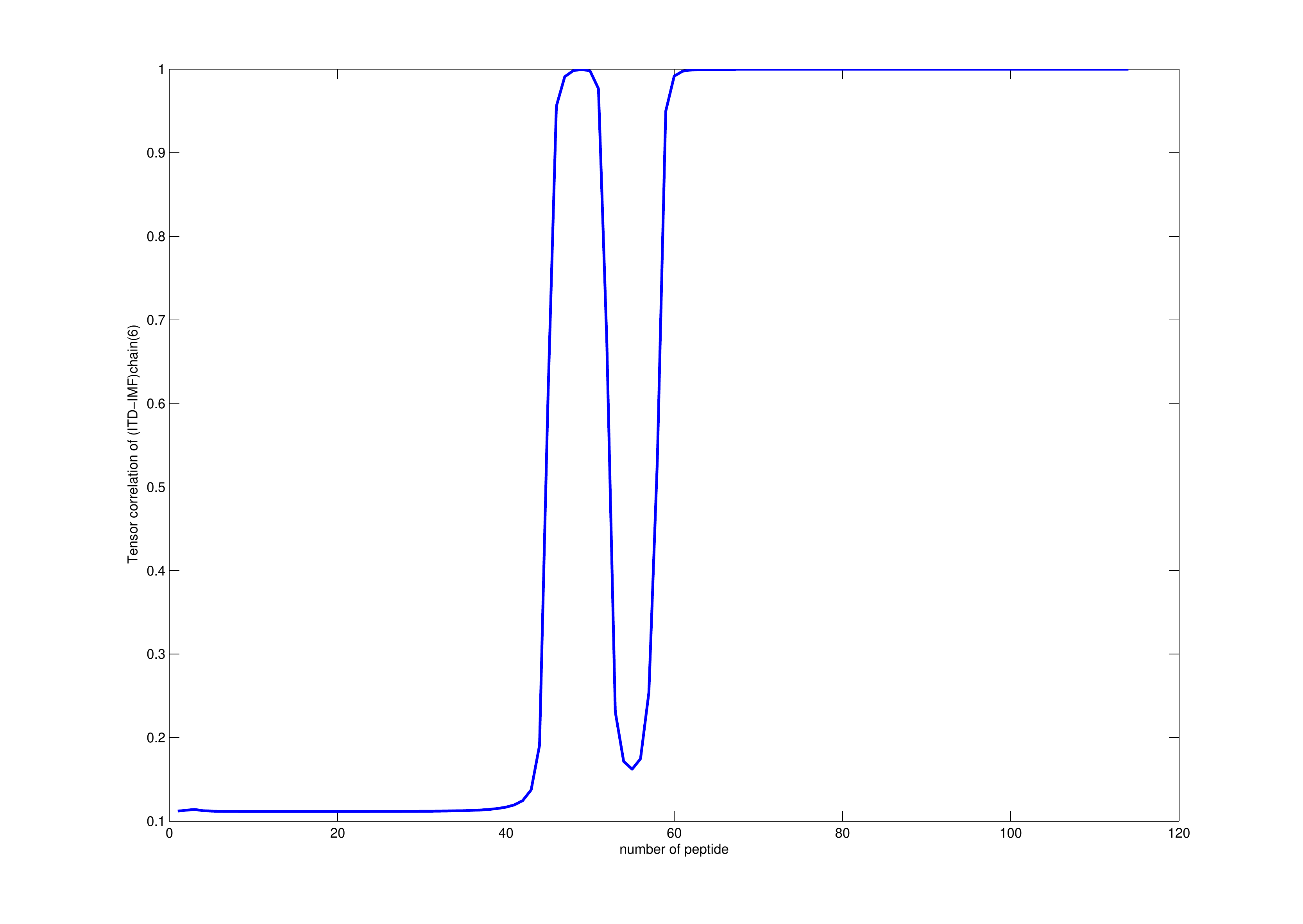}
\caption{
The picture    is a plot of a tensor correlation of the Chern-Simons current of $\ICHAIN(6)$ of the V3 loop region of HIV with 3 different genetic codes.
  }
\label{fig12}\end{figure}

The result of the antigene drift in the V3 gene of HIV is  obtained by using $\ICHAIN(1)-\ICHAIN(6)$.
 One the result of  $\ICHAIN(1)$ of V3 first 120 peptide sequence can be found in Fig. \ref{fig10}
and $\ICHAIN(6)$ of the V3 gene can be found in Fig. \ref{fig11}. The genetic drift is computed by using the tensor correlation of $\ICHAIN(1)-\ICHAIN(6)$. We find that the degree of evolution of the V3 genes is more than the CD gene of the host cell because there exist more local minimum points of the tensor correlation in  $\ICHAIN(1)$ of the V3 gene (Fig. \ref{fig11}) than in the CD gene. The tensor correlation of the Chern-Simons current of $\ICHAIN(1)$ of the V3 loop region of HIV with 3 different genetic codes is plotted in Fig. \ref{fig11_1}.
The tensor correlation of $\ICHAIN(6)$ of the V3 gene can be found in Fig. \ref{fig12}.

\begin{figure}[!t]
\centering
\includegraphics[width=0.75\textwidth]{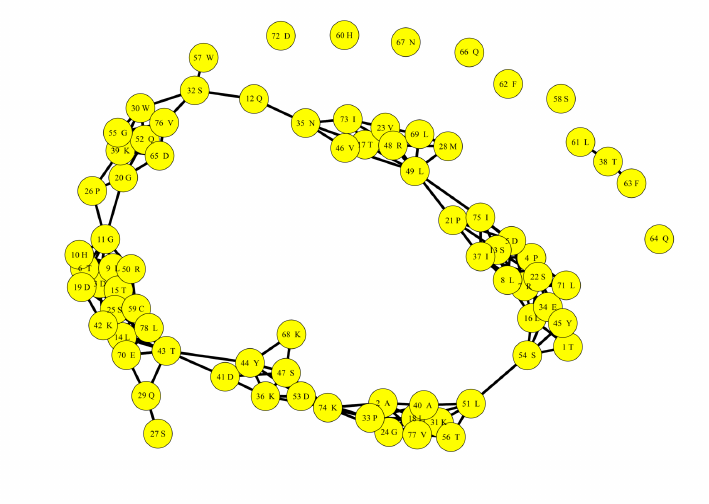}
\caption{In this figure, we show the 1st states of the tensor network of the CD4 coupling with V3 genes from 36 states.  The sequence of the CD4 gene of the Lepus saxatilis gene is TADPDTRLLHGQSLTLTLDGPSVGSPSMQWKSPENKI$\mathbf{T}_{38}\mathbf{K}_{39}\mathbf{A}_{40}$DKTYYVSRLRLQDSGTWSCH$\mathbf{L}_{61}\mathbf{F}_{62}\mathbf{F}_{63}$QDQNKLELDIKIVVL. With the position of the genetic shift at 38, it is connected with 61,63 of the tensor network into the same clusters of CD4 genes of the genetic variation position. }
\label{state1} \end{figure}

\begin{figure}[!t]
\centering
\includegraphics[width=0.48\textwidth]{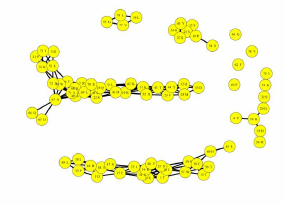}
\includegraphics[width=0.48\textwidth]{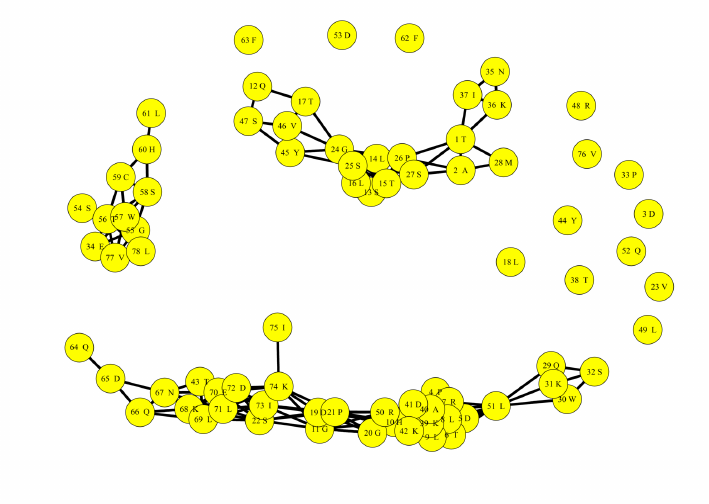}
\caption{On the left panel, we demonstrate the 17th state of the tensor network of the CD4 coupling with V3 genes from 36 states. On the right panel, we demonstrate the 18th state of the tensor network of the CD4 coupling with  V3 genes from all correlated 36 states between 2 genes of all imf of TADPDTRLLHGQSLTLTLDGPSVGSPSMQWKSPENKI$\mathbf{T}_{38}\mathbf{K}_{39}\mathbf{A}_{40}$DKTYYVSRLRLQDSGTWSCHLFFQDQNKLELDIKIVVL. With the position of the genetic shift at 38 through 40, a topology of the tensor network can be broken into 3 different clusters of the CD4 genes.}
\label{fig20} \end{figure}

 We use the gene tensor network to define 36 coupling states of the CD4 gene with the V3 loop gene. We found that in each coupling state, the topology of the gene tensor network is different. There exists in some states only one cluster of the genes and some states are composed of 3 clusters and a small portion of amino acid as a separated cluster. The first gene tensor correlation state is shown in Fig. \ref{state}.
In this state, we found that genes have the genetic variation of different curvature which we measured by using the Chern-Simons current of the gene from different clusters that have no genetic variation. In the 1st coupling state of the V3 gene and the CD4 gene, we found that the amino acids numbered 38, 61 and 63 are connected to each other and form connected components separately from the main connected component. In these components, all genes have genetic shifts with different curvatures in 3 species of genes.
In our time series analysis of the Chern-Simons current for biology, we successfully detected the gene evolution also in 17th-20th coupling state of CD4. The evolution of the genetic shift can be measured by a minimum point of a tensor correlation between 78 genes freely with the $78\times 78$ tensor product of the correlation between the genetic codes. The topology of the tensor network of the peptides numbered 38-40 is shifted from T to I and to I between 3 species  which means that the network of the node number 38 will not be connected with the node number 37 and we must use different clusters (see Figs. \ref{state1} and \ref{fig20}).

In the CD4 gene, we have found in the result of the tensor correlation of $\ICHAIN(1)$ that the minimum point is in the position 39 in a time series data and in the same position in the genetic variation of the CD4 gene from the letter K in the Lepus saxatilis gene to the letter E in the Sylvilagus bachmani gene.

In the equilibrium of the evolution states, there is no change in the curvature of the docking between the host cell receptor and the viral glycoprotein. Therefore the difference between the Chern-Simons currents will be zero in all the positions of docking in the gene sequence. In our data analysis, we found that in CD8 genes,
 the difference in 3 curvatures which we measured by using $\ICHAIN(1)-\ICHAIN(6)$ in 3 species is not zero. It is not in equilibrium states, therefore, we imply that the HIV gene still has an evolution and the coupling between the docking curvature between the CD4 and the V3 loop with the non-equilibrium state is not zero.

\section{Discussion and Conclusions}

It is possible to solve the  central dogma paradox in biology by using a new sheaf cohomology theory  to explain
 why undocking states of proteins induce a docking mechanism  derived from the curvature of a  "gravitational field" analogue  given by the  Chern-Simons current for the gene expression. 
 Specifically, by the  Chern-Simons current,  one can plot all genes in the active area and in the trash area of DNA: this means that    a    data analysis can be performed by standard time series  instead of using probabilistic Bayesian graphical models over a pre-defined flat space of  alphabet scalar field, adopted in standard  computer science algorithms.
 
In this paper,  we give a new definition to  living organism superspace  taking into account a  unoriented supermanifold with sheaf cohomology.
The central dogma is an adjoint functor over the group action with a representation of the alphabet code  with left and right symmetry in a supersymmetric  model. Based on the protein docking in non-equilibrium state, we use the BV-cohomology theory to explain the  trash DNA where are  transposon and retrotransposon components of inactive gene. This area is defied as the transition state of active gene with the triplet state of spin orbit coupling to the transition state in hidden direction of the gene expression. The transposon induces a  loop space in the time series data in which unoriented states of knot and twistors appear from the operation of insert and delete part of the gene. The retrotransposon is an example of HIV reversed transcription process in this model. We give a simple model of moduli state space model for the transition of all states and hidden states of geneon, anti-geneon to transposon and retrotransposon. Furthermore, we construct the wave function of all gene components in the trash DNA and in its active part by using differential 2 forms and modified Dirac operators. It is a Seiberg-Witten equation for  biology that gives rise to a unified model of how living organisms can adapt their behavior in a central dogma model.   The energy states of ghost and anti-ghsot fields can be seen  as the central unit of a sheaf cohomology in a  Grothendieck topology. 
The emergence of Chern-Simons current for active gene and  trash DNA  can be explained  by the retroptransposon state in gene.
 To demonstrate this result, we perform  the data analysis by the  $(IMF-ITD)chain(1,n)$ with tensor correlations between the HIV gene tensor network  and the host receptor gene under a predefined Chern-Simons current. The result of the analysis can detect genetic shift and drift  effects of the retrotransposon in HIV useful  for forecasting the evolution of HIV gene receptor. This process gives rise to a  docking curvature due to the evolution of  spinor field  states in a Diophantine equation derived as the boundary condition for solved it.
This approach is useful for explaining the  memory storage as entanglement states induced from the Hopf fibration over the loop space of transposon and retrotransponson state in the spinor field of  time series of junk DNA. This DNA area is rich of information if we can plot it in time series by the  algebra of Chern-Simons current in complex plane as hidden states of inactive gene.

Furthermore, we derived a master equation of viral gene, attached to the host cell gene, by using the BV-cohomology and Chern-Simons current. 
Specifically, it is possible to  define a new value of codon from the  Chern-Simons current for genetic code. This new definition can be interpreted as the behavior of DNA and RNA interacting between   each other. The result of the new definition is tested wby 3 types of mutations of V3 loop region of glycoprotein of HIV virus to show the alphabet of genetic code as a new time series  construction. With this new time series definition, we can calculate the genetic variation pattern in antigene shift and drift from evolution of gene in traditional graph plot by using our new definition of Chern-Simons current for biology.

In particular, this provides an insight into the time-path of evolution of V3 loop region gene from HIV to another type of HIV. The result of calculation is performed by data analysis of   $\ICHAIN(1)-\ICHAIN(6).$ We found that time series of V3 loop gene of HIV virus has only 6 imf. The last imf is imf6,  a trend of the gene. We analyzed 3 genes coming from the different viral glycoprotein from HIV (see Appendix A) and found that the trend of the same gene of protein, for example V3 loop gene of 3 sample of HIV, have a similar trend. However the different gene from the same virus have different trend. The results are confirmed by the application of Chern-Simons  current for the  detection of similarities in gene expression.  We observed the antigene  drift in time series of Chern-Simons current of V3 loop by using tensor correlation analysis \cite{cohomo7} and also antigene shift in CD4 of 3 samples from rabbits. The evidence of antigene shift and antigene  drift can be modeled by using the BV-cohomogy theory. All states of evolution are a coupling between $\alpha$ and $\beta$ co-cycles while a docking state is not in equilibrium. In future work, we will use the new theory of support spinor machine to forecast and explain the new evolution states in the antigene shift and antigene drift states in time series on cell membrane.
 
As said, we found the genetic variation of genetic drift over V3 HIV viral gene and genetic drift over CD4 gene of host cell of 3 types of rabbit. We notice that viruses have power of evolution more than host cell. Although in our data analysis, we can detect the site of genetic variation over CD4 and V3 with tensor correlation of $(\ICHAIN(1))$. The result is correct when we look into the number of peptides of the  alphabet sequence. This methodology can be useful to detect the pattern matching between DNA of human chromosome in the authentication people by using chromosomes from a  family. It can be also a powerful application for  the disease diagnosis  by classifying the pattern of the plots from the target pattern of classification illness type. We hope  that this new methodology could  be useful for Pubmed to give rise to a visualization of  gene by plotting, instead of showing the  sequence of alphabet,  in which it is difficult to  notice patterns  useful for scientific community as standard methodology in the era of Genbank big data of genome of  various organisms chromosomes. 
The plot of gene tensor network can be  suitable to represent a new time series of genetic code. Here,  we  have shown the simple result of computation. In the genome of various organism, we can use the gene tensor network for representing the time series of genetic code instead of showing a flat alphabet for  writing it in text format.

\section*{Acknowledgment}
\addcontentsline{toc}{section}{Acknowledgment}

This article is based upon work from COST Action CA15117  "Cosmology and Astrophysics Network for Theoretical Advances and Training Actions" (CANTATA), supported by COST (European Cooperation in Science and Technology). S. Capozziello is supported by Istituto Nazionale di Fisica Nucleare (INFN). R. Pincak would like to thank the TH division at CERN for hospitality.  K. Kanjamapornkul is supported by the 100th Anniversary Chulalongkorn University Fund for Doctoral Scholarship.  The work is partly supported by VEGA Grant No. 2/0009/16.  
\addcontentsline{toc}{section}{Acknowledgment}

\newpage

\section*{Appendix A. Time Series of  Genetic Code of V3 loop region of HIV Virus}

The most interesting gene is the V3 loop region  on glycoprotein of HIV virus, the mutation from HIV in March 2013 to HIV antigene shift
is analyzed . We start from gene HIV in 2013 the code download PubMed website and the detail of time series of genetic code is as the followings,

\begin{verbatim}


LOCUS       JN987681                 364 bp    DNA     linear   VRL 05-MAR-2013
DEFINITION  HIV-1 isolate HN46_clR from Viet Nam envelope glycoprotein (env)
            gene, partial cds.
ACCESSION   JN987681
VERSION     JN987681.1
KEYWORDS    .
SOURCE      Human immunodeficiency virus 1 (HIV-1)
ORGANISM  Human immunodeficiency virus 1
            Viruses; Retro-transcribing viruses; Retroviridae;
            Orthoretrovirinae; Lentivirus; Primate lentivirus group.
         /product="envelope glycoprotein"
                     /protein_id="AFC88912.1"
                     /translation="QCTHGIKPVVSTQLLLNAANGSLAEGEIIIRSENLTNNAKTIIV
                     HLNKSVEINCIRPSNNTRTRVTLGPGKVFYRTGDIIGDIRKAYCEINGTKWNEVLGQV
                     AGKLKEHFNKTIIFQPPSG"
ORIGIN
        1 tacaatgtac acatggaatt aagccagtgg tatcaactca attgttactc aatgctgcta
       61 atggcagtct cgcagaagga gagataataa tcagatctga aaatctcaca aacaatgcca
      121 aaaccataat agtgcacctt aataaatctg tagaaatcaa ttgtatcaga ccctccaaca
      181 atacaagaac aagggtaact ctaggaccag gaaaagtatt ctatagaaca ggagacatca
      241 taggagatat aagaaaagca tattgtgaga ttaatggaac aaaatggaat gaagttttag
      301 gacaggtagc tggaaaacta aaagagcact ttaataagac aataatcttt caaccaccct
      361 cagg
//


LOCUS       JN987680                 367 bp    DNA     linear   VRL 05-MAR-2013
DEFINITION  HIV-1 isolate HN46_clQ from Viet Nam envelope glycoprotein (env)
            gene, partial cds.

SOURCE      Human immunodeficiency virus 1 (HIV-1)
  ORGANISM  Human immunodeficiency virus 1
            Viruses; Retro-transcribing viruses; Retroviridae;
            Orthoretrovirinae; Lentivirus; Primate lentivirus group.

                     /product="envelope glycoprotein"
                     /protein_id="AFC88911.1"
                     /translation="QCTHGIKPVVSTQLLLHLLAKWQSCREEIIIRSENLTNNAKTII
                     VHFNRSVEINCTRPSNNTRTSINLGPGKVFYRTGDIIGDIRKAYCEINGTKWNEVLGQ
                     VAGKLKEHFNKTIIFQPPSG"
ORIGIN
        1 tacaatgtac acatggaatt aagccagtgg tatcaactca attgttactt catttgcttg
       61 ctaaatggca gtcttgcaga gaagagataa taatcagatc tgaaaatctc acaaacaatg
      121 ccaaaaccat aatagtgcac tttaatagat ctgtagaaat caattgtacc agaccctcca
      181 acaatacaag aacaagtata aatctaggac caggaaaagt attctataga acaggagaca
      241 tcataggaga tataagaaaa gcatattgtg agattaatgg aacaaaatgg aatgaagttt
      301 taggacaggt agctggaaaa ctaaaagagc actttaataa gacaataatc tttcaaccac
      361 cctcagg



LOCUS       JN987679                 361 bp    DNA     linear   VRL 05-MAR-2013
DEFINITION  HIV-1 isolate HN46_clP from Viet Nam envelope glycoprotein (env)
            gene, partial cds.
ACCESSION   JN987679
VERSION     JN987679.1
KEYWORDS    .
SOURCE      Human immunodeficiency virus 1 (HIV-1)
  ORGANISM  Human immunodeficiency virus 1
            Viruses; Retro-transcribing viruses; Retroviridae;
            Orthoretrovirinae; Lentivirus; Primate lentivirus group.

                     /product="envelope glycoprotein"
                     /protein_id="AFC88910.1"
                     /translation="QCTHGIKPVVSTQLLLSAANGSLAEEEIIIRSENLTDNAKTIIV
                     HLNKSVEINCTRPYNTETKVTRGPGKVYYRTGKITGDIRKAYCEINGTKWNEVLGQVA
                     GKLKEHFNKTIIFQPPSG"
ORIGIN
        1 tacaatgtac acatggaatt aagccagtgg tatcaactca attgttactc agtgctgcta
       61 atggcagtct cgcagaagaa gagataataa tcagatctga aaatctcaca gacaatgcca
      121 aaaccataat agtgcacctt aataaatctg tagaaatcaa ttgtaccaga ccttacaata
      181 cagaaacaaa ggtaactcga ggaccaggaa aagtatacta tagaacagga aaaatcacag
      241 gagatataag gaaagcatat tgtgagatta atggaacaaa atggaatgaa gttttaggac
      301 aggtagctgg aaaactaaaa gagcacttca ataagacaat aatctttcaa ccaccctcag
      361 g
//



LOCUS       KU845559                 234 bp    DNA     linear   MAM 28-MAR-2016
DEFINITION  Lepus saxatilis CD4 (CD4) gene, partial cds.

                     /product="CD4"
                     /protein_id="AMR44297.1"
                     /translation="TADPDTRLLHGQSLTLTLDGPSVGSPSMQWKSPENKITKADKTY
                     YVSRLRLQDSGTWSCHLFFQDQNKLELDIKIVVL"
ORIGIN
        1 actgctgacc cggacacccg cctgctacac ggacagtcac tgaccctaac cttggatggc
       61 ccctctgtgg ggagcccctc catgcaatgg aagagtccag aaaataaaat cacaaaagcc
      121 gacaagactt actacgtgtc caggctgagg ctccaggaca gtggcacctg gtcctgccac
      181 ctgttcttcc aggaccagaa caaactggaa ttagacataa aaatcgtagt attg
//


LOCUS       KU845557                 234 bp    DNA     linear   MAM 28-MAR-2016
DEFINITION  Sylvilagus bachmani CD4 (CD4) gene, partial cds.

SOURCE      Sylvilagus bachmani (brush rabbit)
  ORGANISM  Sylvilagus bachmani
            Eukaryota; Metazoa; Chordata; Craniata; Vertebrata; Euteleostomi;
            Mammalia; Eutheria; Euarchontoglires; Glires; Lagomorpha;
            Leporidae; Sylvilagus.

                     /product="CD4"
                     /protein_id="AMR44295.1"
                     /translation="TAKPDTRLLHGQSLTLTLDGPSVGSPSIQWKSPENKIIEAGKTH
                     SVSKLRLQDSGTWSCHLSFQDQNTLELDIKIVVL"
ORIGIN
        1 actgccaaac cagacacccg cctgctacat ggacagtcac tgaccctaac cttggatggc
       61 ccctctgtgg ggagcccctc catacaatgg aagagtccag aaaataaaat catagaagcc
      121 ggcaagactc actctgtgtc caagctgagg ctccaggaca gtggcacctg gtcctgccac
      181 ctgtccttcc aggaccagaa cacactggag ttagacataa agatcgtagt gttg
//

LOCUS       KU845556                 234 bp    DNA     linear   MAM 28-MAR-2016
DEFINITION  Oryctolagus cuniculus algirus CD4 (CD4) gene, partial cds.
ACCESSION   KU845556
VERSION     KU845556.1
KEYWORDS    .
SOURCE      Oryctolagus cuniculus algirus
  ORGANISM  Oryctolagus cuniculus algirus
            Eukaryota; Metazoa; Chordata; Craniata; Vertebrata; Euteleostomi;
            Mammalia; Eutheria; Euarchontoglires; Glires; Lagomorpha;
            Leporidae; Oryctolagus.

                     /product="CD4"
                     /protein_id="AMR44294.1"
                     /translation="TANPNTRLLHGQSLTLTLEGPSVGSPSVQWKSPENKIIETGKTC
                     SMPKLRLQDSGTWSCHLSFQDQNKLELDIKIVVL"
ORIGIN
        1 actgccaacc cgaacacccg cctgctacat ggacagtcac tgaccctaac cttggaaggc
       61 ccctctgtgg ggagcccctc cgtgcaatgg aagagtccag aaaataaaat catagaaacc
      121 gggaagactt gctccatgcc caagctgagg ctccaggaca gtggcacctg gtcctgccac
      181 ctgtcgttcc aggaccagaa caaactggag ctagacataa aaatcgtagt gttg
//

\end{verbatim}


\begin{thebibliography}{10}
	\providecommand{\url}[1]{#1}
	\csname url@samestyle\endcsname
	\providecommand{\newblock}{\relax}
	\providecommand{\bibinfo}[2]{#2}
	\providecommand{\BIBentrySTDinterwordspacing}{\spaceskip=0pt\relax}
	\providecommand{\BIBentryALTinterwordstretchfactor}{4}
	\providecommand{\BIBentryALTinterwordspacing}{\spaceskip=\fontdimen2\font plus
		\BIBentryALTinterwordstretchfactor\fontdimen3\font minus
		\fontdimen4\font\relax}
	\providecommand{\BIBforeignlanguage}[2]{{%
			\expandafter\ifx\csname l@#1\endcsname\relax
			\typeout{** WARNING: IEEEtran.bst: No hyphenation pattern has been}%
			\typeout{** loaded for the language `#1'. Using the pattern for}%
			\typeout{** the default language instead.}%
			\else
			\language=\csname l@#1\endcsname
			\fi
			#2}}
	\providecommand{\BIBdecl}{\relax}
	\BIBdecl
	
	
	
	\bibitem{Stanley}
	H.~Stanley, V.~Afanasyev, L.~Amaral, S.~Buldyrev, A.~Goldberger, S.~Havlin,
	H.~Leschhorn, P.~Maass, R.~Mantegna, C.-K. Peng, P.~Prince, M.~Salinger,
	M.~Stanley, and G.~Viswanathan, ``Anomalous fluctuations in the dynamics of
	complex systems: From dna and physiology to econophysics,'' \emph{Physica A:
		Statistical Mechanics and its Applications}, vol. 224, no. 1-2, pp. 302--321,
	1996.
	
	\bibitem{genetic_code}
	G.~Gamow, ``Possible relation between deoxyribonucleic acid and protein
	structures [19],'' \emph{Nature}, vol. 173, no. 4398, p. 318, 1954.
	
	\bibitem{virus_a}
	M.~Rossmann and J.~Johnson, ``Icosahedral rna virus structure,'' \emph{Annual
		Review of Biochemistry}, vol.~58, pp. 533--573, 1989.
	
	\bibitem{virus_gene}
	Y.~Bao, Y.~Gao, Y.~Shi, and X.~Cui, ``Dynamic gene expression analysis in a
	h1n1 influenza virus mouse pneumonia model,'' \emph{Virus Genes}, pp. 1--10,
	2017.
	
	\bibitem{simon}
	E.~Witten, ``2 + 1 dimensional gravity as an exactly soluble system,''
	\emph{Nuclear Physics, Section B}, vol. 311, no.~1, pp. 46--78, 1988.
	
	\bibitem{witten_int}
	E. Witten, ``{Notes On Supermanifolds and Integration},'' 2012 , arXiv:1209.2199 [hep-th].
	
	\bibitem{ribo}
	J.~Gebetsberger and R.~Micura, ``Unwinding the twister ribozyme: from structure
	to mechanism,'' \emph{Wiley Interdisciplinary Reviews: RNA}, vol.~8, no.~3,
	2017.
	
	\bibitem{life}
	K.~Osawa, H.~Urata, and H.~Sawai, ``Chiral selection in oligoadenylate
	formation in the presence of a metal ion catalyst or poly(u) template,''
	\emph{Origins of Life and Evolution of the Biosphere}, vol.~35, no.~3, pp.
	213--223, 2005.
	
	\bibitem{ribo2}
	R.~Scarborough and A.~Gatignol, ``Hiv and ribozymes,'' \emph{Advances in
		Experimental Medicine and Biology}, vol. 848, pp. 97--116, 2015.
	
	\bibitem{alphabet}
	T.~Bonfert and C.~Friedel, ``Prediction of poly(a) sites by poly(a) read
	mapping,'' \emph{PLoS ONE}, vol.~12, no.~1, 2017.
	
	\bibitem{code}
	M. Nirenberg and J. Matthaei,  ``The dependence of cell-free protein synthesis
	in e. coli upon naturally occurring or synthetic polyribonucleotides.''
	\emph{Proceedings of the National Academy of Sciences of the United States of
		America}, vol.~47, pp. 1588--1602, 1961.
	
	\bibitem{code2017}
	M.~José, G.~Zamudio, and E.~Morgado, ``A unified model of the standard genetic
	code,'' \emph{Royal Society Open Science}, vol.~4, no.~3, 2017.
	
	\bibitem{protein}
	G. Gamow, A. Rich, and M. Ycas, ``The problem of information transfer from the
	nucleic acids to proteins.'' \emph{Advances in biological and medical
		physics}, vol.~4, pp. 23--68, 1956.
	
	\bibitem{h1n1_variation}
	A.~Jagadesh, A.~Salam, V.~Zadeh, and G.~Arunkumar, ``Genetic analysis of
	neuraminidase gene of influenza a(h1n1)pdm09 virus circulating in southwest
	india from 2009 to 2012,'' \emph{Journal of Medical Virology}, vol.~89,
	no.~2, pp. 202--212, 2017.
	
	\bibitem{h7n9}
	S.~Park, I.~Lee, J.~Kim, J.-Y. Bae, K.~Yoo, J.~Kim, M.~Nam, M.~Park, S.-H. Yun,
	W.~Cho, Y.-S. Kim, Y.~Ko, and M.-S. Park, ``Effects of ha and na
	glycosylation pattern changes on the transmission of avian influenza a(h7n9)
	virus in guinea pigs,'' \emph{Biochemical and Biophysical Research
		Communications}, vol. 479, no.~2, pp. 192--197, 2016.
	
	\bibitem{supermath}
	M.~Shifman, \emph{Felix Berezin: Life and death of the mastermind of
		supermathematics}, 2007.
	
	\bibitem{superla}
	L.~Beyl, ``Super-lagrangians,'' \emph{Physical Review D}, vol.~19, no.~6, pp.
	1732--1745, 1979.
	
	\bibitem{bv}
	E.~Getzler, ``The Batalin-Vilkovisky cohomology of the spinning particle,''
	\emph{Journal of High Energy Physics}, vol. 2016, no.~6, 2016.
	
	\bibitem{g}
	A.~Sepehri and R.~Pincak, ``The birth of the universe in a new g-theory
	approach,'' \emph{Modern Physics Letters A}, 2017.
	
	\bibitem{alireza}
	
	S. Capozziello,  E. N. Saridakis,  K. Bamba, A Sepehri, F. Rahaman, A. Farag Ali, R. Pincak, A. Pradhan, 
	"Cosmic space and Pauli exclusion principle in a system of M0-branes ", 
	\emph{Int. J. Geom. Meth. Mod. Phys.} 14 (2017), 1750095  
	
	\bibitem{amino}
	F.~Antoneli, M.~Forger, P.~Gaviria, and J.~Hornos, ``On amino acid and codon
	assignment in algebraic models for the genetic code,'' \emph{International
		Journal of Modern Physics B}, vol.~24, no.~4, pp. 435--463, 2010.
	
	\bibitem{anomaly}
	E.~Witten, ``{The "Parity" Anomaly On An Unorientable Manifold},'' \emph{Phys.
		Rev.}, vol. B94, no.~19, p. 195150, 2016.
	
	\bibitem{chern}
	P.~Grassi and C.~Maccaferri, ``Chern-simons theory on supermanifolds,''
	\emph{Journal of High Energy Physics}, vol. 2016, no.~9, 2016.
	
	\bibitem{cdot}
	X.~Dong, M.~Moyer, F.~Yang, Y.-P. Sun, and L.~Yang, ``Carbon dots' antiviral
	functions against noroviruses,'' \emph{Scientific Reports}, vol.~7, no.~1,
	2017.
	
	\bibitem{Massey}
	W.~S. Massey, \emph{A Basic Course in Algebraic Topology}, ser. Undergraduate
	Texts in Mathematics.\hskip 1em plus 0.5em minus 0.4em\relax Springer New
	York, 1997.
	
	\bibitem{e8}
	M.~Green and J.~Schwarz, ``Anomaly cancellations in supersymmetric d = 10 gauge
	theory and superstring theory,'' \emph{Physics Letters B}, vol. 149, no. 1-3,
	pp. 117--122, 1984.
	
	\bibitem{Kolmogorov}
	K.~Kanjamapornkul and R.~Pincak, ``Kolmogorov space in time series data,''
	\emph{Mathematical Methods in the Applied Sciences}, vol.~39, no.~15, pp.
	4463--4483, 2016.
	\bibitem{bio}
	A.M. Turing,  "The Chemical Basis of Morphogenesis" {\em   Philosophical Transactions of the Royal 
		Society of London B} (1952) {\bf 237} (641) 37.
	
	\bibitem{grothen}
	M. Artin,  "Grothendieck topologies" p.133
	Harvard University Press. Cambridge, Mass. (1962).
	
	\bibitem{immune}  A. Cogoli and A. Tschopp,  " Lymphocyte reactivity during spaceflight" ,
	{\em Immunology Today } (1985)  6 no. 1.
	
	\bibitem{coxeter}
	H. S. Mac Donald Coxeter,  "Regular Polytopes" (Third ed.). Dover Publications, New York (1973).
	
	\bibitem{mayer}
	R. Bott, L.W. Tu,   "Differential Forms in Algebraic Topology,"  New  Springer-Verlag, Berlin  (1982).
	
	\bibitem{dogma} 
	H. Freeland Judson, " The Eighth Day of Creation: Makers of the Revolution in Biology", Cold Spring Harbor, Cold Spring Harbor Laboratory Press, New York (1996).
	
	\bibitem{alphabet} D.A. Malyshev,  and et. al.,"  A semi-synthetic organism with an expanded genetic alphabet," {\em Nature} (2014) 509 no. 385.
	
	\bibitem{telomere}
	D. Sadava, D. Hillis, C. Heller, and M. Berenbaum,  "Life: The science of biology. (9th ed.) Sunderland, MA: Sinauer Associates Inc. (2011).
	
	\bibitem{rna}  E. Bernstein,  et. al.," Role for a bidentate ribonuclease in the initiation step of RNA interference ," {\em Nature } (2001) 409 no. 368.
	
	
	
	\bibitem{rna2}  I. Macrae  and et. al.,"Structural basis for double-stranded RNA processing by dicer," {\em Science} (2006) 311 (5758) 195.
	
	
	\bibitem{super}
	S. Capozziello, R. Pincak, and K. Kanjamapornkul,  "Anomaly on Superspace of Time Series Data," {\em Zeitschrift fuer Naturforschung A},  (2017). 72(12), pp. 1077-1091. 
	
	
	\bibitem{ssm}
	K. Kanjamapornkul, R. Pincak, S.   Chunithpaisan, and E.  Bartos," Support spinor machine", {\em  Digital Signal Processing} 70 (2017) 59
	
	
	\bibitem{wald} R. M. Wald. "General relativity", The University of Chicago Press, Chicago 1984. 
	
	\bibitem{ghost}
	L.~Faddeev and V.~Popov, ``Feynman diagrams for the yang-mills field,''
	\emph{Physics Letters B}, vol.~25, no.~1, pp. 29--30, 1967.
	
	\bibitem{witten}
	E.~Witten, ``{Notes On Supermanifolds and Integration},'' 2012, arXiv:1209.2199.
	\bibitem{seiberg}
	S.K. Donaldson,  (1996), "The Seiberg-Witten equations and 4-manifold topology.", Bull. Amer. Math. Soc. (N.S.), 33 (1): 45?70, (1996). 
	
	\bibitem{super}
	A.~Cattaneo and F.~Schutz, ``Introduction to supergeometry,'' \emph{Reviews in
		Mathematical Physics}, vol.~23, no.~6, pp. 669--690, 2011.
	
	\bibitem{super2}
	W.~Siegel, ``Superspace duality in low-energy superstrings,'' \emph{Physical
		Review D}, vol.~48, no.~6, pp. 2826--2837, 1993.
	
	\bibitem{super3}
	M.~Denys, T.~Gubiec, R.~Kutner, M.~Jagielski, and H.~Stanley, ``Universality of
	market superstatistics,'' \emph{Physical Review E - Statistical, Nonlinear,
		and Soft Matter Physics}, vol.~94, no.~4, 2016.
	
	\bibitem{superpoint}
	J.~Huerta and U.~Schreiber, ``{M-theory from the Superpoint},'' "2017".
	
	\bibitem{string5}
	E.~Bartos and R.~Pincak, ``Identification of market trends with string and
	d2-brane maps,'' \emph{Physica A: Statistical Mechanics and its
		Applications}, vol. 479, pp. 57--70, 2017.
	
	\bibitem{Pincak10}
	R.~Pincak and E.~Bartos, ``With string model to time series forecasting,''
	\emph{Physica A: Statistical Mechanics and its Applications}, vol. 436, pp.
	135--146, 2015.
	
	\bibitem{pincak11}
	R.~Pincak, ``The string prediction models as invariants of time series in the
	forex market,'' \emph{Physica A: Statistical Mechanics and its Applications},
	vol. 392, no.~24, pp. 6414--6426, 2013.
	
	\bibitem{pincak12}
	M.~Bundzel, T.~Kasanický, and R.~Pincak, ``Using string invariants for
	prediction searching for optimal parameters,'' \emph{Physica A: Statistical
		Mechanics and its Applications}, vol. 444, pp. 680--688, 2016.
	
	\bibitem{Pincak13}
	D.~Horvath and R.~Pincak, ``From the currency rate quotations onto strings and
	brane world scenarios,'' \emph{Physica A: Statistical Mechanics and its
		Applications}, vol. 391, no.~21, pp. 5172--5188, 2012, cited By 6.
	
	\bibitem{cohomo7}
	K.~Kanjamapornkul, R.~Pincak, and E.~Bartos, ``The study of thai stock
	market across the 2008 financial crisis,'' \emph{Physica A: Statistical
		Mechanics and its Applications}, vol. 462, pp. 117--133, 2016.
	
	\bibitem{v3}
	M.~Yokoyama, S.~Naganawa, K.~Yoshimura, S.~Matsushita, and H.~Sato,
	``Structural dynamics of hiv-1 envelope gp120 outer domain with v3 loop,''
	\emph{PLoS ONE}, vol.~7, no.~5, 2012.
	
	\bibitem{viral_replication}
	S.~Artusi, R.~Perrone, S.~Lago, P.~Raffa, E.~Di~Iorio, G.~Pala, and
	S.~Richter, ``Visualization of dna g-quadruplexes in herpes simplex virus
	1-infected cells,'' \emph{Nucleic Acids Research}, vol.~44, no.~21, pp.
	10\,343--10\,353, 2016.
	
	\bibitem{knot}
	D.~Gaiotto and E.~Witten, ``Knot invariants from four-dimensional gauge
	theory,'' \emph{Advances in Theoretical and Mathematical Physics}, vol.~16,
	no.~3, pp. 935--1086, 2012.
	
	\bibitem{cd4_drawing}
	\BIBentryALTinterwordspacing
	S.-Q. Liu, S.-X. Liu, and F.-X. Fu, ``Molecular motions of human HIV-1 gp120
	envelope glycoproteins,'' \emph{Journal of Molecular Modeling}, vol.~14,
	no.~9, pp. 857--870, 2008. [Online]. Available:
	\url{https://www.hiv.lanl.gov/content/sequence/HIV/REVIEWS/Sodroski.html}
	\BIBentrySTDinterwordspacing
	
\end{thebibliography}



\end{document}